\documentclass[12pt,preprint]{aastex}

\usepackage{graphicx}
\usepackage{epsfig}
\usepackage{subfigure}

\usepackage{eqsecnum}

\usepackage{gvbmacro}

\def \aaps{Astron.\ Astrophys.\ Supp.\ }
\def \aap{A\&A }
\def \apjl{ApJ}
\def \apjs{ApJS}
\def \apj{ApJ}

\def \mnras{MNRAS}
\def \prl{Phys.~Rev.~Lett.,}

\def \nat{Nature\ }
\def\msun{{\,M_\odot}}

\def\zsun{{\,Z_\odot}}

\def\alt{\raise0.3ex\hbox{$\;<$\kern-0.75em\raise-1.1ex\hbox{$\sim\;$}}}
\def\agt{\raise0.3ex\hbox{$\;>$\kern-0.75em\raise-1.1ex\hbox{$\sim\;$}}}

\newcommand{\bw}{\begin{widetext}}
\newcommand{\ew}{\end{widetext}}

\newcommand{\lsim}{\,\rlap{\raise 0.35ex\hbox{$<$}}{\lower 0.7ex\hbox{$\sim$}}\,}
\newcommand{\gsim}{\,\rlap{\raise 0.35ex\hbox{$>$}}{\lower 0.7ex\hbox{$\sim$}}\,}

\begin{document}

%

\title{A Unified Model of the Fermi Bubbles, Microwave Haze, and Polarized Radio Lobes: Reverse Shocks in the Galactic Center's Giant Outflows}

\author{\sc
  Roland M.~Crocker\altaffilmark{1,5},
  Geoffrey V.~Bicknell\altaffilmark{1}, 
  Andrew M.~Taylor\altaffilmark{2,6},
and  Ettore Carretti\altaffilmark{3,4}
}
\altaffiltext{1}{Research School of Astronomy and Astrophysics, Australian National University, Canberra, Australia}
\altaffiltext{2}{Dublin Institute of Advanced Studies, Dublin, Ireland}
\altaffiltext{3}{INAF - Osservatorio di Cagliari, Via della Scienza 5, 09047 Selargius, Italy}
\altaffiltext{4}{CSIRO Astronomy \& Space Science, Marsfield, N.S.W., Australia}
\altaffiltext{5}{Future Fellow}
\altaffiltext{6}{Schr{\" o}dinger Fellow}


\date{\today}

\begin{abstract}
The Galactic Center's giant outflows are manifest in three different, non-thermal phenomena: 
i) the  hard-spectrum,  $\gamma$-ray `Fermi Bubbles'  emanating from the nucleus and extending to $|b| \sim 50^\circ$; 
ii) the hard-spectrum, {\it total-intensity} microwave ($\sim 20-40$ GHz) `Haze' extending to $|b| \sim 35^\circ$ in the lower reaches of the Fermi Bubbles; and
iii) the steep spectrum, {\it polarized}, `S-PASS' radio ($\sim$ 2-20 GHz)  Lobes that envelop the Bubbles and  extend to  $|b| \sim 60^\circ$.
We find that the nuclear outflows inflate a genuine bubble in each Galactic hemisphere that
has the classical structure, working outwards, of reverse shock, contact discontinuity, and forward shock.
Expanding into the finite  pressure of the halo and given  appreciable cooling and gravitational losses, 
the contact discontinuity of each bubble is now expanding only very slowly.
We find observational signatures in both hemispheres of
giant, reverse shocks at heights of $\sim$ 1 kpc above the nucleus; their presence ultimately explains all three of the non-thermal
phenomena mentioned above.
Synchrotron emission from shock-reaccelerated cosmic-ray electrons explains the spectrum, morphology, and vertical extent of  the microwave Haze and the polarized radio Lobes.
Collisions between shock-reaccelerated hadrons and denser gas in 
cooling condensations that form inside the contact discontinuity account for most of the Bubbles' $\gamma$-ray emissivity.
Inverse Compton emission from primary electrons contributes at the
10-30\% level.
Our model suggests that the  Bubbles are signatures of a comparatively weak but sustained nuclear outflow 
driven by Galactic center star formation over $\gsim$ few $\times 10^8$ yr.
\end{abstract}


\maketitle


\section{Introduction}
\label{sctn_intro}

\subsection{Overview}

The Fermi Bubbles are enigmatic, giant $\gamma$-ray structures emanating north and south from the  nucleus of the Milky Way \citep{Dobler2010,Su2010,Su2012,Yang2014,Ackermann2014} and extending to heights above the Galactic plane approaching 8 kpc.
Features coincident with the Bubbles or some component of the Bubbles have been discovered in {\it total intensity} radio and microwave emission, {\it polarized} radio and microwave emission, and X-ray 
emission \citep{Finkbeiner2004,Dobler2008,Dobler2012,Ade2012,Carretti2013,Bland-Hawthorn2003,Kataoka2013,Tahara2015}.

The energy source of the Bubbles remains in contention as does the $\gamma$-ray emission process that illuminates them.
Broadly speaking, the Bubbles might be the signatures of a fairly recent (within $\sim$ Myr timescale) explosive outburst  \citep{Su2010,Zubovas2011,Mertsch2011,Zubovas2012,Guo2012,Yang2012,Fujita2013,Barkov2014,Wardle2014,Mou2014} from the central supermassive black hole with their $\gamma$-ray emission originating in inverse Compton (IC) collisions of a non-thermal electron population accelerated in association with this activity.
Alternatively, the Bubbles may result from the integrated effect of more-or-less secular processes associated with the inner Galaxy \citep{Thoudam2013} or Galactic nucleus, e.g., tidal disruption events that occur there every $10^4 -10^5$ year \citep{Cheng2011,Cheng2015} or the concentrated and sustained star-formation activity  \citep[e.g.,][]{Lacki2013} that occurs throughout the nuclear region (inner 200-300 diameter region around the black hole).

Invoking sustained nuclear processes as the origin of the Bubbles complements the possibility that their 
$\gamma$-ray emission originates not from IC emission but from  hadronic collisions experienced by a population of cosmic-ray protons and heavier ions \citep{Crocker2011,Crocker2012,Crocker2014}.  
Such particles
 lose energy much more slowly than the high-energy electrons requisite for IC $\gamma$-ray production, 
thereby allowing the total current energy content of the Bubbles to be accumulated more gradually from relatively less powerful processes such as star-formation 
(the Eddington luminosity of the Galaxy's 4 million solar mass black hole is $5 \times 10^{44}$ erg/s;  in contrast,
the Milky Way's nuclear star formation releases a mechanical power of $\sim 3 \times 10^{40}$ erg/s: see below).

At present, there are strengths and weaknesses to either of these broad scenarios.
IC scenarios share the attractive feature that the very same  electron population that putatively supplies the IC $\gamma$-rays can, for magnetic field strengths approaching 10 $\mu$G, also supply the hard microwave Haze emission via synchrotron radiation \citep{Dobler2010,Su2010,Ackermann2014}.
On the other hand, if described by a single power law, synchrotron emission from this population of electrons would {\it not} explain the comparatively steep radio spectrum 
measured between 2.3 and 23 GHz for the Bubbles' polarized, radio lobe counterparts  \citep{Carretti2013}.
Furthermore,
the fact that the $\gamma$-ray spectrum of the Bubbles does not steepen \citep{Ackermann2014} -- seems, in fact, to harden \citep{Yang2014,Selig2014} -- with increasing Galactic latitude is difficult to explain naturally within IC scenarios given the attenuation of the optical and infrared background relative to the CMB as the distance from the Galaxy increases.

Another apparent weakness of many explosive, AGN-related scenarios is that they require that we are viewing the Bubbles at a privileged time at most a few million years since
the outburst that created them, with the Bubbles currently undergoing rapid expansion at speeds $\gsim 2000$ km/s.
Rapid expansion is required in order either to transport the $\gsim$ TeV electrons (required to produce $\gsim$ GeV $\gamma$-rays via IC on the CMB) 
from the nucleus to the Bubbles' peripheries 
within their $\lsim 10^6$ yr loss times (in which case speeds $\gsim 7000$ km/s $\simeq 0.02 \times c$ are actually required) or
to have sufficiently energetic shocks at the expanding edges of the Bubbles that cascading turbulence injected there may energise similarly energetic electrons in situ \citep{Mertsch2011}.
Yet X-ray observations of the Bubbles have only indicated weak features associated with the Bubble edges \citep{Su2010,Kataoka2013,Fang2014,Tahara2015}, implying rather mild expansion speeds, $\lsim 300$ km/s (close to the sound speed).
Closer to the base of the Bubbles, 
data obtained in
UV absorption observations by \citet{Fox2014} along the $(l,b) =(10.4^\circ, +11.2^\circ)$ sightline (to quasar PDS456) 
reveal cool ($\sim 10^4$ K) gas entrained into a biconical, $\sim 900$ km/s  nuclear outflow 2.5-4.0 Myr ago.

Moreover, there are other indications that the Bubbles are rather long-lived structures:
\begin{enumerate}
\item
The fact that both Bubbles lean away from the $l = 0$ meridian towards Galactic west and, moreover, there are extended plumes of polarized, 2.3 GHz emission leaving north west (south west) from the high-latitude regions of the north (south) bubble \citep{Carretti2013} suggests that the structures are  distorted by a pressure gradient within the Galactic halo set up by  the motion of the Milky Way through the local medium towards Andromeda.
The $\sim 50$ km/s relative speed \citep{van der Marel2012} of this motion implies a rather gentle effect that requires long timescales/low Bubble gas velocities to take effect, 
consistent with the $\sim 100-150$ km/s flow velocities we determine below for  gas in the Bubbles
\citep[cf.][where it was shown that an unreasonably large speed of the Galaxy through the local medium  was required to produce the Bubbles' distortion 
 in the case that they are expanding rapidly due to AGN activity]{Yang2012}.

\item
The $\sim 100$ km/s rotation of the Bubbles' waist in the plane is apparently 
inscribed into giant radio ridges on the near surface of the Bubbles \citep{Carretti2013}, again requiring that the structures not grow too quickly (here meaning $\lsim 10^7$ yr).

\item
The steep spectrum of the Bubbles' polarized radio counterparts between 2.3 and 23 GHz suggests that electrons emitting in this frequency range {\it have} 
had time to cool, mix and accumulate; this requires $\gsim 3 \times 10^7$ year for a reasonable  magnetic field amplitude.

\item
Finally, observations of UV absorption features towards background quasar and AGB stars have established the existence of individual parcels of ionised gas at low Galactic longitude but varying latitude \citep{Keeney2006,Zech2008}.
Kinematic analysis points to this gas participating in a nuclear fountain flow.
Individual gas parcels
seem to  have been launched from the nucleus at different times in the past ranging from 20-50 Myr to more than 800 Myr.
The parcels are all
now
 either rising to or falling from a maximum height of $\sim 12$ kpc,  similar to the vertical extent of the Bubbles.
Some of this material  \citep{Zech2008} appears significantly super-solar in metallicity. 
This phenomenology points to the existence of a long-standing, star-formation-driven, nuclear fountain; as we expand upon later in the paper, we believe this material is condensing in the shell (R3 in Figure~\ref{fig_BubbleSchematic}) leading up to the contact discontinuity in each bubble.

\end{enumerate}

Turning now to the star-formation-driven hadronic model \citep{Crocker2011,Crocker2012,Crocker2014}, this has the appealing feature that the power {\it currently} being delivered into cosmic ray ions 
 by nuclear star-formation is elegantly sufficient to explain the Bubbles' (putatively hadronic) $\gamma$-ray luminosity in steady state.
On the other hand, in the original instantiation of this scenario, the low density of target gas nuclei in the Bubble plasma implied a very long $pp$ loss time of a $\gsim$ few $\times$ Gyr, this being the timescale required to establish a steady state in the first place (i.e., approximately the timescale required to accumulate a sufficient total energy in cosmic ray ions in the Bubbles to power their instantaneous $\gamma$-ray luminosity given the low gas density).
Some of us have recently shown, however, that accumulation of hadronic cosmic rays into gas condensations 
formed by local thermal instability in the Bubbles can significantly reduce the timescale associated with the hadronic scenario to $\sim$ few $\times 10^8$ yr \citep{Crocker2014}.
Another point concerning  the hadronic scenario is that although secondary electrons (and positrons) are a natural concomitant of hadronic $\gamma$-ray production, it is difficult to explain the spectrum of the microwave Haze with such secondaries since IC and synchrotron losses would be expected, in a single zone model, to steepen the steady-state secondary distribution such that it is inconsistent with the hard Haze emission \citep[e.g.,][]{Ackermann2014,Cheng2014b}.
It seems, then, that even the hadronic scenario requires primary electrons at some level to explain the totality of Giant Outflow phenomenology.

Indeed, the power in  cosmic ray {\it primary} electrons accelerated in concert with nuclear star formation is a good match to the Bubbles' 2.3 GHz synchrotron luminosity \citep{Carretti2013}.
The complement to these points is that the nucleus is {\it underluminous} in both diffuse $\gamma$-ray and radio continuum emission given how much star formation occurs there \citep{Crocker2011a,Crocker2011b,Crocker2012} and this seems to be {\it because} both cosmic-ray electrons and ions are advected out of the region in the large-scale outflow before they can radiate.
Thus, these particles radiate on the size scale of the Bubbles into which the nuclear outflow delivers them.
As we show below, however, the passage from nucleus to Bubbles is  not a smooth journey for these particles.

\subsection{Outline of model}

\begin{figure}[hb!]
\centering
\includegraphics[width = 1.2 \textwidth]{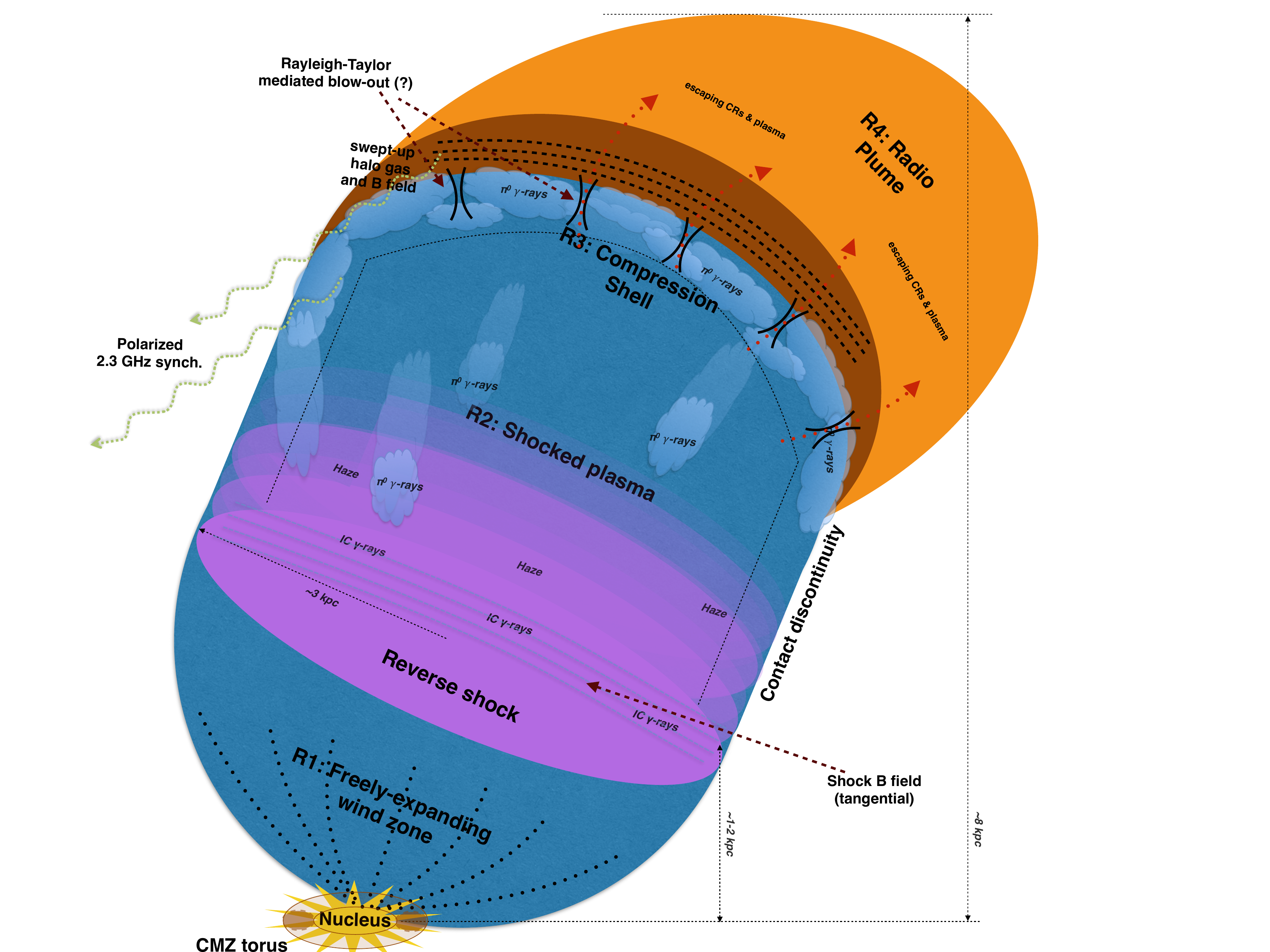}
\caption{Schematic showing the main features of the north bubble according to our model.}
\label{fig_BubbleSchematic}
\end{figure}

Here we briefly explain our model for the Bubbles' structure and non-thermal emission phenomenology.
A schematic showing the configuration of the northern bubble according to our model is shown in Figure~\ref{fig_BubbleSchematic} for reference.

\begin{enumerate}

\item Co-entrained hot plasma, magnetic fields, and cosmic rays are energised within the central $\sim 100$ pc (in radius) region of the Galactic nucleus.
The pressure gradient supplied by these interstellar medium (ISM) phases (with the hot plasma dynamically dominant) accelerates a nuclear wind that quickly 
reaches $\gsim 500$ km/s.
The dense and massive molecular gas torus circumscribing the injection region collimates this outflow and directs it perpendicular to the plane.
Shocks driven into the hot interstellar medium, which is more dense at low latitudes, may account for the X-ray edges observed by \citet{Bland-Hawthorn2003} . 
Radiative shocks driven into atomic and/or molecular gas may also contribute.

\item The Bubbles are inflated by the injection of hot plasma in the nuclear outflow over timescales of at least a few $\times 10^8$ year.
Characteristic scales for the mean mechanical power and mass flux into the bases of the Bubbles are $\sim 10^{40}$ erg/s and $\sim 0.1 \msun$/year.

\item This outflow, however, is not powerful enough -- given the strong gravitational field, the  thermal pressure of the overlying halo gas, and the occurrence of gas cooling -- to break out of the Galaxy.

\item Instead, the outflow generates the two Fermi Bubbles and associated phenomena.
 The Bubbles inflate against an atmosphere of non-negligible pressure.
 An important ingredient of our scenario is that recent work indicates the gas density and pressure
in the halo, especially in the vicinity of the Bubbles, is larger than has often been assumed in the past \citep[e.g.,][]{Stocke2013,Miller2014}\footnote{A further consideration is that it is 
increasingly likely that the  gas around the Bubbles is itself in a flow driven into the halo by star-formation occurring in the Milky Way's molecular ring (at Galactocentric radius of $\sim$ 4 kpc) and/or the Bar \citep{Breitschwerdt1991,Breitschwerdt2002,Everett2008,Everett2010,Kretschmer2013,de Boer2014};
halo density models that assume hydrostatic equilibrium may not convey the full story in the inner Galaxy.}.

\item The Bubbles develop the classical wind-blown bubble structure of (starting from the smallest scales and working outwards): 
 i) a reverse shock, ii) a contact discontinuity (CD), and a forward shock; see Figure~\ref{f:bubble}.

\item 
The contact discontinuity 
separates  ISM material swept-up and shocked in the downstream of the forward shock from shocked Bubble material in the downstream of the internal reverse shock.
(In our model, the $\gamma$-ray edge of the Bubbles is identified with the contact discontinuity.)\footnote{We tentatively identify some of the $\gamma$-ray substructure in the base of the Bubbles -- e.g., the cocoon \citep{Su2012,Ackermann2014} -- as IC emission 
from upstream cosmic-ray electrons interacting with the dense photon fields close to the disk before reaching the shocks and 
from downstream electrons that have been reaccelerated upon reaching one of the shocks (but are not able to reach very far past them).}

\item The Bubbles quickly (within $\sim10^7$ year) reach pressure equilibrium with their environment; at this point (neglecting, as we do in our current model, cooling in the region between forward shock and CD) the forward shock degenerates into a sound wave.
The  CD decelerates even faster and is subsonic after only $\lsim 10^6$ year; at  present it is expanding at only $\sim$ few km/s.

\item Thermal plasma and cosmic rays initially suffer adiabatic and radiative energy losses  as they are blown out of the nucleus on the conically-expanding outflow 
(R1 in fig~\ref{fig_BubbleSchematic}).
However, they encounter giant reverse shocks (one at $\sim$1 kpc height in each Bubble) where the ram pressure of this flow equilibrates with the thermal pressure of the shocked, downstream gas 
(cf. the model of Lacki (2013) where the ram pressure of the expanding outflow equilibrates with the thermal pressure of the background Galactic halo atmosphere).
At these shocks,
the plasma and cosmic-rays are re-heated/re-accelerated. (In the remainder of this paper, unless otherwise noted, the shocks we refer to generically are these reverse shocks interior to the contact discontinuity.)

\item In the immediate downstream of the shocks, there is a systematic sub-sonic flow of $v_\mathrm{down }\sim 1/4 v_\mathrm{w} \sim $ 100-150 km/s 
(with $v_\mathrm{w}$ the wind speed immediately upstream of the shock)
There is  energy-independent transport of the non-thermal particles into the Bubbles
mediated by this flow in combination with cosmic-ray streaming at approximately the Alfven speed.
High-energy cosmic-ray electrons produce inverse-Compton and synchrotron emission on photon and magnetic fields in the shock-downstream.
Given there is a systematic flow (such that, at any particular distance from the shock, all electrons are the same `age'), the spectrum of these electrons 
(and their resultant emission)  is hard (mirroring the injection distribution) up to some time (and thus distance)-dependent maximal energy above which the spectrum is exponentially suppressed.
Given the rather slow transport speeds and relatively short cooling times, microwave-synchrotron-emitting electrons reach only $\sim$ 1-2 kpc past the shocks 
and $\gsim$ GeV inverse-Compton-emitting electrons only reach few $\times 100$ pc.
(In contrast,  cosmic-ray protons and 2.3-GHz-synchrotron-emitting electrons survive to reach the full extent of the Bubbles.)
Altogether, we show below that
 the hard spectrum synchrotron radiation 
 from cosmic-ray electrons reaccelerated on the shocks and transported downstream provides an excellent match to the luminosity, spectrum, distribution, and morphology of the microwave Haze.

\item As the Bubbles age and their volume slowly increases (more than compensating for their density decline), the radiative losses of the plasma gradually become more significant.
In fact, there is a steady-state limit (at constant radius and mass) where the Bubbles reach both pressure equilibrium with the environment and the cooling radiation from their interior plasma completely saturates the injected mechanical energy.
As this limit is approached, an increasingly large fraction of the hot plasma mass injected into the Bubbles is lost by drop-out as cooling leads to local thermal instability that causes individual parcels of gas to cool, collapse, lose buoyancy, and fall out of the Bubbles under gravity.

\item Because of the density-squared dependence of cooling, this mass drop-out process  dominantly occurs in the somewhat over-dense shell of 
material that forms  {\it inside} of the CD where the slow flow of plasma away from the reverse shock is gradually arrested (R3 in Figure~\ref{fig_BubbleSchematic}).

\item While the plasma evolves radiatively in the shell (as it cools down into localised, over-dense condensations), cosmic ray hadrons and magnetic fields, in constrast, continue to evolve adiabatically.
Their compression in the shell material leads to adiabatic energy gain; they are compressed until they equilibrate with the thermal pressure of the volume-filling plasma in R2 between the shell and the reverse shock.
In other words, adiabatic evolution of these non-thermal phases leads them to become the dominant dynamic agent in the shell.

\item
The ultimate extent and mass of the shell R3 is determined by pressure balance condition explained above and the further condition that the shell's cooling radiation  saturates the enthalpy flux represented
by the slow flow of hot plasma into it from R2.

\item Given that both the mass of shell target material and the hadronic cosmic-ray energy density in the shell can be determined from these considerations, we can calculate the hadronic emission from the shell; 
we find that this essentially saturates at the observed $\gamma$-ray luminosity of the Bubbles for reasonable parameter values.
The fact that the dominant hadronic emission occurs mostly in the shell explains
why the volumetric emission from the Bubbles peaks towards their edges \cite[as required to reproduce their flat {\it projected} surface brightness:][]{Su2010}.

\item As for the protons and heavier ions, the lower-energy cosmic-ray electrons that live long enough to reach the shell are also adiabatically compressed and reenergised there.
At the same time, collisions between the adiabatically compressed hadron population and the shell gas inject secondary electrons (+ positrons) into R3.
Unlike for the region downstream of the reverse-shocks, however, in the shells there is a mixture of electron populations with different ages and thus different characteristic energies. 
This leads to the integrated, steady-state shell electron spectrum being steepened by the canonical difference in spectral index of 1 associated with synchrotron/IC losses.
This cooled  primary + locally-produced secondary electron population synchrotron-radiates on the adiabatically-compressed shell magnetic field, explaining the steep-spectrum, 
polarized emission measured between 2.3 and 23 GHz.

\item The shell/CD does not constitute a perfect boundary, however: observationally, in 2.3 GHz polarisation data, there are plumes of radio emission that extend north-west and south-west
beyond the $\gamma$-ray boundaries of the Bubbles (R4).
Recent X-ray evidence \citep{Tahara2015} seems to indicate hot plasma venting from the north bubbles, roughly coincident with the northern radio plume.
We speculate that these plumes are symptomatic of the break-out 
of the light fluid of co-mixed cosmic rays, plasma, and magnetic fields through the heavy shell, mediated by the Rayleigh-Taylor instability.
However, the contact discontinuity generates the (required) $\gamma$-ray edge because, 
even if some cosmic-ray protons also escape through this boundary, this delimits the region of dense target gas accumulated into the shell from the interior flow.

\end{enumerate}

\subsection{Outline of paper}

The remainder of this paper is structured as follows.
We describe the input data that frame the problem in \S \ref{sctn_Preliminaries}.
In  \S \ref{s:sctn_ReverseShocks} we show that, incorporating the effects of gravity, the Bubbles should contain giant, reverse shocks at heights of $\sim 1$ kpc into the outflows north and south from the nucleus.
In \S \ref{s:sctn_BubbleModel} 
we set out a semi-analytic calculation describing the evolution of a bubble fed by an outflow of fixed mechanical power and mass flux expanding into an atmosphere of finite, non-negligible pressure.
We show that, adopting values for these quantities apposite to the Fermi Bubbles, such a bubble reaches the size scale of the Fermi Bubbles after a few $\times 10^8$ yr
and that the asymptotic radius of such a bubble for these parameters is not much larger than that of the observed Fermi Bubbles.
This finding removes any `Why now?' problem and  justifies our subsequent steady-state treatment of the Bubbles' non-thermal emission.
In \S \ref{sctn_Scan} we describe our procedure for delimiting the region of 
the parameter space defined by the energy and mass fluxes ($\dot E_0$ and $\dot M$ respectively) injected at the nucleus
that describe the present day Bubbles.
In \S \ref{sctn_NTEmission} we demonstrate explicitly that, in the small
region of parameter space consistent with prior constraints,
the Bubbles' i) predicted hadronic $\gamma$-ray emission;
ii) predicted microwave synchrotron emission; and 
iii) radio continuum synchrotron emission
all match 
the observed values.
In other words, we show that our scenario 
consistently explains, in a unified model, 
the non-thermal phenomena described as 
the `Fermi Bubbles', the `Microwave Haze', and the `S-PASS Lobes'.
In \S \ref{sctn_Evidence} we review specific observations that seem to support (or, in one case, may challenge) our model.
Finally, in \S \ref{sctn_Discussion} we set out implications of our model and future work.


\section{Preliminaries}
\label{sctn_Preliminaries}

\subsection{Geometry}

Adopting a distance to the Galactic Center of 8 kpc, geometrically the Bubbles can be approximated as a pair of spheres with radii $\sim 3.2$ kpc 
with 
a total volume of $V_\textrm{\tiny{FB}} \simeq  8.4 \times 10^{66}$ cm$^3$
and
a combined surface area of $A_\textrm{\tiny{FB}} \simeq 2.5 \times 10^{45}$ cm$^2$.
A better characterisation of the Bubbles' real geometry seems to be that they are bi-cones emanating north and south from the nucleus to  heights of 2-3 kpc above the plane above which they collimate into cylinders \citep{Bland-Hawthorn2003,Carretti2013}.
The Fermi Bubbles reach $\sim 7$ kpc maximal height above the Galactic center accounting for projection effects; the polarized radio lobes extend up to $\sim 8$ kpc.
A schematic showing the configuration of the northern bubble according to our model is shown in Figure ~\ref{fig_BubbleSchematic}.

\subsection{Other Parameters}
\label{sctn_OtherParams}

\citet{Kataoka2013} have performed scans across the limbs of both the northern and southern bubbles at high latitudes with the {\it Suzaku} X-ray satellite.
In neither scan do these authors find evidence for a change of temperature between the exterior and interior plasma; rather the plasma temperature seems constant at $\sim (3.0-3.5) \times 10^6$ K.
For the northern scan, starting in the bright North Polar Spur feature (located east in projection from the northern bubble) and scanning inwards across the Bubble edge,
these authors  find a drop in emission measure $EM$ at the edge 
from $ (5-6) \times 10^\mathrm{-2}$ cm$^\mathrm{-6}$ pc outside
to  $ (3-4) \times 10^\mathrm{-2}$ cm$^\mathrm{-6}$ pc inside.
In the south, the emission measure is approximately constant at $ (1-2) \times 10^\mathrm{-2}$ cm$^\mathrm{-6}$ pc across the scan.
Adopting a total cross-sectional area of $A_\mathrm{FB,X} = 7.0 \times 10^\mathrm{44}$ cm$^2$, 
we find an average line-of-sight distance through the Bubbles $l_\mathrm{LOS} \equiv  V_\mathrm{FB}/A_\mathrm{FB,X} \simeq 3.7$ kpc.
With this distance scale and assuming {\it for the moment} a uniform density distribution for the Bubble plasma, 
a characteristic 
 ionised hydrogen number density in the Bubbles can be estimated
as $\langle n_\mathrm{H^+}^\mathrm{FB}\rangle \sim \sqrt{EM/l_\mathrm{LOS}} \simeq (1-3) \times 10^\mathrm{-3}$ cm$^\mathrm{-3}$.
Using the volumetric density estimate, the total plasma mass of the Bubbles is then estimated as
\begin{equation}
M_\textrm{\tiny{FB}} \simeq  1.4 \ m_p \ n_H \ V_\textrm{\tiny{FB}} = 2.1 \times 10^7 \ \msun \left(\frac{\langle n_\mathrm{H^+}^\mathrm{FB} \rangle}{0.003 \ \textrm{cm}^{-3}}\right) \, .
\end{equation}

In reality, we show below that most of the emission measure 
through the Bubbles (and, therefore, most of their observed thermal X-ray emission) 
is  contributed by gas in a shell
whose real
density is larger than the volumetric average density estimated above by a factor
$\langle n_\mathrm{H^+}^\mathrm{shell}\rangle/\langle n_\mathrm{H^+}^\mathrm{FB} \rangle 
\sim (2 \Delta r_\mathrm{shell}/l_\mathrm{LOS})^{-1/2}  
$, 
and thickness (conservatively)
 $r_\mathrm{shell} \lsim$ 2 kpc,
inside the contact discontinuity.
The fitted X-ray temperature is thus likely to be characteristic of this shell; 
we expect the volume-filling, shocked plasma between the shell and the reverse shock to be hotter and more rarefied

Whatever the exact configuration of plasma within the Bubbles, consistent with our modeling, 
they are not strongly-over pressured with respect to the surrounding halo \citep[consistent with the absence of obvious X-ray limb brightening in their higher reaches:][]{Kataoka2013,Tahara2015}.
Again working with the characteristic, volumetric-average
number density for the moment,
the pressure in the Bubbles 
$ P_\mathrm{FB} = n_\mathrm{tot}^\mathrm{FB} \ k_B \ T_\mathrm{FB} \  \simeq 2.0$ eV cm$^{-3} \ n_\mathrm{H}^\mathrm{FB}/(0.003$ cm$^{-3}) \ T_\mathrm{FB}/(3.5 \times 10^6$ K)  
with the total plasma number density $n_\mathrm{tot}^\mathrm{FB} \simeq 2.3 \ n_\mathrm{H}^\mathrm{FB}$ 
(adopting a universal $n_\mathrm{He} = 0.1 n_\mathrm{H}$ in a totally ionized, electrically neutral plasma and ignoring metals for the moment).

This Bubble pressure should be compared with the determination of halo pressures derived from
the recent modeling of \citet{Miller2014} of OVIII emission line data.
Adopting halo metallicities of $\sim 0.3 \zsun$ as suggested by independent evidence, 
these authors determine a halo density profile that gives $n_e \simeq 2 \times 10^{-3}$ at halo height of $z = 8$kpc and
$n_e \simeq 6 \times 10^{-3}$ at 4 kpc.
Together with their determination of a fairly constant halo gas temperature of  $\sim 2 \times 10^6$ K,
these results suggest a total atmospheric pressure around the Bubbles of  $P_\mathrm{ext} = n_\mathrm{tot}^\mathrm{halo}  T^\mathrm{halo} \sim 2.3/1.2 \times (6 \times 10^{-3}$ cm$^{-3})$ $\times \ 2 \times 10^6$ Kelvin = 
3.2 $\times 10^{-12}$ dyn cm$^{-2} \ = \
2.0$ eV cm$^{-3}$ at their $\sim 4$ kpc
half-height.

\subsection{Discussion of Total Energetics and Other Parameters}
\label{sctn_TotalEnergetics}

In the limit that radiative losses can be neglected and the atmospheric density treated as constant, the total enthalpy of a slowly inflated bubble is
\begin{equation}
H \equiv \frac{\gamma}{\gamma - 1} P_\mathrm{ext} V  = \dot{E} \ t
\end{equation}
where $\dot{E}$ is the average mechanical power fed into the bubble over time $t$ and $\gamma$ is the effective adiabatic index of the bubble fluid.
Employing the atmospheric pressure of $P_\textrm{\tiny{ext}} = 2.0$ eV cm$^{-3}$  motivated above, this suggests $H_\textrm{\tiny{FB}} \simeq 6.7 \times 10^{55}$ erg for the Bubbles 
(probably a lower limit as it neglects the z-dependence of the halo density and pressure).
  Accounting only for core-collapse supernovae, the minimum mechanical power injected by nuclear star-formation
can be estimated from the GC star-formation rate $\rm SFR_\textrm{\tiny GC}$ as
\begin{equation}
\dot{E}_\textrm{\tiny GC SF} \gsim {\rm SFR}_\textrm{\tiny GC} \times 1 \ {\rm SN}/(90 \msun) \times 10^{51} {\rm erg}/{\rm SN} = 2.8 \times 10^{40} \> {\rm erg s^{-1}} \left(\frac{\rm SFR_\textrm{\tiny GC}}{0.08 \msun \> \rm yr^{-1}} \right)
\label{e:power}
\end{equation}
where a single core-collapse supernova (SN) requires the formation of $\sim 90 \msun$ in stars (for a \citet{Kroupa2001} initial mass function (IMF) if the least massive star to explode 
has zero-age main sequence mass of $8 \msun$),
we assume
$10^{51}$ erg mechanical energy release per supernova, and we have normalised to a nuclear star-formation rate of SFR$_\textrm{\tiny{GC}} = 0.08 \msun$ yr$^{-1}$ \citep{Crocker2012}.

Note that there are other significant star-formation-related sources of mechanical energy including stellar winds and proto-stellar outflows that could, in a full accounting, 
double
this estimate. 
However, 
we adopt equation~(\ref{e:power}) as our fiducial value for the injected power
\citep[cf.][]{Lacki2013}.
With this power, a conservative minimum timescale  to inflate the Bubbles (derived assuming an unrealistically high 100\%  
of  mechanical energy injected in the nucleus is transferred to the Bubbles and neglecting cooling and gravitational losses, which we show below are both actually important) 
is $t = H_\textrm{\tiny{FB}} / \dot{E}_\textrm{\tiny{GC SF}} \simeq 8 \times 10^7$ 
years.
Given the current surface area of the Bubbles, the expansion speed where $\dot{H}_\textrm{\tiny{FB}}$ saturates the star-formation mechanical power is $\sim 14$ km/s; this is an upper limit on the current expansion velocity of the Bubbles' contact discontinuity.

The total plasma mass in the Bubbles of $M_\mathrm{FB} \sim 2.1 \times 10^7 \msun$,
if fed by the nuclear outflow at a mass inflow rate, $\dot{M}$, similar to the SFR, requires a {\it minimum} timescale to establish of 
$M_\mathrm{FB}/\dot{M} = 2.6 \times 10^8$ year $\times (0.08 \ \msun/$yr$)/\dot{M}$.
Normalizing to volumetric average values for plasma temperature and density obtained in \S\ref{sctn_OtherParams},
the cooling time of the plasma in the Bubbles is $t_{cool} = 3/2 \  k_B T/(\Lambda[T] \ n_\mathrm{tot}) \simeq 5 \times 10^8$ year $\times T_\textrm{\tiny{FB}}/(3.5 \times 10^6 $K) $ \times (0.007$ 
cm$^{-3}/n_\mathrm{tot}^\textrm{\tiny{FB}}$)[\footnote{This timescale is a factor $\sim2$ longer than that found by \citet{Crocker2014}; 
the revised number is founded on the latest cooling curve calculation using the MAPPINGS IV v 4.2.5 NEQ cooling functions: Ralph Sutherland, private communication; also see \citet{Dopita2013,sutherland13a}.}] where $\Lambda[T]$ is the cooling function \citep{Dopita2013,sutherland13a} for plasma at temperature $T$.
Thus even the minimal inflation time is comparable to the radiative cooling time indicating that {\it an adiabatic treatment is inadequate}. 
Note the confluence of timescales around the formation/maintenance of the Bubbles at order $\sim$ few  $\times 10^8$ year \citep[cf.][]{Crocker2014}.

We show 
 below (\S~\ref{s:sctn_BubbleModel}) that the Bubbles approach a steady 
 state such 
 that their radii do not increase and that the freshly-heated plasma mass flowing into them is matched by the drop-out of cooling plasma, which condenses under a local thermal instability.

In such a steady state, 
material is advected from the  cylindrical acceleration zone in the nucleus, of volume $V_\mathrm{GC}$ and radius $r_0$, by the wind, with speed $v_\mathrm{w}$, in a timescale $t_\mathrm{w}$,
implying a rate of mass loss of 
\begin{equation}
\dot{M}_\mathrm{GC} \equiv \mu m \, n_\mathrm{tot}^\mathrm{GC}  \ V_\mathrm{GC}/t_\mathrm{w} = 
\mu m  \, n_\mathrm{tot}^\mathrm{ GC} \ 2 v_\mathrm{w} \ \pi r_0^2
\end{equation}
 (where $m$ is the atomic mass unit and $\mu  m$ is the mean mass per particle in the plasma) is identical to the rate of mass injection into the Bubbles which is, in turn, identical to the rate at which mass is cooling out of the structures, 
 \begin{equation}
 \dot{M}_{\rm cool} = \mu m \, 
n_\textrm{\tiny tot}^\textrm {\tiny FB}  V_\textrm{\tiny FB}/t_{\rm cool} \, .
\end{equation}
Hence, we estimate
the radius of the nuclear region from which the outflow feeding the Bubbles emerges 
as
\begin{equation}
r_0 \simeq \sqrt{\frac{V_\textrm{\tiny FB} \, n_\textrm{\tiny tot}^\textrm{\tiny FB}}{2\pi \, n_\textrm{\tiny tot}^\textrm{\tiny GC} \,  v_{\rm wind} \, t_{\rm cool}}} \simeq 70  \, \left(\frac{v_{\rm wind}}{500 \>\rm km \> s^{-1}} \right)^{-1/2} \left(\frac{n_\textrm{\tiny tot}^\textrm{\tiny FB}}{0.004 \> \rm cm^{-3}} \right)^{1/2}
\left(\frac{n_\textrm{\tiny tot}^\textrm{\tiny GC}}{0.1 \> \rm cm^{-3}} \right)^{-1/2} \> {\rm pc,}
\end{equation}
\noindent
where we normalise $n_\textrm{\tiny tot}^\textrm{\tiny GC}$ to a value informed by GC X-ray observations \citep{Muno2004}.
This is close to the radius of the Central Molecular Zone (CMZ) molecular torus, 80-100~pc \citep{Molinari2011}.
The CMZ is the region of enhanced molecular gas density and star-formation activity surrounding the central supermassive black hole and the good agreement of our estimate of the outflow radius and the radius of this region supports our model that the Bubbles are inflated by star-formation over this region.


\section{Location of reverse shocks}
\label{s:sctn_ReverseShocks} 

In section~\ref{s:sctn_BubbleModel}, we describe a numerical model for the expansion of a radiative bubble. 
An important feature of any bubble model is the location of the reverse shock and, in the case of the Fermi Bubbles, the presence of such shocks  has ramifications for the production of microwave and radio emission. 
We also show, in this section, that the reverse shock position affects our estimate of the energy flux into the Bubbles, which is an important parameter in our later numerical modeling.

\subsection{Condition for reverse shock}

Let $\rho_{\rm w}$ and $v_{\rm w}$ be the density and velocity respectively of the Galactic Center wind. The location of the reverse shock is determined from the condition that the ram pressure of the wind and the internal bubble pressure, $p_{\rm b}$ are related by the Rankine-Hugoniot condition for a strong 
shock\footnote{Our modeling below demonstrates that the real shocks are, as assumed here, relatively strong with the upstream flow having a Mach number in the range 6-8.}
\begin{equation}
p_{\rm b} = \frac{2}{\gamma +1} \rho_{\rm w} v_{\rm w}^2
\label{e:reverse}
\end{equation}
where $\gamma=5/3$. On each side of the nucleus, we approximate the wind as conically expanding with solid angle $\Omega_{\rm w}$ so that the mass flux is
\begin{equation}
\dot M = \Omega_{\rm w} R^2 \rho_{\rm w} v_{\rm w}  
\label{e:mdot}
\end{equation}
where $R$ is the spherical radius. Hence the location of the shock is given by:
\begin{equation}
R_{\rm shock} = \left( \frac {2}{\gamma+1} \frac {\dot M v_{\rm w}}{p_{\rm b} \Omega} \right)^{1/2} 
\end{equation}

Normally for a starburst wind, with initial power $\dot E$ and mass flux $\dot M$, we would assume that the wind velocity,  $v_{\rm w} \simeq (2 \dot E / \dot M)^{1/2}$ is constant. In this case the shock location would be given by:
\begin{equation}
R_{\rm shock} = \frac {2^{3/4}}{(\gamma + 1)^{1/2} } 
\frac {( \dot E \dot M )^{1/4}} {(p_{\rm b} \Omega )^{1/2}}
\end{equation}

However, the Galactic Center wind, with a kinetic power $\sim  10^{40} \> \rm ergs \> s^{-1}$ is relatively weak, compared to say, the $\sim 10^{42} \> \rm  erg \> s^{-1}$  wind from the \emph{dwarf} starburst galaxy M82 \citep{Shopbell1998,Cooper2008} and, in the case of the Galactic Center, gravity reduces the velocity of the wind on scales $\sim \rm kpc$.

\subsection {Effect of gravitational potential on the wind energy flux and velocity}
\label{s:gravitation}

In order to estimate the effect of gravity on the energy flux and the wind velocity, we begin with the equation for the conservation of energy. In the following $\epsilon_{\rm w}$ 
is the internal energy density of the wind, $\phi$ is the gravitational potential and $h_{\rm w}$ is the specific enthalpy. We have:
\begin{equation}
\pd{}{t} \left[ \frac{1}{2} \rho_{\rm w} v_{\rm w}^2 + \epsilon_{\rm w} + \rho_{\rm w} \phi\right] +
\pd{}{x_i} \left[ \left(\frac{1}{2} v_{\rm w}^2 + h_{\rm w} + \phi \right) \rho_{\rm w} v_{{\rm w},i} \right] 
=0
\end{equation}
Integrating over a volume bounded by streamlines and capped at the top by the reverse shock ($S$) and below by a surface near the base of the flow ($S_0$), we have, for a stationary flow, a conserved energy flux,

\begin{equation}
F_E = \int_S \left(\frac{1}{2} v_{\rm w}^2 + h_{\rm w} + \phi \right) 
\rho_{\rm w} v_{{\rm w},i} n_i \> dS
= \int_{S_0} \left(\frac{1}{2} v_{\rm w,0}^2 + h_{\rm w,0} + \phi_0 \right) 
\rho_{\rm w,0} v_{{\rm w},i,0} n_{i,0} \> dS_0
\end{equation}

Denoting the hydrodynamic part of the energy flux by
\begin{equation}
\dot E =  \int_S \left(\frac{1}{2} v_{\rm w}^2 + h_{\rm w} \right) 
\rho_{\rm w} v_{{\rm w},i} n_i \> dS
\label{e:Edot1}
\end{equation}
we have that
\begin{equation}
\dot E = \dot E_0 - \int_S \rho_{\rm w} \phi v_{{\rm w},i} n_i \> dS 
+ \int_{S_0} \rho_{\rm w,0} \phi_0 v_{{\rm w,0},i} n_{i,0} \> dS_0 
\end{equation}
Assuming that the bounding surfaces are approximately equipotential, and taking $\Delta \phi = \phi - \phi_0$ as the difference in potential between the two surfaces, we have
\begin{equation}
\dot E = \dot E_0 - \dot M \Delta \phi 
= \dot E_0 \, \left( 1 - \frac {\dot M}{\dot E_0} \Delta \phi \right)
\label{e:Edot2}
\end{equation}
 The factor $ 1 - {\dot M} \Delta \phi / {\dot E_0}$ determines the reduction in hydrodynamic power resulting from the gravitational field. 

We now use Bernoulli's equation to relate the velocity, specific enthalpy and potential on $S$ with the corresponding 
\begin{equation}
\frac {1}{2} v_{\rm w}^2 + h_{\rm w} + \phi = \frac {1}{2} v_{\rm w,0}^2 + h_{\rm w,0} + 
\phi_0
\end{equation}
and approximate $\dot E_0$ by
\begin{equation}
\dot E_0 \simeq \dot M \times \left( \frac {1}{2} v_{\rm w,0}^2 + h_{\rm w,0} \right)
\end{equation}
This implies that
\begin{equation}
v_{\rm w}^2 + 2 h_{\rm w} \simeq 2 \frac {\dot E_0}{\dot M} - 2 \Delta \phi
\end{equation}
As the flow expands, the enthalpy term diminishes in importance compared to the kinetic term (i.e. $h_{\rm w} \ll v_{\rm w}^2/2$). Hence, 
\begin{equation}
v_{\rm w} \simeq \left(\frac {2 \dot E_0}{\dot M} \right)^{1/2} \, 
\left( 1 - \frac {\dot M}{\dot E_0} \Delta \phi \right)^{1/2}
\label{e:vw}
\end{equation}
Without the gravitational field, the asymptotic wind speed would be $(2 \dot E_0/\dot M)^{1/2} \simeq 630 (\dot E_0/5 \times 10^{39} \> \rm ergs \, s^{-1})^{1/2} (\dot M/ 0.04 \msun \, \rm y^{-1})^{-1/2} \> \rm km \> s^{-1}$. The factor $\left( 1 - {\dot M} \Delta \phi / {\dot E_0} \right)^{1/2}$ determines the reduction in wind speed resulting from the gravitational potential.

\subsection{Location of the reverse shocks in the presence of gravity}
\label{s:shockPosition}

The factor $\left( 1 - {\dot M} \Delta \phi / {\dot E_0} \right)$, appearing in equation~(\ref{e:vw}), depends on the location of the  reverse shock through the dependence of the potential difference $\Delta \phi$ on cylindrical coordinates, based on the Galactic Center. We now describe how we determine the approximate location of the reverse shock and the hydrodynamic energy flux there.

Using equations~(\ref{e:reverse}), relating the bubble pressure ($p_{\rm b}$) and the wind ram pressure and (\ref{e:mdot}) for the mass flux, together with equation~(\ref{e:vw}) for the wind velocity, we have for the radial location of a spherical shock in a conically expanding wind:
\begin{equation}
r_\mathrm{shk} = \left( \frac {2}{\gamma+1} \right)^{1/2} \,
\frac {(2 \dot E_0 \dot M)^{1/4}}{\left(p_{\rm b} \Omega \right)^{1/2}} \,
\left( 1 - \frac {\dot M}{\dot E_0} \Delta \phi \right)^{1/4}
\label{e:Rshock}
\end{equation}

For the potential of the Galaxy, we use the analytic forms and parameters provided by \citet{Breitschwerdt1991}. For the combined  disk and bulge potential difference from the center of the Galaxy, expressed in $(\rm km \> s^{-1})^2$, as a function of cylindrical coordinates $r$ and $z$, both in kpc, this gives
\begin{eqnarray}
\Delta \phi [r,z] &\simeq&  1.60 \times 10^5 \, -\frac{8.82 \times 10^4}{\sqrt{r^2+z^2+0.245}}
-\frac{1.10\times 10^6}{\sqrt{r^2+\left(\sqrt{z^2+0.270}+7.26\right)^2}}   \nonumber \\ 
&& + \frac{5.81 \times 10^5}{13 + \sqrt{r^2+z^2}} + 4.47 \times 10^4 \ln \left[13 + \sqrt{r^2 + z^2}\right] \quad
(\hbox{km s}^{-1})^2
\end{eqnarray} 
Notwithstanding the expression for $\Delta \phi$ being in terms of cylindrical coordinates, the contours of $\Delta \phi$ are approximately spherical in the region of interest ($r_\mathrm{shk}  \la 1.5 \> \rm kpc$). Hence, in solving equation~(\ref{e:Rshock}) we have approximated $\Delta \phi$ by its value on the $z$-axis, i.e. $\Delta \phi [0,r_\mathrm{shk} ]$.

\begin{figure}[hb!]
\centering
\includegraphics[width = 0.7 \textwidth]{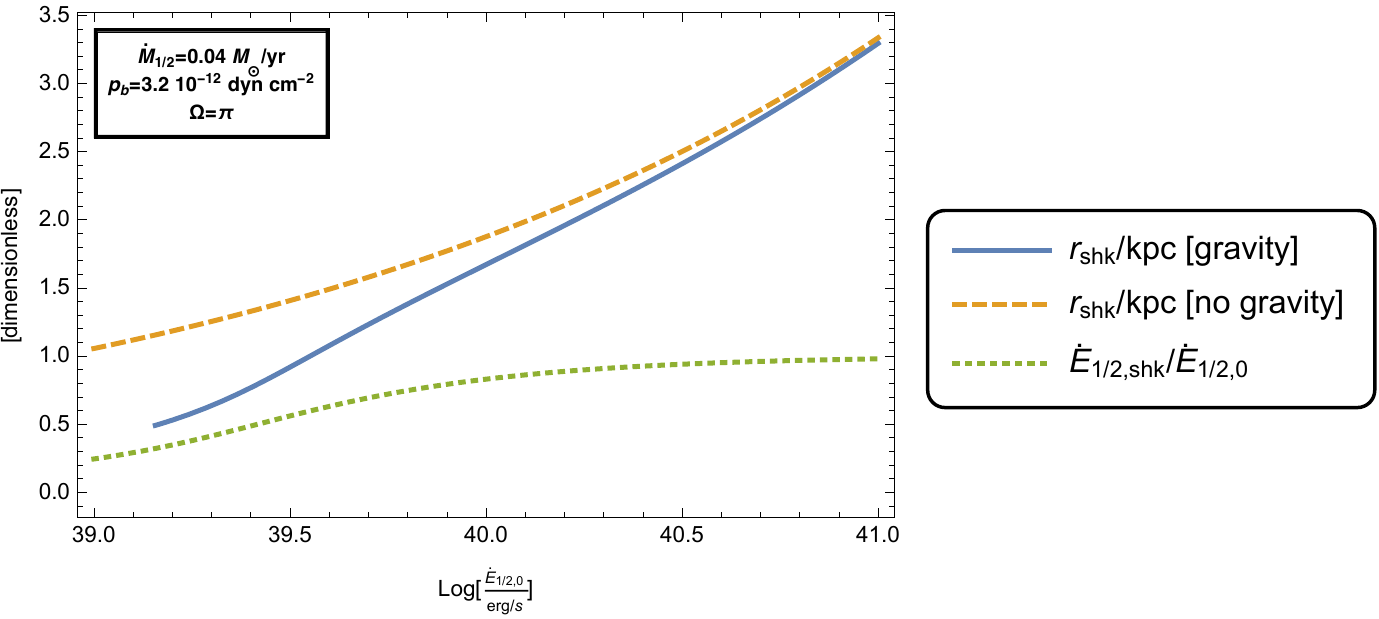}
\caption{{\bf Distance to the shock and relative mechanical power at the shock} as a function of mechanical power injected at the nucleus into one hemisphere, $\dot{E}_{1/2,0}$. 
The downstream thermal pressure in the bubble is 2 eV cm$^{-3} \simeq 3.2 \times 10^{-12}$ dyn cm$^{-2}$, the mass flux into the flow is 0.04 $\msun$/yr and we adopt  $\Omega \simeq \pi$ \citep{Lacki2013}, corresponding to an expansion cone angle $\sim 120^\circ$.}
\label{f:rshk}
\end{figure}

In Figure~\ref{f:rshk} we show, as a function of the mechanical power injected at the nucleus into one hemisphere (for the parameters specified in the caption), our numerical solution for i) the distance to the shock in kpc (with and without gravitational deceleration of the flow) 
and ii) the relative mechanical power available at the shock (given gravitational deceleration).
It is evident that for the chosen parameter values, which are apposite to the nuclear outflow that feeds the Fermi Bubbles, the flow can lose 30-50\% of its mechanical power to gravity before reaching the shock.

\section{Expansion of a Radiative Bubble}
\label{s:sctn_BubbleModel}

We now describe our calculation of the expansion of a bubble into an atmosphere of finite pressure incorporating the dynamical effect of energy losses due to cooling radiation emitted by the shocked material located between reverse shock and contact discontinuity.
We demonstrate here that, for parameters relevant to the Fermi Bubbles, we expect the contact discontinuity to be very slowly expanding (in line with the X-ray data)
at only $\sim$ few km/s
 and for the structures to have reached a size very similar to that observed. 

\subsection{Spherical bubble model}

\begin{figure}[hb!]
\centering
\includegraphics[width = 0.7 \textwidth,angle=-90]{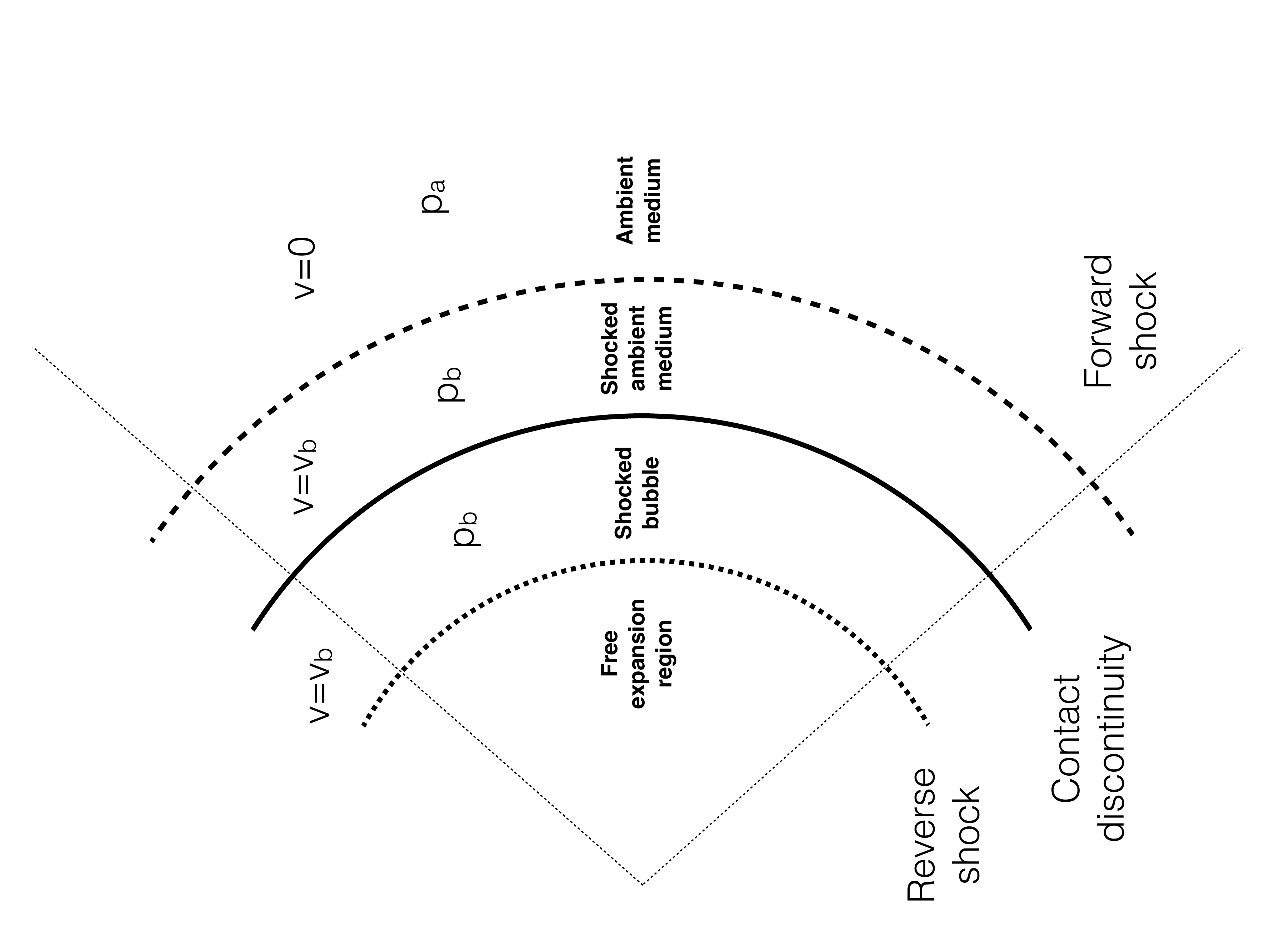}
\caption{Illustration of the key features of a bubble.}
\label{f:bubble}
\end{figure}

As a first approximation to modeling the Fermi Bubbles, we treat them as two hemispheres of a  single spherical bubble driven by a spherically symmetric wind propagating into a constant density, constant pressure medium.
In the observed Fermi Bubbles the morphology close to the Galactic plane is probably constrained by the larger density and pressure there. 
In comparing the radius of this model with the extent of the Fermi Bubbles, we calculate an equivalent spherical radius $r_{\rm FB} = \left[3/(4 \pi) V_{\rm FB} \right]^{1/3}\simeq 4.1 \> \rm kpc$, where $V_{\rm FB} = 285 \ \rm kpc^3$ is the estimated volume.   
Our model generalizes the usual treatment of a wind-driven bubble \citep[e.g.][]{weaver77a}, in that the bubble is {\it not} assumed to be strongly over-pressured with respect to the ambient medium and is also subject to significant optically thin radiative cooling.
The model {\it in this section} does not incorporate a gravitational potential (so that there is no internal or external pressure stratification) and, in common with other bubble models, neglects the small but finite fraction of the total bubble volume in the region upstream of the internal reverse shock.

Fig.~\ref{f:bubble} lays out the main features of such a bubble: There is a reverse shock interior to the bubble where the decreasing ram pressure of the spherically-expanding, supersonic nuclear wind is balanced by the internal pressure of the bubble as described in \S\ref{s:sctn_ReverseShocks}, and a region of shock-heated bubble material downstream of the reverse shock, in which the internal energy dominates the kinetic energy. 
This region is bounded by a contact discontinuity, external to which is a shell of shocked, ambient interstellar medium between the forward shock, which is expanding into the external atmosphere, and contact discontinuity. 
We approximate this region as one of constant spatial pressure and velocity and neglect cooling there. 
We also {\it in this section} approximate the region between contact discontinuity and reverse shock as constant in density.
The bubble is fed by mechanical energy injected at a rate $\dot{E}$ and mass at a rate $\dot{M}$. 
We assume a constant $\gamma$-law equation of state, with the pressure, $p$ and internal energy density, $\epsilon$ related by $p = (\gamma -1) \epsilon$ and $\gamma=5/3$. 
It is important to note that the growth of the total mass of the bubble $M_{\rm b}$ does {\it not} satisfy $\dot{M}_{\rm b} = \dot{M}$ because of the phenomenon of mass dropout driven by cooling \citep{Crocker2014}. We estimate the rate of mass dropout by equating the post-reverse shock enthalpy flux of that mass with the cooling inside the bubble.

The equations governing the pressure, radius, mass and density of the bubble are derived in the appendix, using energy, mass, and momentum conservation integrated over the various regions identified above. We initiate the evolution of the bubble at $t=10^5 \> \rm yr$ with initial values defined by an adiabatic bubble (see the appendix for details). 

Cooling sets up local thermal instability of the bubble plasma leading it to collapse into over-dense condensations that, having lost buoyancy (in a real situation that includes a gravitational potential), fall out of the bubble \citep[see][and references therein]{Crocker2014}.

As the bubble's pressure decreases as a result of the combined effect of the work done by expansion and the radiant loss of energy, it can reach a steady state where its pressure equilibrates with the surrounding atmosphere, the energy input is balanced by thermal radiation, and the mass flux is balanced by mass dropout. 
We show below that, for parameters appropriate to the Fermi Bubbles, the system is approaching this limit.

\subsection{Evolution of a spherical bubble with parameters relevant to the Galactic Center}
\label{s:speherical_bubble}

On the basis of the numbers presented in \S\ref{sctn_TotalEnergetics} we choose as representative values for the {\it total} mass and mechanical power injected at the nucleus ~$(\dot{E}_0,\dot{M}) \sim (10^{40} \ {\rm erg \> s^{-1}}, 0.08 \msun \> \rm y^{-1})$ (so that $\dot{E}_{1/2,0} \equiv 1/2 \times \dot{E}_0$ and $\dot{M}_{1/2} \equiv 1/2 \times  \dot{M}$ is injected into the flow into each hemisphere);
the exact parameter range allowable for these quantities is delimited in  \S \ref{sctn_Scan}.
One must remember, however, that in \S \ref{s:shockPosition} we showed that, for parameter values in this range, 
the flow into each Bubble can lose $\sim$30-50\% of its mechanical power to gravity before reaching the shocks.
For purposes of comparing our current modeling - which neglects gravity -- with our previous (and subsequent) results, 
 we now  therefore choose an injected mechanical power that is reduced by 40\% with respect to the fiducial value nominated above, namely $6 \times 10^{39}$ erg/s.
Our purpose here is to show that, for parameters relevant to the Fermi Bubbles, the expansion of a radiative bubble (into an atmosphere of pressure similar to that of the Galactic halo)
quickly slows down (consistent with the lack of X-ray brightening at the edges of the Fermi Bubbles) and, moreover, there is an asymptotic radius for such a  bubble that is not much larger than the equivalent radius of the observed Fermi Bubbles.

\begin{figure}
\centering
\includegraphics[width=1.0\textwidth]{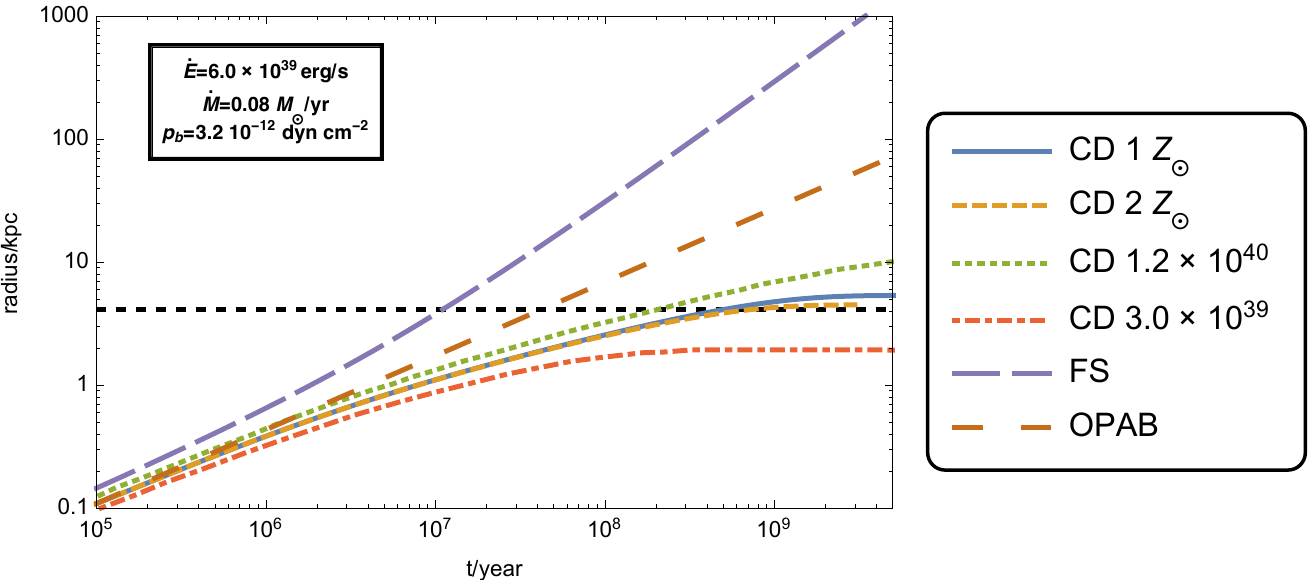}
\caption{{\bf Radius of the bubble contact discontinuity (CD)  and forward shock (FS)} vs. time for $\dot{E} = 6 \times 10^{39}$ erg/s and solar metallicity (unless otherwise marked) and $\dot{M} = 0.08 \msun$/yr (fixed); 
the atmosphere has pressure 2.0 eV cm$^{-3} \simeq 3.2 \times 10^{-12}$ dyn cm$^{-2}$.
Contact discontinuity (CD) curves are: 
i)  solid blue   for $\dot{E} = 6.0 \times 10^{39}$ erg/s;
ii) dashed orange for $\dot{E} = 6.0 \times 10^{39}$ erg/s and cooling with twice solar metallicity ($Z = 2 Z_\odot$);
iii) dotted green for $\dot{E} = 1.2 \times 10^{40}$ erg/s; and
iv) dot-dashed orange for $\dot{E} = 3.0 \times 10^{39}$ erg/s.
The long dashed violet curve is for the forward shock.
The brown long-dashed curve shows the analytical result for the expansion of an over-pressured, adiabatic bubble (OPAB; see Appendix); this very substantially
over-estimates the bubble radius for $t \gsim$ few $\times 10^7$ yr.
The horizontal black dashed line represents the equivalent spherical radius of 4.1 kpc corresponding to the total volume of the Fermi Bubbles; for the fiducial model the
bubble has attained this size by $\sim 5 \times 10^8$ yr.  
}
\label{f:radius}
\end{figure}

As justified above, another relevant parameter is an ambient pressure of $2.0 \> \rm eV \> cm^{-3} (= 3.2 \times 10^{-12} \> \rm dyn \> cm^{-2})$.
Note that we assume the plasma has solar metallicity but verify that, given the metallicity affects the cooling, assuming twice solar metallicity does not change our results significantly (cf. Figure~\ref{f:radius}).

\begin{figure}
\centering
\includegraphics[width=0.8\textwidth]{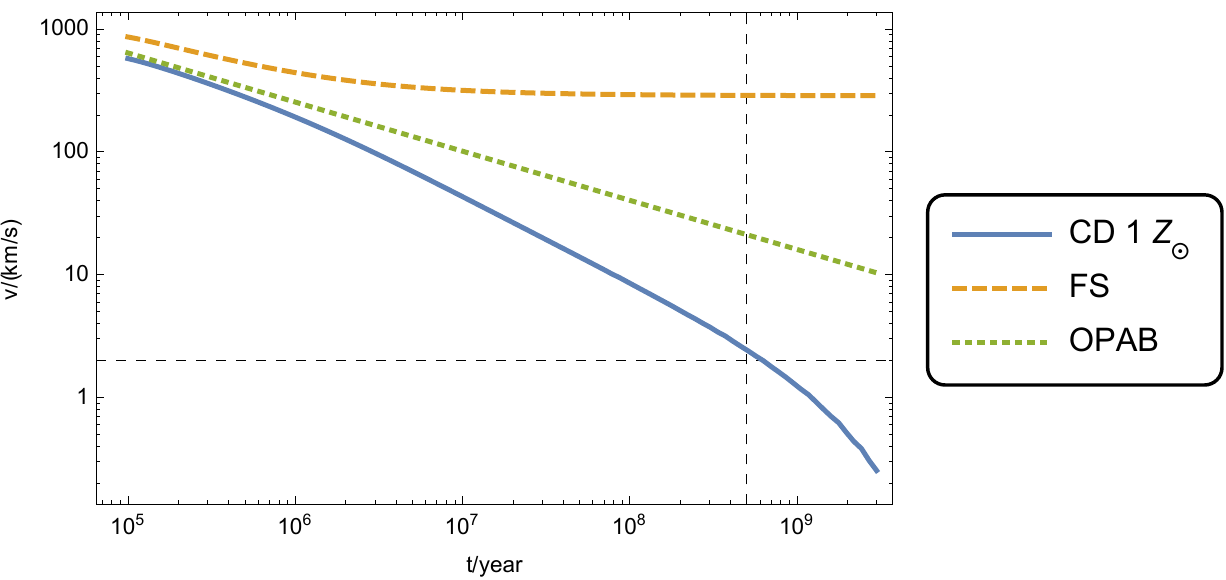}
\caption{{\bf Velocity of the bubble contact discontinuity (CD) and forward shock (FS) versus time} for cooling with $Z = Z_\odot$. Parameters as for Figure~\ref{f:radius}.
The dotted (green) curve shows the analytical result for the expansion of the contact discontinuity of an over-pressured, adiabatic bubble (OPAB).
The expansion velocity of the forward shock asymptotes to the sound speed as it degenerates into a sound wave.
For the chosen parameters, at the age ($\sim 5 \times 10^8$ yr indicated by the vertical line) where the bubble has attained a volume equivalent to the observed Fermi Bubbles (cf.~Figure \ref{f:radius}), its contact discontinuity is expanding at 2-3 km/s.
}
\label{f:velocity}
\end{figure}

\begin{figure}[h!]
\centering
\includegraphics[width=0.8\textwidth]{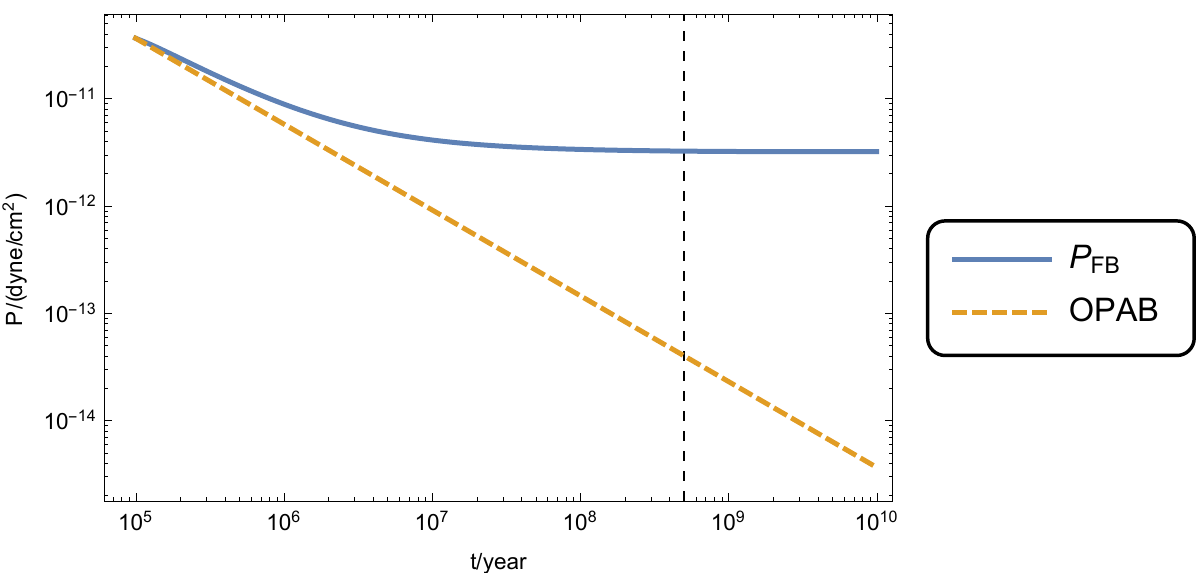}

\caption{{\bf Pressure interior to the contact discontinuity} (solid blue); this asymptotes to the assumed atmospheric pressure $3.2 \times 10^{-12}$ dyn cm$^{-2}$.
Parameters as for Figure~\ref{f:radius}. 
The vertical line indicates when the bubble has attained a volume equivalent to the observed Fermi Bubbles (cf. Figure~\ref{f:radius}).
The dashed orange curve is for an over-pressured, adiabatic bubble.
}
\label{f:pressure}
\end{figure}

Figure~\ref{f:radius} shows the evolution of the bubble   radius (i.e., contact discontinuity) for  input parameters $\dot{E} = 6.0 \times 10^{39}$ erg/s and $\dot{M} = 0.08 \msun$/yr 
(i.e., $1.2 \times 10^{15}$ erg/g as the energy per mass content of the outflow), as well as the locus of the forward shock. 
The radius of a highly over-pressured bubble and the equivalent spherical radius of the Fermi Bubbles is also shown for comparison. 
Clearly, for the fiducial parameters the bubble radius asymptotes to approximately 5 kpc after about a Gyr and from about $3 \times 10^8 \> \rm yr$ onwards, the bubble radius is not far from the spherical equivalent radius 
$\simeq 4.1 \> \rm kpc$ of the observed Bubbles. 
For comparison, for fixed $\dot{M} = 0.08 \msun$/yr, cases of $\dot{E} = 1.2 \times 10^{40}$ erg/s ($2.4 \times 10^{15}$ erg/g) and $\dot{E} = 3.0 \times 10^{39}$ erg/s ($6.0 \times 10^{14}$ erg/g) are also shown in Figure~\ref{f:radius}. 
The former does not asymptote to a final radius of $\sim 10$ kpc until 10 Gyr (with volume $\sim 15 \times V_\mathrm{FB}$) and it spends only a brief time in the vicinity of the currently-observed size, tending to suggest a return to the ``Why now?" problem encountered by explosive scenarios.
The latter never reaches the size (or minimum temperature) of the observed Fermi Bubbles.

The expansion of the bubble decelerates from  approximately $2 \times 10^6 \> \rm yr$ on, and this is clearly shown in 
Figure~\ref{f:velocity}. 
From about $10^8 \> \rm yr$ on, the bubble expansion velocity decelerates to less than $10~\rm km \> s^{-1}$ as it approaches pressure equilibrium with the environment. The approach to pressure equilibrium is shown in Fig.~\ref{f:pressure}. 
At the same time the forward shock becomes a sound wave propagating into the background medium at the sound speed $\simeq 280 \> \rm km \> s^{-1}$ (again see Fig.~\ref{f:velocity}).

\begin{figure}
\centerline{\includegraphics[width=0.8\textwidth]{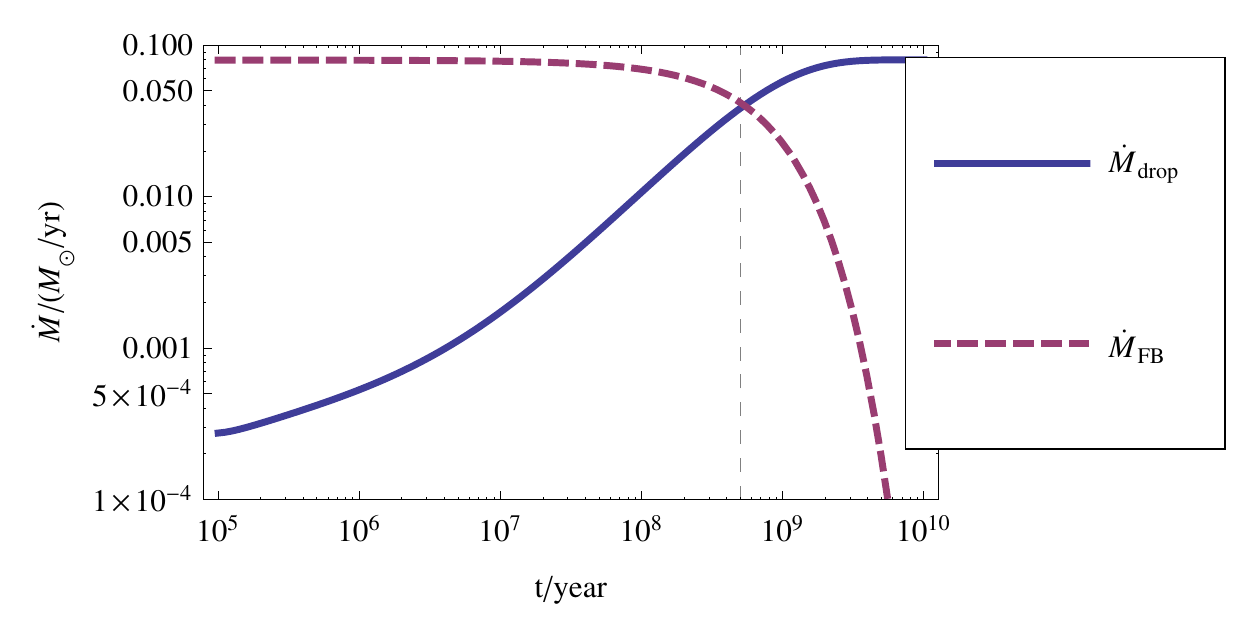}}
\caption{{\bf Mass drop out} (solid blue) and mass growth of Bubbles (dashed purple). 
Parameters as for Figure~\ref{f:radius}.
The vertical line indicates when the bubble has attained a volume equivalent to the observed Fermi Bubbles (cf. Figure~\ref{f:radius}).
At times $\gsim$ Gyr mass drop out by itself saturates the injection rate of freshly-heated plasma. 
}
\label{f:massDropOut}
\end{figure}

When the bubble's growth saturates, its total mass and energy content are fixed, with the post-reverse shock enthalpy flux balanced by the radiative luminosity. 
Figure~\ref{f:massDropOut} shows that the mass dropout saturates at the injection rate of $0.08 \> \msun \> \rm y^{-1}$ and that, consistent with this, the nett mass injection into the bubbles saturates at zero maintaining the total mass interior to the bubbles at a constant value.
%

For comparison with other sections, we can estimate the final, asymptotic volume, $V_{\rm f}$ in this circumstance, with the input hydrodynamic energy flux, $\dot E \equiv 5/2 \dot{M}/(\mu m) k_B T$ arriving at and processed through the shocks, balanced by radiative cooling and with the pressure in the downstream bubble region in equilibrium with the halo atmosphere. 
Here the post-reverse shock bubble temperature is $T = 3/ 16 \, \mu m/k_{\rm B} \, v_{\rm w}^2 \simeq 3.5 \times 10^6 (v_{\rm w}/500 \> \rm km \> s^{-1})^2 \> K$; the
total post-shock number density in the bubble is $n_\mathrm{tot} = 4 \dot M / (\mu m \Omega R_{\rm sh}^2 v_{\rm w})$ (see \S~\ref{s:sctn_ReverseShocks} for the definitions of these symbols).
Balancing the energy injected into the bubble with thermal cooling and normalizing to fiducial values for temperature and number density (such that
pressure equilibrium with the halo is satisfied) ,
we then have that  the final volume of the bubble is
\begin{eqnarray}
V_{\rm f} &\simeq& \frac {\dot E}{n_\mathrm{tot}^2 \Lambda[T]} = \frac{5/2 \ \dot{M}/(\mu m) \ k_B T_\mathrm{FB}}{n_\mathrm{tot}^2 \Lambda[T]} \nonumber \\
 &=& 
2.0 \times 10^{67} \, {\rm cm^3} \> 
\left(  \frac {T_\mathrm{FB}}{3.5 \times 10^{6} \, \rm K }\right) \, 
\left( \frac {\dot M}{0.08 \msun \, \rm yr^{-1}}\right) \,
\nonumber \\
&& \times  
\left( \frac {n_{\rm tot}}{\rm 0.0066 \ cm^{-3}}\right)^{-2} \,  
\left( \frac {\Lambda[T]}{10^{-23.15} \rm \ erg \ cm^3 \ s^{-1}}\right)^{-1} \, ,
\label{eq_rf}
\end{eqnarray}
where we normalize to a $\dot{M}$ favored by the modeling presented in  \S\ref{sctn_Scan}.
This is $\sim  2.3 \ \times$ the observationally-estimated volume of the Fermi Bubbles, $V_\mathrm{FB} = 8.2 \times 10^{66}$ cm$^{-3}$, consistent with our expectation that the Bubbles are still slowly expanding.
Note that the estimated volume would be smaller if the increase in density of cooling material in the shell region (R3) were taken into account. 
 
\section{Scan over parameter space}
\label{sctn_Scan}

We have shown in the previous section that i) the Bubbles are expected to contain giant, reverse shocks at heights of $\sim$1 kpc above the nucleus and ii) the Bubbles are likely to be slowly expanding structures close to pressure equilibrium with the surrounding halo.
We now explore the implications of this general picture more quantitatively.
In particular, in this section we determine the allowable ranges for the rate at which mechanical energy and mass is fed into the Bubbles and thermalized at the reverse shocks; in the subsequent section we examine acceleration of cosmic rays at these shocks and their attendant, non-thermal radiation.

\subsection{Procedure}

\label{sctn_Procedure}

We now explain our procedure for scanning over the steady mass and mechanical energy flux, $\dot{M}$ and $\dot{E}_0$ respectively, into the central acceleration region.
As shown above  these parameters determine the  temperature reached within the central acceleration region following thermalisation of the mechanical energy.
The high pressure in this region accelerates the plasma into a flow that is collimated into a biconical wind transverse to the Galactic disk by the surrounding dense molecular torus of the Central Molecular Zone.
In this expanding flow, the plasma cools adiabatically until it reaches the reverse shock in each bubble where it is shock (re)heated.
For every $\{\dot{E}_0, \dot{M} \}$ in the parameter space, the asymptotic velocity of the wind, $v_\textrm{\tiny{asym}}$ is determined.
Note that $v_\textrm{\tiny{asym}} $ is reached soon after the flow escapes from the central acceleration zone; subsequent to reaching $v_\textrm{\tiny{asym}}$ 
the adiabatically-cooling flow decelerates in response to the gravitational potential, $\phi$ \citep{Breitschwerdt1991}, as described above.
Given the high (i.e., supersonic) upstream flow speeds in which we are interested, we treat the flow evolution in the ballistic limit 
where we ignore small corrections  due to the temperature evolution of the upstream gas.

In our modeling, for every sampled $\{\dot{E}_0, \dot{M} \}$ point, we determine i) the height of the reverse shocks $R_\mathrm{sh}$
and ii) the temperature of the shocked material, $T_\mathrm{sh}$.
Given that, observationally, the Bubbles do not  expand latitudinally very much at heights greater than the expected position  of each shock, we neglect further possible 
adiabatic plasma temperature losses with height above the shocks.
We also assume a uniform  plasma temperature in R2 given by that for a strong shock with upstream flow velocity $v_\mathrm{w}$ and  $\gamma = 5/3$.
Hence
$T_\mathrm{FB} = T_\mathrm{sh} = 3/16 \ \mu  m \ v_\mathrm{w}^2/k_B $.

The shocks deliver a fraction, $\epsilon_{CR}$, of their total mechanical power (accounting for gravitational losses), $\dot{E}$, into freshly-accelerated cosmic rays: $L_{CR} \equiv \epsilon_{CR} \times \dot{E}$;
the spectral index of the cosmic rays' power-law distribution is controlled by the Mach number of the shocks.
We shall see subsequently  that, {\it a posteriori}, we are warranted in assuming strong shocks: the hard, non-thermal 
spectra of both the $\gamma$-ray emission from the Fermi Bubbles and the microwave Haze require shock Mach numbers of $M \gsim 4$ (see \S\ref{sctn_NTEmission}).
We also find  (see \S\ref{sctn_HadronicGammaEmission}) that,
given non-thermal ISM phases dominate the dynamics of the compression shell R3, its size is dependent on the efficiency with which the shocks accelerate cosmic rays, 
$\epsilon_{CR}$ (cf.~eq.~\ref{eq_shellVolume}).
We have explored a range of values for this parameter.
Also important are 
the pressure of the atmosphere into which the Bubbles are expanding and the  effective temperature of the compression shell.
For the former 
our fiducial value is $P_\textrm{\tiny{halo}} = 2$ eV cm$^{-3} = 3.2 \times 10^{-12}$ erg cm$^{-2}$ adopted from \S \ref{sctn_OtherParams} and, for the latter,
following \S \ref{sctn_OtherParams} we fix $T_\mathrm{shell} = 3.5 \times 10^6$ K (with $T_\mathrm{FB}$ somewhat larger in general).

\subsection{Restriction of Parameter Space}
\label{sctn_Results}

\begin{figure}
\centerline{\includegraphics[width=1.0\textwidth]{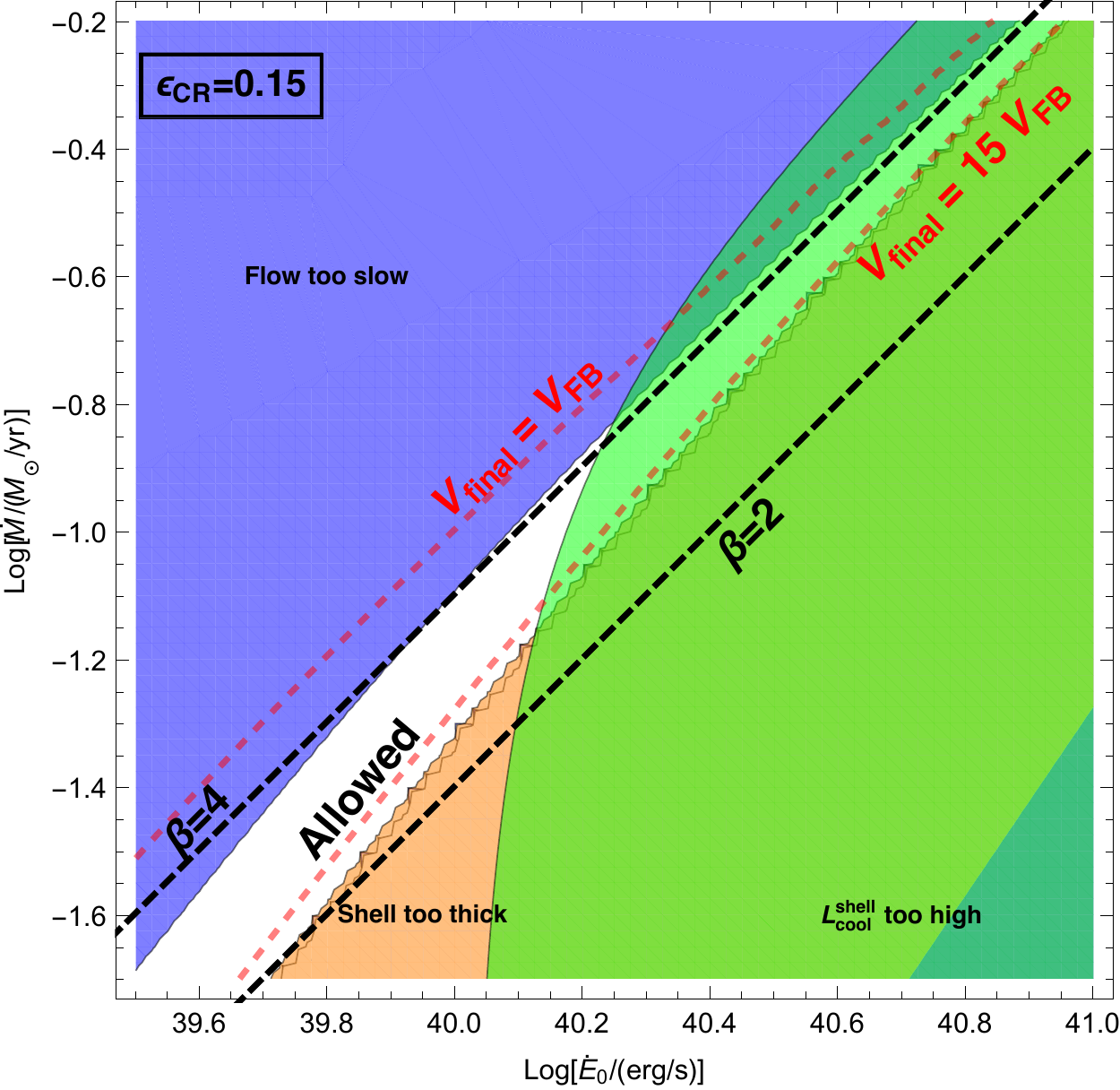}}
\caption{
{\bf Explored $\{\dot{E}_0,\dot{M} \}$ parameter space and excluded regions} for the case that $P_\mathrm{halo}^{FB} = 2 $ eV cm$^{-3}$, $\epsilon_{CR} = 0.15$ and $T_\textrm{\tiny{shell}} = 3.5 \times 10^6$ K.
White space shows the allowed region.
The region masked out in 
green predicts a bolometric cooling luminosity ($> 10^{40}$ erg/s) from the compression shell (R3) that is too high to be compatible with X-ray constraints from ROSAT \citep{Snowden1997,Almy2000}. 
The region masked out in 
orange predicts  a compression shell that is too thick ($> 1.4$ kpc).
The region masked out in blue predicts that the flow from the interior region into the shell is too slow, i.e., 
the transport time from R2 to R3 is longer than the plasma cooling time in R2 so the plasma cools before it reaches the shell at variance with the assumptions of the model.
The thick black dashed curves delineate the star-formation-driven mass efflux vs.~mechanical energy input into the giant outflows
for mass loading rates  of $\beta$ = 2  and 4, scaling the results of \citet[][]{Strickland2009}.
The dashed red lines trace the locus of $\dot{E}_0/\dot{M}$ points that generate asymptotic volumes of, respectively, $V_\mathrm{FB}$
 and $15 \times V_\mathrm{FB}$ obtained from eq.~\ref{eq_rf}.
}
\label{fig_plotNTPw}
\end{figure}

With the downstream temperature and shock height corresponding to every $\{\dot{E}_0, \dot{M} \}$ point determined, we can determine a number of subsequent quantities including the Mach number of the shock, the bolometric thermal luminosity of the Bubbles, and the thickness of cooling material in the compression shell leading up to the contact discontinuity from the inside (region R3 in Figure~\ref{fig_BubbleSchematic}).
By confronting these modeled values with physical or observational constraints we can exclude much of the $\{\dot{E}_0,\dot{M} \}$ parameter space; cf. Figure~\ref{fig_plotNTPw}.
In this figure we show  the case
that the downstream, shocked plasma  in region R2 has reached pressure equilibrium with an exterior atmosphere of
 $P_\mathrm{halo} = 2 $ eV cm$^{-3}$, 
 the fraction of mechanical power available at the shocks going into cosmic rays is
 $\epsilon_{CR} = 0.15$  
 and the compression shell at R3 has an effective temperature of $T_\textrm{\tiny{shell}} = 3.5 \times 10^6$ K.
Most of the potential parameter space for these parameters can be excluded by only three considerations:
\begin{enumerate}
\item The compression shell should not be too thick; 
the requirement of a flat, projected surface brightness in $\gamma$-rays leads to a geometric restriction that 
the shell thickness, $d_\mathrm{shell}$, satisfies $d_\mathrm{shell} < 0.456 r_B \sim 1.4$ kpc (see \S\ref{sctn_SrfcB}).
This excludes the green region in Figure~\ref{fig_plotNTPw}.
Note that the shell thickness  is determined by by the generalization of eq.~\ref{eq_shellThickness} below.

\item The time for the gas to flow from the shocked interior (R2) into the compression shell (R3) (at speed given by eq.~\ref{eq_v_shell} below) should be shorter than the radiative cooling time in R2 (otherwise cooling takes place in R2, not in R3, at variance with the assumptions of the model).
\item The total cooling luminosity, $L_\textrm{\tiny{cool}}^\textrm{\tiny{shell}}$, of the compression shell, R3, should not exceed the upper limit on the bolometric luminosity of the X-ray bulge which, conservatively, is $10^{40}$ erg/s \citep{Snowden1997,Almy2000}.
\end{enumerate}
Note that while, in principle, the allowed parameter space seems to extend off the plot towards the lower left, there are two considerations that exclude  $\dot{E}_0 \lsim 3 \times 10^{39}$ erg/s, 
at least in the context of the current work: 
i) within a hadronic model, the minimum (thick target) cosmic-ray luminosity necessary to sustain the Bubbles' $\gamma$-ray luminosity is $\sim 3 \times 10^{38}$ erg/s \citep{Crocker2011}; 
a mechanical power at least a factor of $\sim$3 (for high cosmic-ray acceleration efficiency) larger than this has to be be delivered {\it at the shocks} in order to sustain their emission and one must then factor in gravitational losses (cf. Figure~\ref{f:rshk});
ii) we show below  (cf. Figure~\ref{plotSynchrotronGiantShocks}) that, in order to reproduce the giant outflows' synchrotron luminosity, 
an unreasonably large fraction of the total cosmic-ray luminosity of the shocks has to go into cosmic-ray electrons for  $\dot{E}_0 \lsim 3 \times 10^{39}$ erg/s.

We show below that, {\it within the restricted parameter space} we have mapped out, our model well reproduces the $\gamma$-ray, microwave, and radio phenomenology of the Bubbles.
First, however, we remark on three interesting findings.

First,  the $\dot{M}/\dot{E}_0$ ratio indicated by this restricted parameter space  is again entirely consistent with -- and, indeed, indicative of -- the Bubbles' being driven by nuclear star-formation.
In Figure~\ref{fig_plotNTPw} the over-plotted black curves delineate  expected star-formation-driven mass flux vs.~star-formation-driven mechanical energy input for mass loading rates, as marked, of $\beta$ = 2, and 4 that bracket our allowable parameter space \citep[scaling results of][]{Strickland2009}.
Here
mass loading is defined with respect to the mass injection rate from supernova and stellar winds, $\dot{M}_{SN + SW}$,
 that accompanies a given rate of star formation, $SFR$, not with respect to this star formation rate itself: 
 $\dot{M}_{SN + SW} =  \dot{M}_{SN + SW}[SFR]$ with $\dot{M}_{tot} \equiv \beta \times \dot{M}_{SN + SW}$ \citep{Strickland2009}.
From modeling of the diffuse, hard X-ray emission emanating from the starburst exemplar M82, \citet{Strickland2009} find -- for centralised mass loading {\it only} -- 
mass loading rates in  the range $1 < \beta < 2.8$ with a practical upper limit at $\beta \sim 10$ for a very hot, 30-70 $\times 10^6$ K, wind.
The near coincidence of our allowable parameter space with a scale for the $\beta$ parameter derived from modeling other star-formation-driven outflows \citep[][and references therein]{Strickland2009}
supports the notion that the outflow is star-formation-driven; {\it a priori} there is no reason for an AGN-driven outflow to be anywhere close to this $\beta$ range.

Note that there is, in general, the possibility for an outflow to also be subjected to {\it distributed} (in addition to centralised) mass loading.
The population of HI clouds entrained into the nuclear outflow detected by \citet{McClure-Griffiths2013} and the high-speed, warm ionised material found by \citet{Fox2014} 
 together
constitute plausible evidence that exactly such distributed mass loading is taking place in the nuclear outflow, presumably as it scrapes along the inside of the CMZ molecular gas torus. 
The fact that our allowable parameter space extends into regions corresponding to very slightly higher values of $\beta$ than determined by \citet{Strickland2009} for M82 may reflect the occurrence of  both centralised and distributed mass loading in generating the final effective $\beta$.
Note that the pressure in the CMZ $H_2$ is very close to that in the central X-ray-emitting plasma; the molecular gas is thus able to collimate this plasma and cause it to escape down the steepest density gradient, into the Galactic halo\footnote{At the same time, the CMZ torus has the necessary weight to anchor the Fermi Bubble field lines.}.

A second interesting finding comes from consideration of the dashed red curves in Figure~\ref{fig_plotNTPw}.
These show where the  final volumes (obtained from the conditions of pressure equilibrium and balance between cooling radiation and incoming enthalpy flux according to eq.~\ref{eq_rf}) of  
spherical, radiative bubbles fed with energy and mass at the nominated rates are, respectively, equal to the current estimated volume of the observed Fermi Bubbles and $15 \times$ the current volume.
Left and upwards of the $V_\mathrm{FB}$ curve, radiative bubbles never reach the current  size of the observed Fermi Bubbles also, therefore, also ruling out this region of parameter space.
With increasing distance right and downwards of the  $15 V_\mathrm{FB}$ curve, there is an increasingly severe temporal fine-tuning problem as  
radiative bubbles spend only a brief time over their entire evolution close to the current observed size. 
We emphasise, then,  that for 
the region of parameter $\{\dot{E}_0,\dot{M}\}$ parameter space delimited by the independent considerations set out above, the Fermi Bubbles are naturally found to have a size close to that observed.
In other words, there is no ``Why now?" problem in this model.

A final interesting finding is  that allowable values of $P_\mathrm{halo}$ and $\epsilon_\mathrm{CR} = 0.15$  are quite tightly constrained within our model: we are not able to reproduce the observed $\gamma$-ray luminosity of the Bubbles (see \S\ref{sctn_NTEmission} below) for $P_\mathrm{halo} \lsim 1.5$ eV cm$^{-3}$; for $\epsilon_\mathrm{CR} \gsim 0.3$ we find that the parameter space allowed by the three considerations listed above closes off completely, and, for   $\epsilon_\mathrm{CR} \lsim 0.05$ we cannot simultaneously reproduce both the observed $\gamma$-ray and radio + microwave luminosities of the giant outflows (see below).
This allowable range for $\epsilon_\mathrm{CR}$
is consistent with 
independent determinations of the fraction of shock mechanical power that goes into such particles in other systems \citep[e.g.,][on the nuclear outflow from local starburst NGC253]{Zirakashvili2006}.

We lastly note that, in the allowed parameter space, cooling in the shell does not saturate the  mechanical energy injected in the nucleus; this means that the contact discontinuity must be expanding and thereby doing work  on its surroundings at a rate $P \dot{V}$ that saturates the residual mechanical energy.
The Bubbles, therefore, have not yet reached their asymptotic, final volume.
In fact, we find that the contact discontinuity must be expanding at about 2 km/s to saturate the injected power not lost to cooling radiation in the shell.
Comparing against the results from \S \ref{s:sctn_BubbleModel}, the spherical radiative bubble is expanding this quickly at an age of $5 \times 10^8$ yr when, self-consistently,  it has reached a volume equivalent to the observed Fermi Bubbles (see Figure~\ref{f:velocity}); thus the model hangs together in this parameter region.

\section{Non-Thermal Particles Accelerated at the Shocks and their Emission}
\label{sctn_NTEmission}

In the previous section we delimited the range of allowable $\{\dot{E}_0,\dot{M}\}$ parameter space for the GC's giant outflows.
In this section we want to determine what our model predicts for the non-thermal emission associated with these outflows, particularly the emission associated with the cosmic ray electron and cosmic ray proton and heavier ion populations accelerated at the giant, reverse shocks and subsequently carried into the Bubbles' shocked downstream regions (R2) and, if they survive long enough, further into the compression shell leading up to the contact discontinuity (R3).
We will determine the hadronic luminosity of the non-thermal protons and heavier ions colliding with gas in R2 and, dominantly, R3, and the radio + microwave synchrotron and $\gamma$-ray inverse-Compton emission from the non-thermal electrons within each point of the $\{\dot{E}_0,\dot{M}\}$ parameter space.
 Remarkably we show below that -- within the range of  $\dot{E}_0$ and $\dot{M}$ allowable according to the analysis of \S\ref{sctn_Scan}-- our model reproduces the radio, microwave, and $\gamma$-ray non-thermal phenomenology of the giant outflows.

\subsection{Non-thermal particle spectra and energetics}

In the test particle limit (where the dynamics are completely dominated by thermal particles), 
the spectral index of the non-thermal particle population accelerated at a shock is given by \citep{Bell1978}
\begin{equation}
\gamma_\textrm{\tiny{CR}} = \frac{R + 2}{R - 1} = \frac{(\gamma + 1)M^2}{(\gamma - 1)M^2 + 2} \, ,
\label{eqn_CRSPIN}
\end{equation}
where $R$ is the shock compression ratio.
Observations of $\gamma$-ray and microwave emission reviewed below indicate populations of non-thermal protons and electrons with power-law distributions, $dN/dE  \propto E^{-\gamma}$ with spectral indices of $\gamma < 2.3$ requiring Mach numbers $M \gsim 4$.
Consistent with this,  our modeling favors upstream flows at the giant shocks with Mach numbers  in the range 6-9.
This will generate non-thermal particle populations with spectral indices of $\sim 2.1$, close to the central values required to explain both the Bubbles' $\gamma$-ray  and the Haze's 
hard microwave emission.

Energy scales for individual, non-thermal particles of interest to us can be reckoned as follows.
Adopting a fiducial magnetic field of 7 $\mu$G downstream of the shock (\S \ref{sssctn_B}) the characteristic energy of electrons synchrotron radiating at $\nu =$ 23 GHz is 
$E_c = 18$ GeV $(B/7 \ \mu$G$)^{-1/2} \ (\nu/23 \ $GHz$)^{1/2}$.
For IC emission, $\gamma$-rays at $E_\gamma = 1$ GeV are up-scattered by primary electrons of energy 
$E_e \simeq 640$ GeV $(E_\gamma/1$ GeV$)^{1/2} \ (E_{\gamma,0}/6.3 \times 10^{-4}$ eV$)^{-1/2}$ adopting an energy for the photon background characteristic of the CMB.
Finally, proton primaries of hadronic $\gamma$-rays in the observed range of 1-100 GeV have energies covering the approximate range 10-1000 GeV.

These characteristic energies should be compared with upper limits to the energies that the giant shocks might accelerate non-thermal particles.
The Hillas criterion that a particle's acceleration time be at least equal to the maximal diffusive escape time from an acceleration region, of fiducial size $r_\textrm{\tiny{acc}} \sim$ 100~pc[\footnote{This scale corresponds to the approximate size of linear features found in the 2.3 GHz polarization map that, below, we  identify with plausible signatures of the shocks.}], sets a maximum energy 
limit of $E_\textrm{\tiny{max}} \simeq 10^{15}$~eV $r_\textrm{\tiny{acc}}/(100$~pc) $B/(7 \ \mu$G) v/(500 km/s).
For synchrotron loss-limited acceleration of electrons over a timescale $t_\textrm{\tiny{acc}} \equiv 4 D/v^2$ where $D$ is the diffusion coefficient near the shock and $v$ is a characteristic velocity, in the most optimistic case of Bohm diffusion $D_\textrm{\tiny{Bohm}} = 1/3 \ r_g c$, with $r_g$ the particle gyroradius, we find a maximum upper limit to the accelerated electron energy of $E_e^\textrm{\tiny{max}} \simeq 4 \times 10^{13}$ eV  $(B/(7 \ \mu$G))$^{-1/2} \ (v/(500 $ km/s)).
In this case, the shocks can easily accelerate electrons up to the requisite energies to produce the observed microwave emission and even the IC $\gamma$-rays by up-scattering the CMB (though whether  such high energy electrons can reach the top of the Bubbles is another matter).

In our modeling we have explored different values for the fraction, $\epsilon_{\mathrm{CR}}$ of shock mechanical power that is injected into freshly-accelerated, non-thermal particles, $L_{\mathrm{CR}}$ finding that ranges of $\sim 5-20$ \% are acceptable.
For the magnetic field amplitudes explored below, we find that electrons are required to take up a fraction 5-30\% of $L_{\mathrm{CR}}$ in order to explain the radio and microwave emission from the Bubbles, 
with the remainder going into ions, dominantly protons.

\subsection{$\gamma$-ray Emission from the Bubbles}

\subsubsection{Observations}

The measured $\gamma$-ray spectrum of the Bubbles is rather hard, approximately $F_\gamma \propto E_\gamma^{-2.1}$ between 1-100 GeV \citep{Su2010}
but with a hardening below $\sim 1$ GeV.
The Bubbles have a rather flat {\it projected} surface brightness \citep{Su2010}.
A recent paper from the {\it Fermi} collaboration   \citep{Ackermann2014} gives  the 100 MeV - 500 GeV luminosity of the Bubbles as  
$4.4\pm0.1$[stat]$^{+2.4}_{-0.9}$[sys]$ \times 10^{37}$ erg/s.
This publication also reports that the Bubbles' $\gamma$-ray spectrum cuts off above $\sim 100$ GeV.

\subsubsection{Expected Hadronic $\gamma$-ray Emission}
\label{sctn_HadronicGammaEmission}

A successful model of the Bubbles' $\gamma$-ray phenomenology needs to  explain their luminosity, spectrum, and morphology (and, most generally, spectral morphology).
Generic to any hadronic model is a $\sim$ GeV-scale low energy down-turn feature in the spectral energy distribution, 
whose position in energy is fixed by the mass of the neutral 
pion.
This is broadly consistent with the spectral phenomenology alluded to above \citep{Crocker2011}.

We show immediately below that our model provides a good match to the overall  $\gamma$-ray luminosity of the Bubbles.
In addition, for acceptable regions of the $\{\dot{E}_0,\dot{M}\}$ parameter space we find moderately high
Mach numbers (6-9) at the giant shocks,  naturally accounting for the hard 
1-100 GeV hard spectrum of the Bubbles (and the hard, non-thermal microwave Haze as we later explain).
We also show below that our model predicts the {\it volumetric} emission from the Bubbles should peak towards their edges; 
this is consistent with  their flat (projected) surface brightness.
We do not explicitly deal with the observed high-energy cut-off here; this may indicate either a high-energy cut-off in the accelerated proton spectrum emerging from the shocks 
or a loss of confinement of the cosmic ray hadrons in the Bubbles at $\sim$TeV energies.

\subsubsection{Hadronic $\gamma$-ray Luminosity}
\label{sctn_HadronicGammaLuminosity}

Adapting the calculations presented
by \citet{Crocker2014} we can calculate the hadronic luminosity of the Bubbles.
Our essential argument is that, given the expansion timescale, the evolution of the plasma in the Bubbles cannot be treated as adiabatic but rather the implication of plasma cooling must be considered.
This cooling leads to local thermal instability and the formation of over-dense, cool condensations that, having lost buoyancy, fall down through the Bubbles under gravity. 
(In the steady state limit the drop-out of these condensations completely saturates the injection rate of fresh plasma mass.)
Globally there is a slow flow in the Bubbles of material downstream of the reverse shock and towards the contact discontinuity.
As the CD is approached, this flow must be decelerated with a consequent increase in plasma density.
Given the $n^2$ dependence of the cooling, it is in this  over-dense shell region (R3 in Figure~\ref{fig_BubbleSchematic})  that thermal instability will preferentially form the condensations.
Thus there will  be a flow of gas at a rate (saturating $\dot{M}$ in the steady state limit) of shocked plasma from the  interior volume of the Bubbles behind the reverse shocks, R2, into this  shell, R3, with subsequent thermal collapse, and drop-out under gravity.
We review the observational evidence for formation of cool gas condensations near the edges of the Bubbles in \S \ref{ssctn_Cndsnts}.

In our model, in the interior region, R2, the gas evolves adiabatically whereas in the denser shell, R3, it evolves radiatively.
However, a crucial point is that {\it cosmic rays and magnetic fields  evolve adiabatically in both shell and volume gas}.
Adiabatic compression of these non-thermal ISM components thus raises their {\it relative} pressure within the shell.
In fact,
compression of the shell by the  plasma in   R2 is arrested when the adiabatic energy-gain of magnetic fields and cosmic rays results in their reaching pressure equilibrium with the interior plasma.
The size of the shell is dually controlled by this together with the requirement that the cooling radiation taking place within it 
(strictly, in the steady state limit) balances the enthalpy flux represented by the flow of hot plasma from the hot interior.

As we now show, we can use this pressure equilibrium between the adiabatically-compressed cosmic rays and magnetic fields in R3 and the thermal plasma in R2 
and the size of the shell determined by the enthalpy flux argument
to determine
the hadronic $\gamma$-ray luminosity of the Bubbles.
This is dominated by collisions in the shell between the adiabatically compressed cosmic rays and the compressed, cooling gas which can be calculated as shown below.
The {\it Fermi}-LAT band hadronic (`pp') luminosity of the shell is 
\begin{eqnarray}
L_\gamma^\mathrm{pp} &\simeq& \frac{3}{2} \times \frac{1}{3} \times f_\mathrm{bol} \ u_\mathrm{p}^{\mathrm{shell}} \ V_\mathrm{shell}  \ n_\mathrm{H}^\mathrm{shell} \ \sigma_\mathrm{pp} \ \kappa_\mathrm{pp} \ c \nonumber \\
& = & \frac{f_\mathrm{bol} }{2} \ u_\mathrm{p} ^{\mathrm{shell}} \ \frac{M_{\mathrm{shell}}}{\mu m} \ \frac{n_\mathrm{H}^\mathrm{shell} }{n_{\mathrm{tot}}^\mathrm{shell} }  \ \sigma_\mathrm{pp} \ \kappa_\mathrm{pp} \ c\, .
\end{eqnarray}
Here, in the first line, the first 3/2 pre-factor (approximately) corrects for the presence of heavy ions amongst beam and target nuclei \citep{Mori1997}, $1/3$ comes from the relative multiplicity of $\pi^0$ amongst all daughter pions,  $f_\mathrm{bol}$ is the fraction of the bolometric luminosity emitted in the relevant $\gamma$-ray band, $u_\mathrm{p} ^{\mathrm{shell}}$ is the energy density of relativistic protons in the shell,  $\sigma_\mathrm{pp} \simeq 4 \times 10^{-26}$ cm$^2$ is the total hadronic cross-section and $\kappa_\mathrm{pp} \simeq 0.5$ is the hadronic inelasticity.
In the second line $n_\mathrm{H}^\mathrm{shell}/n_{\mathrm{tot}}^\mathrm{shell} = 0.44578$ is the $H^+$ to total plasma number density ratio in the shell at solar metallicity, $m$ is the atomic mass unit, and $\mu = 0.6039$ is the mean mass of the particles in a solar metallicity plasma.

We now assume pressure balance between the relativistic shell components (`p' for protons and heavier ions and `B' for the magnetic field), 
$P_\mathrm{p}^{\mathrm{shell}} + P_\mathrm{B}^{\mathrm{shell}}$, and the thermal plasma in the interior, $P_\mathrm{FB}$,
as justified above \citep[and also adopt cosmic ray/magnetic field equipartition in the shell as justified observationally by][$P_\mathrm{p}^{\mathrm{shell}} \simeq P_\mathrm{B}^{\mathrm{shell}}$]{Carretti2013} so that the pressures represented by these various components are related by
\begin{equation}
P_\mathrm{p}^{\mathrm{shell}} \equiv \frac{1}{3} u_\mathrm{p} ^{\mathrm{shell}} =  \frac{1}{3} u_\mathrm{B}^{\mathrm{shell}} = \frac{1}{2} P_\mathrm{FB} \, .
\label{eq_P_balance}
\end{equation}
We now calculate $M_{\mathrm{shell}}$ as follows:
mass conservation implies that the flux of mass from interior volume, of number density $n_{\mathrm{tot}}^{\mathrm{FB}}$,
 into the shell is equal (in the steady state limit) to the mass flux into the Bubbles, $\dot{M}$.
Thus the effective speed, $v$ with which plasma is carried across the the surface, of area $A_{\mathrm{shell}}$ separating the interior region R2 from the shell R3 is
\begin{equation}
v \equiv \frac{\dot{M}}{\mu  m \ A_{\mathrm{shell}} \ n_{\mathrm{tot}}^{\mathrm{FB}} } \simeq 13 \  \textrm{km/s}  \ \left(\frac{\dot{M}}{\msun/\textrm{year}} \right) 
\ \left(\frac{n_{\mathrm{tot}}^{\mathrm{FB}}}{0.002 \ \textrm{cm}^{-3}} \right)^{-1}  \ \left(\frac{A_{\mathrm{shell}}}{2.5 \times 10^{45} \ \textrm{cm}^2} \right)^{-1} \, .
\label{eq_v_shell}
\end{equation}
The enthalpy flux represented by this flow is dissipated in cooling radiation from the shell, thus
\begin{equation}
\frac{5}{2} \ n_{\mathrm{tot}}^{\mathrm{FB}} \ k_B \ T_{\mathrm{FB}} \ v \ A_{\mathrm{shell}} = (n_{\mathrm{tot}}^{\mathrm{shell}})^2 \ \Lambda[T_{\mathrm{shell}}] \ V_{\mathrm{shell}} 
\label{eq_enthalpy}
\end{equation}
where $\Lambda[T_{\mathrm{shell}}]$ is the cooling function at the shell temperature and metallicity (the latter assumed to be solar as for the hot plasma) and, geometrically, we assume a thin shell.
As for the shell temperature, as we have emphasised, the X-ray spectrum across the edge of the Bubbles, encompassing the region of the shell, 
has been successfully fit with a single temperature, collisional ionisation equilibrium plasma model of temperature $\sim 3.5 \times 10^6$ K and we therefore 
adopt $T_{\mathrm{shell}} = 3.5 \times 10^6$ K (so that the temperature in the interior volume is necessarily $> 3.5 \times 10^6$ K).
In reality, 
the gas distribution in the shell is more complicated as the gas cascades  
from the high injection temperature, where it starts to cool, down to optical emission line temperatures. 
Such a
multi-temperature distribution is inferred, e.g., from the coincident X-ray and H$\alpha$ emission observed from
filaments around the bright central galaxies in clusters \citep[e.g.,][]{Fabian2011} 
and is common to cooling flows.
In any case, here we ignore the complication of the multi-temperature nature of the shell gas 
and assume the effective temperature apposite to the calculation of shell cooling losses is  precisely that found from fitting to the observed X-ray radiation for a single-temperature model.

Rearranging and combining eqs.~\ref{eq_v_shell} and \ref{eq_enthalpy} we find
\begin{equation}
V_{\mathrm{shell}}   = \frac{5}{2} \frac{k_B \ T_{\mathrm{FB}} \ \dot{M}}{ (n_{\mathrm{tot}}^{\mathrm{shell}})^2 \ \Lambda[T_{\mathrm{shell}}] \ \mu m} = 
\frac{5}{2} \frac{k_B \ T_{\mathrm{FB}} \dot{M}}{ (n_{\mathrm{tot}}^{\mathrm{FB}})^2 \ \Lambda[T_{\mathrm{shell}}]    \ \mu m} 
\left(\frac{n_{\mathrm{tot}}^{\mathrm{FB}}}{n_{\mathrm{tot}}^{\mathrm{shell}}}\right)^2
\label{eqn_Rshell}
\end{equation}
Following eq.~\ref{eq_P_balance} specifiying pressure balance between the adiabatically compressed shell magnetic field and cosmic rays and the thermal gas in the interior 
and given adiabatic evolution of the cosmic rays: 
\begin{equation}
u_{\mathrm{p}}^{\mathrm{shell}} = \left(\frac{n_{\mathrm{tot}}^{\mathrm{shell}}}{n_{\mathrm{tot}}^{\mathrm{FB}}}\right)^{4/3} u_{\mathrm{p}}^{\mathrm{FB}} \, ,
\end{equation}
then for $u_{\mathrm{therm}}^{\mathrm{FB}}$ as the internal energy density of the plasma in the Bubbles, 
$L_{\mathrm{CR}}$ and $\dot{E}_{\mathrm{therm}}$  
the power fed, respectively, into freshly accelerated cosmic rays and freshly heated plasma at the reverse shocks,
we have that 
\begin{equation}
\frac{n_{\mathrm{tot}}^{\mathrm{FB}}}{n_{\mathrm{tot}}^{\mathrm{shell}}} = \left(\frac{u_{\mathrm{p}}^{\mathrm{FB}}}{u_{\mathrm{therm}}^{\mathrm{FB}}}\right)^{3/4} \simeq \left(\frac{L_{\mathrm{CR}}}{\dot{E}_{\mathrm{therm}}}\right)^{3/4} \simeq \epsilon_{\mathrm{CR}}^{3/4} 
\end{equation}
where the last two near-equalities rely on adiabatic evolution of both cosmic rays and thermal plasma 
downstream of the shocks {\it in the interior volume} R2.
We thus find, implicitly defining the thickness of the shell, $d_\mathrm{shell}$, (assumed to be composed of  two concentric shells, one around each spherical bubble): 
\begin{eqnarray}
V_{\mathrm{shell}} & \equiv & V_{\mathrm{FB}} - 2 \frac{4 \pi}{3} (r_\mathrm{FB} - d_\mathrm{shell})^3  \nonumber \\
& \simeq & \frac{5}{2} \frac{k_B \ T_{\mathrm{FB}} \dot{M}}{ (n_{\mathrm{tot}}^{\mathrm{FB}})^2 \ \Lambda[T_{\mathrm{shell}}]    \ \mu m} \left(\frac{L_{\mathrm{CR}}^{\mathrm{FB}}}{\dot{E}_{\mathrm{therm}}^{\mathrm{FB}}}\right)^{3/2}
\label{eq_shellVolume}
\end{eqnarray}
which implies a characteristic thickness of the shell in the geometrically thin limit
\begin{eqnarray}
d_\mathrm{shell} & \simeq & 510 \  \mathrm{pc} \ \left( \frac{T_{\mathrm{FB}}}{6.0 \times 10^6 \ \textrm{K}} \right) \left(\frac{\dot{M}}{0.1 \msun/\mathrm{yr}}\right) \left( \frac{n_{\mathrm{tot}}^{\mathrm{FB}}}{0.004 \ \textrm{cm}^{-3}} \right)^{-2} \left(\frac{L_{\mathrm{CR}}^{\mathrm{FB}}}{\dot{E}_{\mathrm{therm}}^{\mathrm{FB}}}/0.1\right)^{3/2} 
\label{eq_shellThickness}
\end{eqnarray}
where we set $A_{\mathrm{shell}} = 2.5 \times 10^{45}$ cm$^2$.
The total mass of the shell, in all generality, is
\begin{eqnarray}
M_{\mathrm{shell}} 
&\simeq &\frac{5}{2} \frac{k_B T_{\mathrm{FB}} \dot{M}}{n_{\mathrm{tot}}^{\mathrm{FB}} \ \Lambda[T_{\mathrm{shell}}]} 
\left(\frac{L_{\mathrm{CR}}^{\mathrm{FB}}}{\dot{E}_{\mathrm{therm}}^{\mathrm{FB}}}\right)^{3/4} 
\nonumber \\
& \simeq & 2.5 \times 10^7 \ \msun \left( \frac{T_{\mathrm{FB}}}{6.0 \times 10^6 \ \textrm{K}} \right) \left(\frac{\dot{M}}{0.1 \msun/\mathrm{yr}}\right) \left( \frac{n_{\mathrm{tot}}^{\mathrm{FB}}}{0.004 \ \textrm{cm}^{-3}} \right)^{-1} \left(\frac{L_{\mathrm{CR}}^{\mathrm{FB}}}{\dot{E}_{\mathrm{therm}}^{\mathrm{FB}}}/0.1\right)^{3/4} \nonumber \\
\, 
\end{eqnarray}
and we may finally write the 1-100 GeV hadronic luminosity of the shell as 
\begin{eqnarray}
L_\gamma^\mathrm{pp} &\simeq& \frac{15 \ f_\mathrm{bol} }{8} 
\frac{(k_B T_{\mathrm{FB}})^2 \dot{M}}{\mu m \Lambda[T_{\mathrm{shell}}]} 
\frac{n_H}{n_{\mathrm{tot}}} \ \left(\frac{L_{\mathrm{CR}}^{\mathrm{FB}}}{\dot{E}_{\mathrm{therm}}^{\mathrm{FB}}}\right)^{3/4}
\sigma_\mathrm{pp} \ \kappa_\mathrm{pp} \ c 
\nonumber \\
&\simeq& 2.4 \times 10^{37} \  \mathrm{erg/s} \left(\frac{T_{\mathrm{FB}}}{6.0 \times 10^6 \ \mathrm{K}}\right)^2 \left(\frac{\dot{M}}{0.1 \msun/\mathrm{yr}}\right) \left(\frac{L_{\mathrm{CR}}^{\mathrm{FB}}}{\dot{E}_{\mathrm{therm}}^{\mathrm{FB}}}/0.1\right)^{3/4}
\end{eqnarray}
which compares favourably to the $\sim 2 \times 10^{37}$ erg/s 1-100 GeV luminosity of the Bubbles \citep{Su2012} if $\gtrsim 10$\% of the mechanical power at the shock is delivered to cosmic rays.
Note that in the interior region, we are normalising to a somewhat hotter temperature than the $3.5 \times 10^6$ K measured in X-rays which we assume is characteristic of the cooling shell and that here $f_\mathrm{bol} \simeq 0.4$.

\begin{figure}
\centerline{\includegraphics[width=1.0\textwidth]{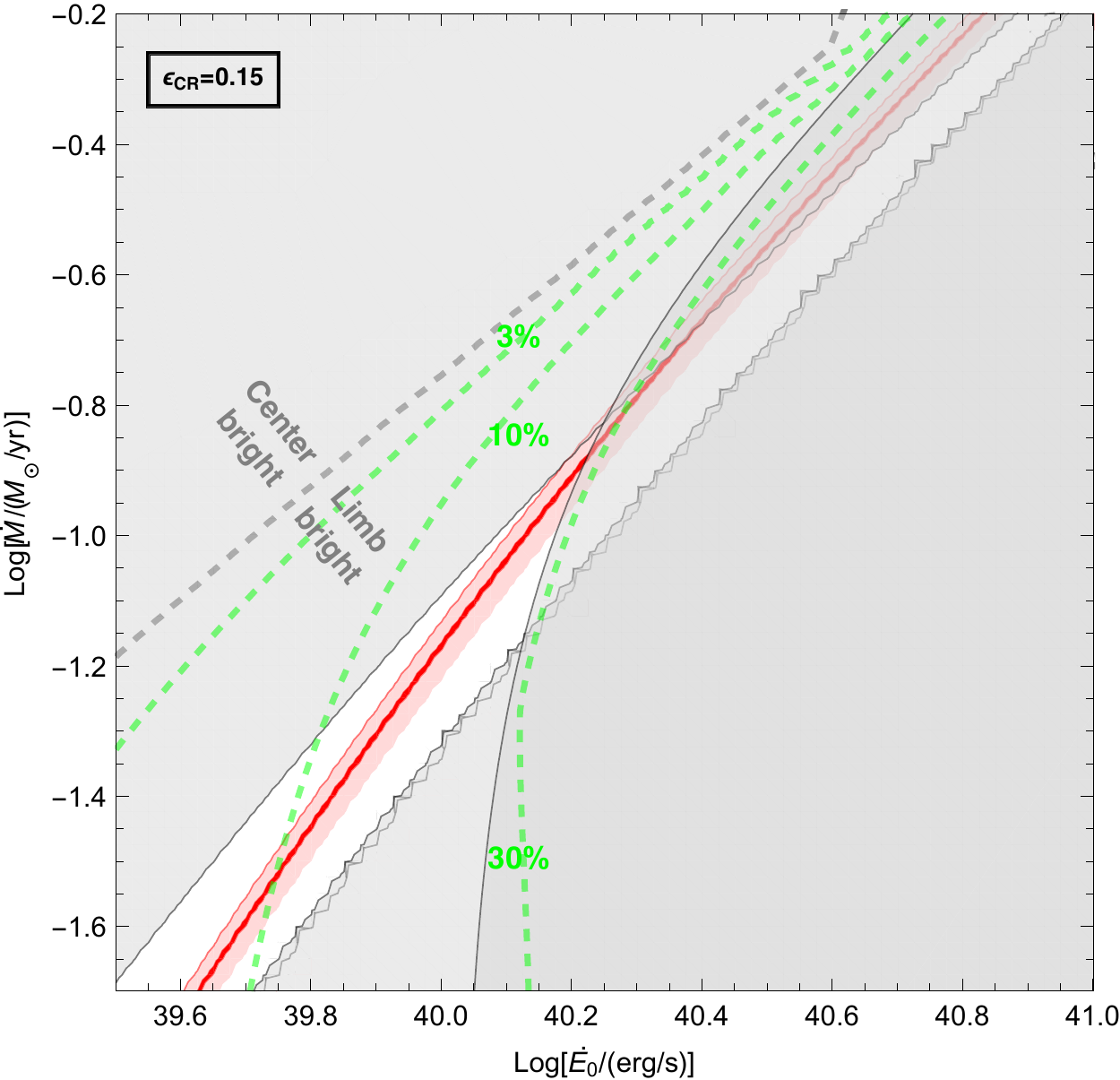}}
\caption{{\bf Predicted hadronic and IC $\gamma$-ray luminosity within the allowed parameter space} for the case that $P_{\mathrm{halo}}= 2 $ eV cm$^{-3}$, $\epsilon_{\mathrm{CR}} = 0.15$ and $T_\mathrm{shell} = 3.5 \times 10^6$ K.
Regions that are masked out in gray   do not obey all the
physical or observational constraints as in Figure~\ref{fig_plotNTPw}.
The pink/red band is the locus of parameter space points for which the  predicted hadronic luminosity is close to that observed (central value of $4.4 \times 10^{37}$ erg/s 
over 100 MeV $< E_\gamma <$ 500 GeV
with 1-$\sigma$ statistical+systematic errors shown as the pink regions: \citealt{Ackermann2014}).
The green dashed curves report the total 100 MeV-500 GeV inverse Compton (IC) luminosity as a proportion (3\%,10\%, or 30\% as labeled)
 of $4.4 \times 10^{37}$ erg/s.
The IC curves assume that, of the 15\% of the  mechanical energy at each shock that goes into freshly-accelerated cosmic rays,  10\%  goes into primary electrons, $L_e = 0.1 \ L_\mathrm{CR} = 0.1 \times 0.15 \times \dot{E}$.
The gray, dashed curve shows where the naive flat surface brightness condition (eq.~\ref{eqn_Flat}) is exactly satisfied.
}
\label{plotMaskedNTemission}

\end{figure}

The results of a full calculation of the hadronic luminosity of the filaments in the downstream is plotted in Figure~\ref{plotMaskedNTemission}.
Note that there is a sub-dominant component of the Bubbles' hadronic $\gamma$-ray emission due to collisions between the cosmic-ray protons and the rarefied, volume-filling plasma but, for fiducial numbers, this provides only 5-20\% of the integrated shell luminosity.

\subsubsection{Inverse Compton $\gamma$-ray Emission}

As has been noted, the same population of non-thermal electrons synchrotron radiating at microwave frequencies to explain the Haze may emit IC $\gamma$-rays in the Fermi band with the correct Bubble spectrum.
Within our scenario, however, we find that this process is sub-dominant to the hadronic $\gamma$-ray emission explaining $\sim$ 10-30\% of the total observed 100 MeV-500 GeV $\gamma$-ray luminosity of the Bubbles for the fiducial case that the
power fed into freshly accelerated electrons is 10\% of the power fed into all CRs (which is, itself, assumed, to be 15\% of the mechanical power at the shock in this figure): $L_e = 0.1 \ L_\mathrm{CR} = 0.1 \times 0.15 \times \dot{E}$; see fig~\ref{plotMaskedNTemission}.
Of course, this conclusion is contingent on the assumed $L_e/L_\mathrm{CR}$ but significantly higher values of this ratio can be rejected on two considerations:
\begin{enumerate}

\item Within our model, IC cannot reproduce the observed morphology of the Bubbles' overall $\gamma$-ray emission alone 
as the high-energy electrons never reach more than a  $\sim$ 300 pc downstream of the shocks (cf. Figure~\ref{plotDownstreamCRRange}).
On the other hand, this expected concentration of  IC emission does imply that, within a few degrees of the shocks, it can contribute easily more than half 
of the total 100 MeV-500 GeV $\gamma$-ray intensity even for $L_e/L_\mathrm{CR} \sim 0.1$ 
and we thus expect $\gamma$-ray substructure around the shocks; {\it preliminarily, we find evidence for such} (see \S\ref{sctn_blah}).

\item Values $L_e/L_\mathrm{CR} \gsim 0.1$ are disfavored on theoretical grounds \citep{Bell1978}, on the basis of analysis of the Galaxy's global cosmic ray and non-thermal radiation budget \citep{Strong2010}, and from observations and analysis 
of non-thermal radiation associated with giant cluster shocks \citep[and references therein]{Pinzke2013,Brunetti2014}.

\end{enumerate}

\subsubsection{Projected Surface Brightness}
\label{sctn_SrfcB}

As has been remarked, the Bubbles exhibit a rather flat surface brightness.
We consider a toy, two-zone geometric model of a spherical shell of luminosity $L_\mathrm{shell} \equiv \epsilon_\mathrm{shell}  \ V_\mathrm{shell} $  enveloping a spherical interior region of luminosity $L_\mathrm{interior} \equiv \epsilon_\mathrm{interior}  \ V_\mathrm{interior} $.
The outer radius of the shell is $r_B $ and the inner radius of the shell/outer radius of the interior is $r_\mathrm{shell}$. 
In the far field limit, we find from elementary geometric considerations that a roughly flat projected surface brightness  for such a two zone model requires that the volumetric emissivities are related as
\begin{equation}
\left.\frac{\epsilon_\mathrm{shell}}{\epsilon_\mathrm{interior}}\right|_\mathrm{flat} \simeq \left[ 1+ \left(\frac{r_B}{r_\mathrm{shell}} \right)^3 \left(\sqrt{1 - \left( \frac{r_\mathrm{shell}}{r_B} \right)^2 } -1 \right)\right]^{-1} \, .
\label{eqn_Flat}
\end{equation}
This implies the geometric restriction that $r_\mathrm{shell} > 0.544 \ r_B \sim 1.7$ kpc (or $d_\mathrm{shell} \lsim 1.4$ kpc; cf.~orange exclusion region in Figure\ref{fig_plotNTPw}).
Over the range  
$0.544 \ r_B < r_\mathrm{shell} \leq r_B$, one finds
$\epsilon_\mathrm{shell}/\epsilon_\mathrm{interior} |_\mathrm{flat} > 4.3$.
In Figure~\ref{plotMaskedNTemission} the gray dashed curve shows where eq.~\ref{eqn_Flat} is satisfied exactly; left and upwards of the  curve, the overall $\gamma$-ray surface brightness morphology tends to center brightened, and right and downwards of the curve, the morphology tends to limb brightened according to our simplistic analysis.
A constant volume emissivity model would produce a centre-brightened morphology and the original instantiation of the hadronic model \citep{Crocker2011} was criticised on this ground.
The updated hadronic model presented here, if anything, tends towards limb-brightening in the favoured parameter space according to  eq.~\ref{eqn_Flat}.
However, we do not believe that this is a problem for our model for three reasons: 
i) the observed surface brightness is not exactly flat, indeed there is, e.g., a large region  close to the east edge of the southern bubble that appears brightened \citep[`the cocoon':][]{Su2012,Ackermann2014}; 
ii) we have assumed an unrealistically simple geometry in deriving eq.~\ref{eqn_Flat}; 
and 
iii) we expect, in reality, some relative diminution of shell hadronic emission with a compensating increase  in interior emission due to condensations falling down through the Bubbles under gravity, still emitting $\gamma$-rays (cf.~Figure~\ref{fig_BubbleSchematic}).
A more accurate treatment of this geometrical aspect of the  problem requires numerical modeling.

\subsection{Synchrotron Emission}

\subsubsection{Observations}

The 23-44 GHz Microwave Haze is found \citep{Finkbeiner2004,Dobler2008,Dobler2012,Ade2012} in total intensity data up to Galactic latitudes $|b|  \lsim 35^\circ$ 
above which it becomes severely attenuated though there has been a claim \citep{Dobler2012b} that faint, Haze-like emission is detected up to the latitudinal extremities of the Bubbles.
Here we adopt  0.4 sr for the solid angle of the Haze emission.
The Haze has a hard spectrum $F_\nu \propto \nu^{-0.55}$ \citep{Ade2012}, characteristic of synchrotron emission from 
a population of electrons with spectrum $dN_e/dE_e \propto E_e^{-2.1}$ freshly-accelerated at a strong shock.

It has also been claimed \citep{Gold2011} that the Haze exhibits little coincident polarized emission.
On the other hand, regions of enhanced, polarized emission coincident with sub-structure within the Bubbles {\it have} actually been found, in particular, a bright region coincident with part of the the eastern edge of the north bubble \citep{Jones2012} and the counterparts of the strongly magnetised ridges found in polarized intensity at 2.3 GHz.
Indeed, there is diffuse, polarized 23 GHz emission evident over most of the solid angle of the Bubbles \citep[cf.~Figure~3 of]{Carretti2013}.

The polarization ridges appear to wrap around the near surface of the roughly biconical structure circumscribing the Bubbles in a fashion consistent with the rotation of the Galaxy.
There is diffuse, polarized 2.3 GHz emission enveloping the entire Bubbles, stretching  beyond them, in fact, at high latitudes in two plumes that extend towards Galactic north west and south west.
In contrast to the Haze detected in total intensity at microwave frequencies, 
the spectrum of {\it polarized} intensity between 2.3 and 23 GHz is steep, $F_\nu \propto \nu^{-1.1}$ within the solid angle of the Bubbles \citep{Carretti2013} consistent with an electron population of mixed ages that has been cooled by synchrotron and/or IC emission; beyond the Bubbles and extending into the plumes, the spectrum further steepens  \citep[cf.~Figure~S4 for the Supplementary Material of][]{Carretti2013}.

\subsubsection{Synchrotron Emission in our Model}

Before proceeding to  detailed discussion, we explain briefly how our model accounts for the complicated synchrotron emission phenomenology of the GC's giant outflows.
For clarity, note that in general we shall refer to 23-44 GHz band as `microwave' and 2.3 GHz emission as `radio'.

The nucleus directly accelerates a population of hard-spectrum primary electrons that is advected from the region before it can lose much energy radiatively in situ \citep{Crocker2011a,Crocker2011b,Crocker2012,Carretti2013}.
These electrons cool adiabatically and radiatively in the expanding outflow (R1) until they encounter the giant reverse shock in either bubble where they are reaccelerated.

At latitudes in the Bubbles higher than the shocks, there are then effectively three zones for synchrotron emission in our model:
\begin{enumerate}

\item A zone downstream of each reverse shock, partially encompassing R2, where the electrons are being conveyed systematically 
downstream by the combination of the downstream flow and their streaming along magnetic field lines.
In this zone, there is a 1-to-1 correspondence between distance from the shock and electron `age'.
The frequency-dependent extent of this region is given approximately by the requirement that the transport time
from the shock be less than the
 total (synchrotron + IC) cosmic ray electron loss time.
The synchrotron spectrum within this zone is thus `uncooled',  $F_\nu \sim \nu^{-0.5}$,
reflecting that the electron spectrum itself is
simply the hard injection spectrum produced at the shock, $dN_e/dE_e \sim E_e^{-2}$ up to some (position- and frequency-dependent) maximum energy above which  it is exponentially cut-off
(cf. figs.~\ref{plotDownstreamCRRange} \& \ref{plotDownstreamSpctrm}).
The hard-spectrum microwave Haze is largely attributable to the synchrotron radiation from this electron population that is radiating on the volume-filling magnetic field in the base of R2. 

\item A zone corresponding to the compression shell that covers, in projection, the entire solid angle of the $\gamma$-ray Bubbles out to the contact discontinuity, i.e., R3.
In this zone, the adiabatically amplified magnetic field pervading the cooling gas is rather strong and regular.
Synchrotron emission  here
 is by i) the adiabatically-compressed, lower-energy primary electrons that have long enough loss time that they can travel all the way from the shock and ii) secondary electrons that result from hadronic collisions occurring in the denser shell gas.

Two important considerations for synchrotron emission from R3 are that 
i) given relative transport and loss times {\it microwave-emitting primary electrons accelerated at the shock never reach this zone} and 
ii) in this zone, there is a {\it mixture of electrons of different ages} as all shell electrons simply cool in situ 
(with the qualification that a fraction of the electrons apparently leak from R3 into the radio plumes, R4: see below).
The latter implies that, above some break energy, $E_\mathrm{bk}$, the integrated, steady-state shell electron spectrum of R3 is  
$dN_e/dE_e \sim E_e^{-3}$ \citep{Kardashev1962} implying a synchrotron spectrum above $\nu_\mathrm{bk}$ of $F_\nu \sim \nu^{-1}$.
This explains the steep, polarized, and rather uniform spectrum measured between 2.3 and 23 GHz over the solid angle of the radio lobes.
The break energy is implicitly defined by $t_\mathrm{c}(E_\mathrm{bk}) = t_\mathrm{FB}$ where $t_\mathrm{c}$ is the energy-dependent electron cooling time
(with cooling processes dominantly synchrotron radiation and IC for the relevant electron energy range) and
$t_\mathrm{FB}$ 
is the age of the Bubbles over which the electrons have been accumulated in the shell.
At even higher energies (for typical parameters) the spectrum suffers a further quasi-exponential cut-off due to the fact that higher-energy primaries can never reach this zone; 
this cut-off is softened by the presence of secondary electrons produced in situ which can supply microwaves even at these large distances from the shock \citep[cf.][]{Dobler2012b}.

\item A zone identified with the polarized radio plumes that extend beyond the boundaries of the north and south bubbles at high latitudes.
For this zone, a 1-to-1 correspondence between distance from the Bubbles' edges (the contact discontinuity) and electron age can be re-established as the electrons are systematically advected into the plumes.
We thus expect, in principle,
the position-dependent electron distribution take the form of the injection spectrum (already steepened to $\sim E_e^{-3}$) with an exponential cut-off that moves to progressively lower electron 
energy with increasing distance from the edges.
With only two spectral data points currently available (at 2.3 and 23 GHz), this cut-off translates into a steepening with distance into the plumes (matching observations).
\end{enumerate}

\subsubsection{Magnetic fields}
\label{sssctn_B}

To determine the intensity of synchrotron emission from the Bubbles we need to know the magnetic field amplitude in two environments: in the volume-filling plasma phase interior to the Bubbles, R2 in Figure~\ref{fig_BubbleSchematic}, and in the denser shell formed by local thermal instability, R3.
For the latter, as mentioned, we assume that the pressure of the adiabatically-compressed, non-thermal ISM components, i.e., cosmic rays + magnetic fields, reaches equilibrium with the  plasma pressure in the interior, $P_{\mathrm{FB}} = 2.0$ eV cm$^{-3}$ for fiducial parameters.
Given the finding \citep{Carretti2013} that CRs and magnetic fields are in rough equipartition in the Bubbles (and, therefore, in the shell adopting an adiabatic index for the tangled magnetic fields of $\gamma_B = 4/3 = \gamma_{\mathrm{CR}}$ ), then $U_\mathrm{B}^\mathrm{shell} \simeq 3/2  \ P_{\mathrm{FB}} \simeq $ 3 eV cm$^{-3}$ so that $B_\mathrm{shell}  \simeq 11 \ \mu$G.
This is in good agreement with the 
 magnetic field amplitude of $\sim 12 \ \mu$G suggested by an equipartition analysis of the radio data \citep[and supported by an independent broadband analysis that does not assume equipartition:][]{Carretti2013} to pertain in a
200-300 pc deep `sheath' enveloping the Bubbles.

With this determination, we may work backwards to arrive at the adiabatically `decompressed' field pertaining in the R2 
plasma, $B_\textrm{\tiny{FB}} =(n_\textrm{\tiny{FB}}/n_\textrm{\tiny{fil}})^{2/3}  B_\textrm{\tiny{fil}} \simeq   (L_{\mathrm{CR}}/\dot{E}_\textrm{\tiny{cool}}) \ B_\textrm{\tiny{fil}} \sim (5-7) \ \mu$G.
Again, this is consistent with a determination for the volume-filling field amplitude made on the basis on an equipartition analysis of 2.3 GHz radio polarisation data \citep{Carretti2013}.

\subsubsection{Synchrotron Emission from the Shock Downstream}

cosmic-ray transport downstream of the shock is a combination of the $\sim 100 -150$ km/s flow of the shocked, downstream plasma and cosmic-ray streaming along the field lines at the 
Alfven velocity \citep{Kulsrud1969,Skilling1971} of $\sim240$ km/s.
The range microwave-emitting electrons can reach in their loss time is then a function of magnetic field amplitude which controls both synchrotron losses and the Alfven velocity
and reaches a maximum for $B \simeq 10 \ \mu$G: see Figure~\ref{plotDownstreamCRRange}.
Here the cooling time, $t_c$, is the time taken for an electron injected at the accelerator with the maximal possible energy, $E_{max}$, to be cooled to the energy 
at which it
synchrotron radiates into the frequency range of concern, $E_c \equiv m_e \ c^2 \ \sqrt{\nu/\nu_g[B]}$, where $\nu_g[B]$ is the electron gyro radius in magnetic field $B$.
The energy of an electron injected at $E_0$ into a constant magnetic field $B$ evolves according to
\begin{equation}
E[t]  
=\frac{E_0}{1 + t/\tau[E_{0}]}\, ,
\end{equation}
where $\tau[E] = 1.3\times10^{8} (B/{\rm 10~\mu G})^{-2} ~(E/ {\rm ~GeV})^{-1} {\rm ~yrs}$,  ignoring non-synchrotron losses (particularly IC) for illustrative purposes
and  assuming pitch angle scattering on timescales short with respect to the loss times \citep[e.g.,][]{Reynolds2009}.

With this description of the energy evolution of the electrons, one obtains that an electron spectrum injected as a power law by the accelerator, $dN_e[E_0]/dE \equiv K E_0^{-s}$, evolves according to 
\begin{equation}
\frac{dN_e[E,t]}{dE} =  K E^{-s} (1-t/\tau[E])^{s-2}\, .
\label{eq_downstreamSpctrmElectrons}
\end{equation}
The evolution of a cosmic ray electron spectrum, injected with $s=2.1$, as the electrons are conveyed downstream of a shock and, simultaneously, suffer synchrotron + IC losses is illustrated in Figure~\ref{plotDownstreamSpctrm}.

Once the transport time exceeds the cooling time $t_c$  a cutoff feature is introduced into the electron spectrum. 
Applied to the outflow population, GeV IC emitting electrons only reach a few $\times 100$ pc past the shocks and
microwave-emitting electrons reach a few kpc.
 
\subsubsection{Synchrotron Emission from the Compression Shell}
  
\begin{figure}
\centerline{\includegraphics[width=1.0\textwidth]{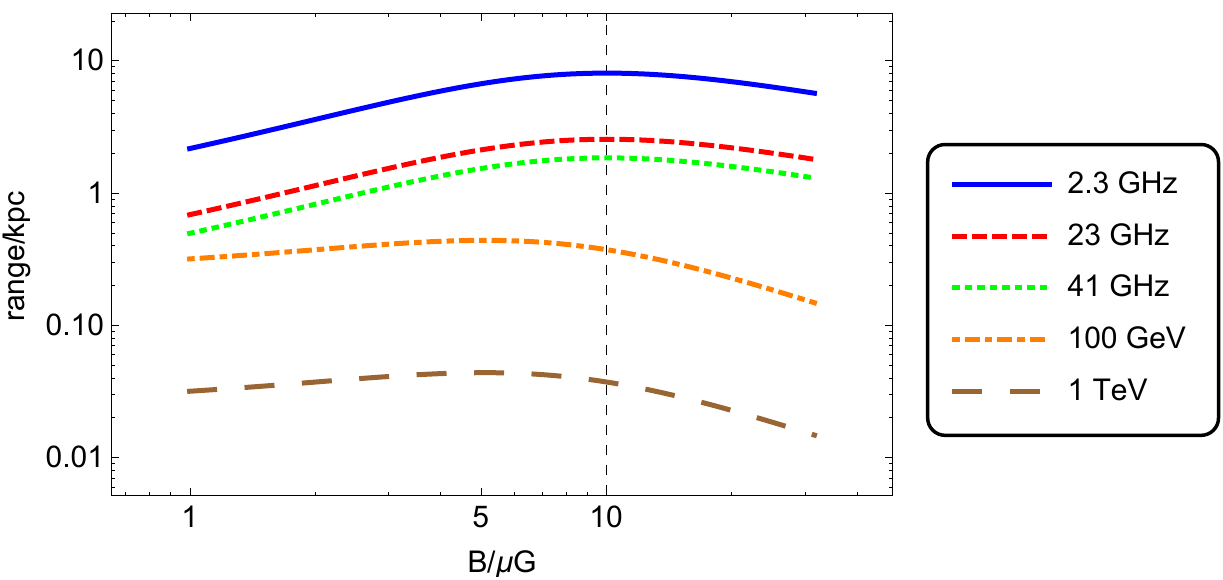}}
\caption{{\bf Ranges of cosmic-ray electrons downstream of the shock}  that emit synchrotron radiation into the specified characteristic frequency or with energies 100 GeV
and 1 TeV (dominating IC emission) as a function of downstream region magnetic field.
The range is determined by the condition that $t_\mathrm{transport} = t_\mathrm{c}$ with both synchrotron and IC losses accounted for.
Here the downstream gas velocity is assumed to be a fixed 112 km/s ($\sim 1/4 \ v_\mathrm{w}$) and the cosmic rays are taken to stream at  the Alfven speed in the specified field and for a density of $\rho = 1.4 \ n_H \ m_p$ where we adopt  $n_H = 3 \times 10^{-3}$ cm$^{-3}$.
Note that Haze electrons reach a maximum $\sim 3$ kpc downstream from the shock and 2.3 GHz electrons can reach a maximum $\sim 10$ kpc; electrons IC 
emitting 1-100 GeV $\gamma$-rays are concentrated close to the shock.
The vertical line shows the $\sim 10 \ \mu$G field where the range of the radio and microwave-emitting electrons is approximately maximised.
The total vertical extent of the Bubbles is $\sim8$ kpc with the shocks forming at heights $\sim$ 1 kpc; note that 2.3-GHz-emitting electrons can reach the edges of the Bubbles whereas microwave-emitting electrons do not; from this analysis the vertical extent of the microwave should be $\sim R_\mathrm{sh} +$ range[23 GHz] $\sim 3$ kpc; this is consistent with observations that show a drop-off in the intensity of the microwave emission at $\sim 35^\circ$ corresponding to the distance of 2.7 kpc for the near-surface of the Bubbles at 5 kpc distance.
}
\label{plotDownstreamCRRange}

\end{figure}

\begin{figure}
\centerline{\includegraphics[width=1.0\textwidth]{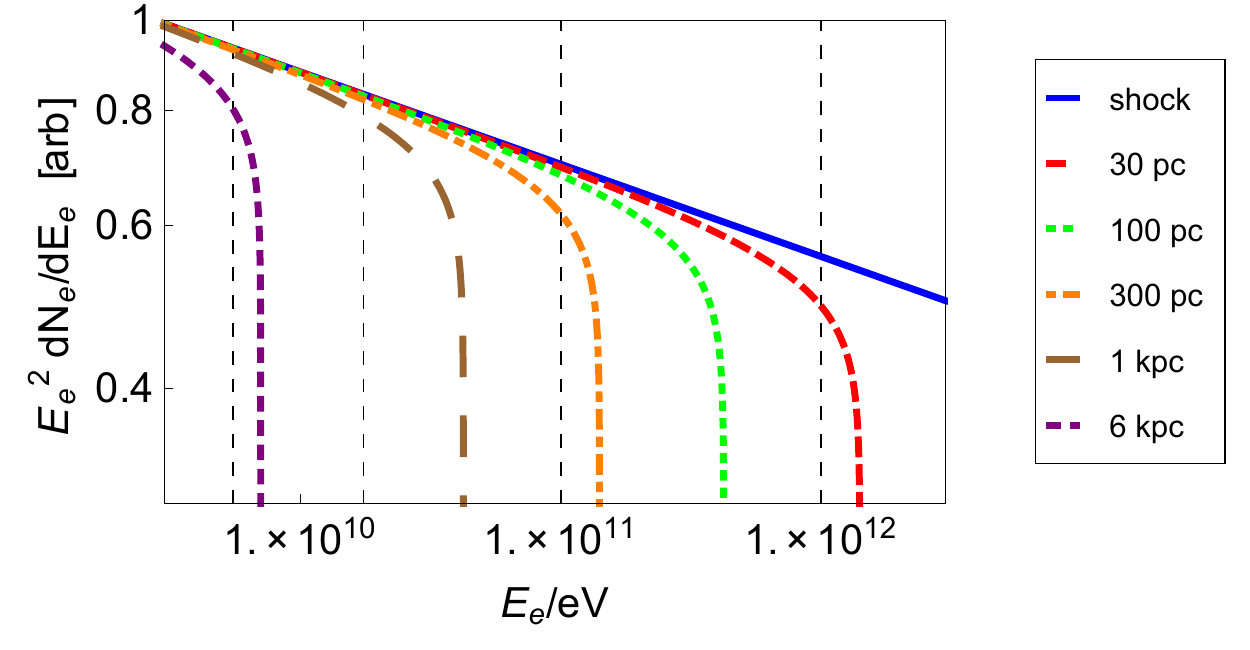}}
\caption{{\bf Spectra of cosmic-ray electrons at various distances downstream of a shock.} 
The electrons are injected at the shock with an assumed distribution $dN_e/dE_e \propto E_e^{-2.1}$ and are conveyed downstream at a total speed given by the downstream flow speed (112 km/s) + streaming at the Alfven speed (240 km/s).
At the same time, they suffer synchrotron  (with $B = 7 \mu$G) and IC losses ($U_{ISRF} = 1.5$ eV cm$^{-3}$); the position-dependent spectrum evolves according to Eq.~\ref{eq_downstreamSpctrmElectrons}.
From left to right, the vertical dashed lines indicate the primary electron energy corresponding to i)  synchrotron emission at 2.3 GHz; ii) synchrotron emission at 23 GHz; iii) 100 GeV IC emission off  starlight; and iv) 10 GeV 
 IC emission off the CMB.
 Note that, for these (realistic) parameters, IC emission is concentrated within $<$ kpc of the shock; in contrast 23-GHz emitting electrons extend $\sim$ 2 kpc, and 2.3-GHz emitting electrons can reach the full extent of the Bubbles.
Note there is a one-to-one correspondence between distance from shock and time since acceleration (electron `age') with the implication that, provided the emission is resolved, 
nowhere does the spectrum steepen by the canonical -1 due to the synchrotron + IC losses, rather it maintains its hard injection spectral index up to some distance/time-dependent energy above which  it is cut-off.
}
\label{plotDownstreamSpctrm}

\end{figure}

In contrast, lower-energy $\sim$GHz-emitting electrons survive longer, reach the shell before the contact discontinuity and are  accumulated inside it 
 (cf.~Figure~\ref{plotDownstreamCRRange}).
Compression of the cooling plasma in this shell leads to adiabatic amplification of the frozen-in magnetic field and cosmic rays, implying a second re-energization of those electrons that actually survive to this point.
The spatially-coincident, enhanced gas and cosmic ray ion populations close to the Bubbles' edges or in the filaments also lead, however, to significant hadronic collisions which, in addition to producing hadronic $\gamma$-ray emission, also inject secondary electrons (and positrons) via the decay  of collisionally-produced charged mesons.
The edges/filaments are thus populated by both enhanced primary and secondary electron populations in addition to adiabatically-compressed magnetic fields. 
The electrons do not escape from these regions but lose their energy in situ to synchrotron radiation.
Accumulation and trapping of electrons in this region leads to mixing of cosmic ray  electron populations of various ages.
Cooling + mixing (in projection) of differently-aged electron populations will be enhanced if there are downdrafts towards the edges of the Bubbles or convection within it.
Mixing steepens the overall steady-state population into an $\propto E_e^{-3}$-type distribution \citep{Kardashev1962}.
The steep-spectrum, polarized emission detected at 2.3 GHz \citep{Carretti2013} is largely attributable to this population that radiates 
in the enhanced magnetic field region associated with the cooling shell towards the edges of the Bubbles (and possibly also swept up halo magnetic field outside the CD).
We can thus evaluate the 2.3 GHz emission from the Bubbles in a thick target, single-zone model; the emission so calculated is displayed in   Figure~\ref{plotSynchrotronGiantShocks} (green curves).

A prediction of our model is the existence of a break in the shell radio spectrum corresponding to where the  synchrotron + IC cooling of shell electrons 
becomes less than the age over which they have been accumulated, roughly the age of the Bubbles
$t_\mathrm{FB}$.
The break occurs at a frequency of
\begin{equation}
\nu_{\rm br} = 1.96 \times 10^8 \, {\rm Hz} \left( \frac {B}{10\  \mu \rm G} \right)^{-3} 
\, \left( \frac {t_\mathrm{FB}}{10^8 \, \rm yr} \right)^{-2} \nonumber \, .
\end{equation}
The fact that the 23 to 2.3 GHz spectrum of the radio lobes is $\propto \nu^{-1.1}$ with the break necessarily {\it below} 2.3 GHz implies corroborative evidence 
that the outflow structures have a  age of $> 2.9 \times 10^7$ yr $(B_\mathrm{shell}/(10 \ \mu G) )^{-3/2} \ (\nu_{\rm br}/(2.3 \ \rm GHz))^{-1/2}$.

\subsubsection{Synchrotron Emission from the Radio Plumes}

Within the solid angle of the cooling shell that covers the Bubbles, mixing of electrons of different ages implies a rather uniform, cooled spectrum.
Note, however, that if some low-energy electrons are able to escape from the Bubbles entirely, leaking out in a systematic flow, 
a position dependent steepening can be re-established.
This expectation is borne out by the radio + microwave phenomenology: the spectrum of {\it polarized} intensity between 2.3 and 23 GHz is rather constant at $F_\nu \propto \nu^{-1.1}$ within the solid angle of the Bubbles (accounting for depolarisation effects at 2.3 GHz close to the plane) but steepens going into the radio plumes
that extend beyond the edges of the Bubbles.

\subsubsection{Energetics of Synchrotron Emission and Secondary Electrons}

\begin{figure}
\centerline{\includegraphics[width=1.0\textwidth]{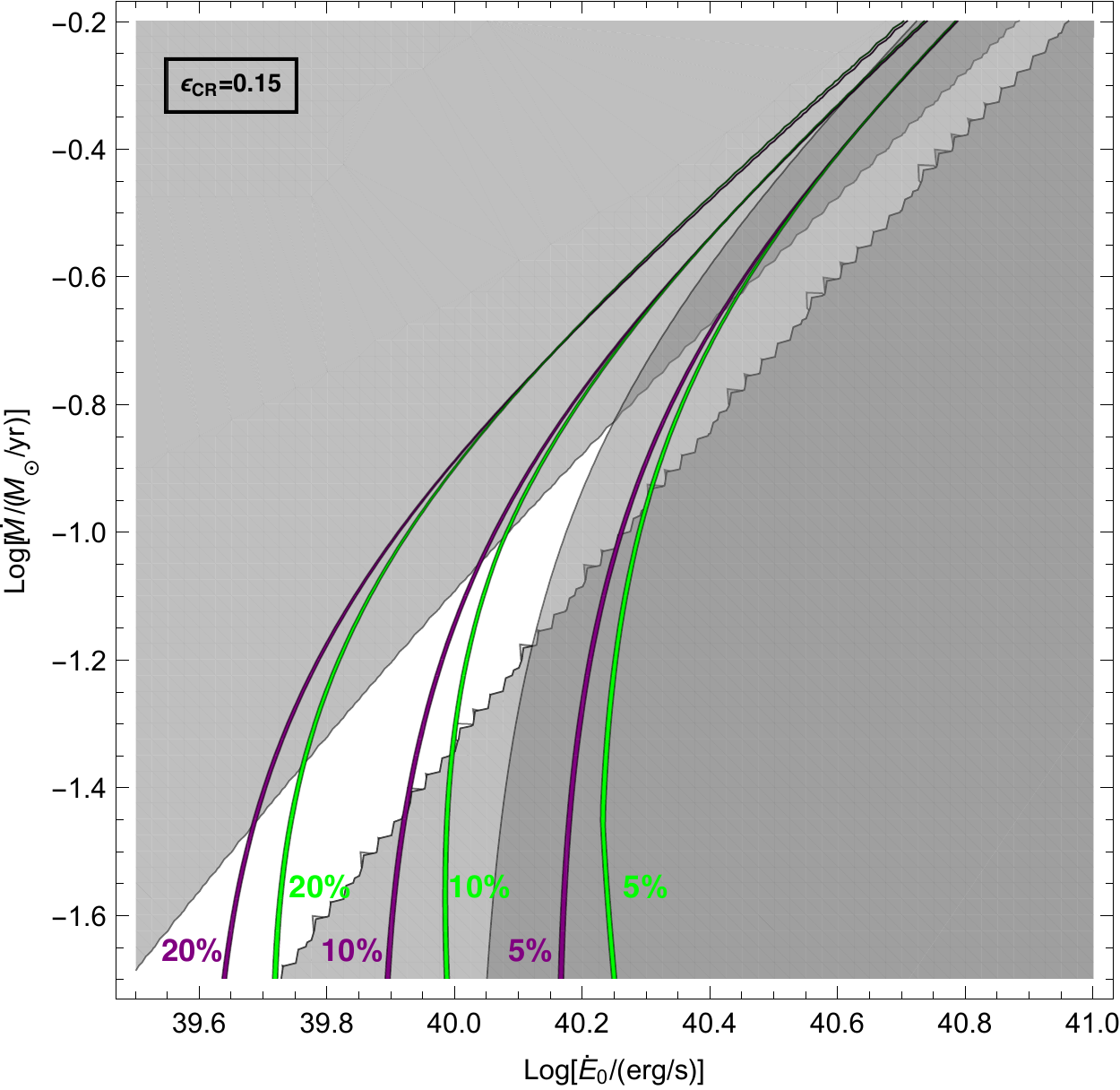}}
\caption{{\bf Predicted synchrotron emission} for  $P_{\mathrm{halo}}= 2 $ eV cm$^{-3}$, $\epsilon_{\mathrm{CR}} = 0.15$ and $T_\mathrm{shell} = 3.5 \times 10^6$ K. 
Regions masked out in gray   do not obey all the
physical constraints as described in Figure~\ref{fig_plotNTPw}.
Purple curves refer to the microwave Haze and green curves to the giant, 2.3 GHz radio lobes.
The curves are labelled according to where the 
modelled
2.3 GHz and 23 GHz intensity matches that observed for the nominated percentage 
of total cosmic-ray power going into freshly accelerated electrons, viz. $L_e = \{5\%,10\%,20\%\} \times L_{\mathrm{CR}}$ where $L_{\mathrm{CR}} \equiv \epsilon_{\mathrm{CR}} \dot{E}$ with $\dot{E}$ the mechanical power arriving at the shocks.
The curves account for both primary and secondary electron synchrotron;  here, secondary synchrotron accounts for 5-20\% of the total 2.3 GHz intensity
over the favored parameter space.
The purple curve for emission over the Haze region assumes 
a volume-filling magnetic field of  7 $\mu$G and an interstellar radiation field calculated at 3 kpc above the plane 
\citep[$U_\textrm{\tiny{ISRF}} = 1.7$ eV cm$^{-3}$:][]{Porter2008}; we adopt reference values for the Haze intensity  at 23 GHz of 1200 Jy/sr \citep{Su2010} and 18.5 kJy/sr  for the total intensity at 2.3 GHz  \citep{Carretti2013}.
}
\label{plotSynchrotronGiantShocks}

\end{figure}

For the magnetic field amplitudes investigated,
our calculations of synchrotron emission show that
a power going into freshly accelerated electrons of
5-20\% the total cosmic-ray power at the shock is required to reproduce the observed radio and microwave emission, i.e., $L_e \simeq (0.05-0.2) \times L_{\mathrm{CR}}$.
 Here $L_{\mathrm{CR}}$, the power going into all cosmic rays at the shock, is assumed to be a fixed fraction of 15\% of the total mechanical power available at the shock.
Fig.~\ref{plotSynchrotronGiantShocks} shows the 2.3 GHz and microwave synchrotron emission of both primary and secondary electrons. 
Within our favored parameter space, secondary electrons generated in the filaments explain 5-20\% of the Bubbles' 2.3 GHz luminosity.

\section{Connections to Observation}
\label{sctn_Evidence}

\subsection{Observational Evidence for Giant Shocks in the Bubbles}

\subsubsection{South Bubble}
\label{sctn_blah}

The S-PASS 2.3 GHz polarisation observations reveal a large-scale, linear depolarization
structure (strictly, a region of low polarisation intensity) in the southern bubble.
The feature extends in a south-east direction from the vicinity of the western edge of the south bubble at $(l,b) \sim (350^\circ,-17^\circ)$ to the vicinity of the eastern edge of this bubble at $\sim(9^\circ,-32^\circ)$ \citep[see the supplementary material of][section 4.1 and  Figure~S6]{Carretti2013}.
There are counterparts to this feature evident in $\gamma$-rays \citep[where the feature appears as an edge particularly evident in  10-30 GeV Fermi data; cf.][Figure~3a also Ruizhi Yang, private communication]{de Boer2014} and in H-$\alpha$ 
emission\footnote{Alex Hill, private communication, 2014.}.
At 2.3 GHz, this feature is likely the result of cancellation between intrinsically polarized synchrotron emission on two perpendicular, line-of-sight-superposed magnetic field structures.
One of these is the front side sheath of the south bubble whose gross magnetic field structure extends in a roughly south west direction as evident from the 2.3 and 23 GHz polarisation data 
\citep[reflecting the rotation of the bubble base in the same sense as the Galaxy and conservation of angular momentum in the expanding outflow:][]{Carretti2013}.
The cancelling field must run south east along the long axis of the depolarisation feature.

We believe that a compelling explanation of this feature is that it represents the alignment of magnetic field lines with the shock front in the southern bubble (i.e., there is a perpendicular magnetic field configuration at the shock).
Adopting a rough distance to the near side of the bubble of 5 kpc, the z-height of the depolarisation feature is 1-3 kpc, similar to the shock location
suggested by our analysis in \S\ref{s:sctn_ReverseShocks}.

\subsubsection{North Bubble}
\label{sctn_NB}

\begin{figure}
\centerline{\includegraphics[width=0.7\textwidth]{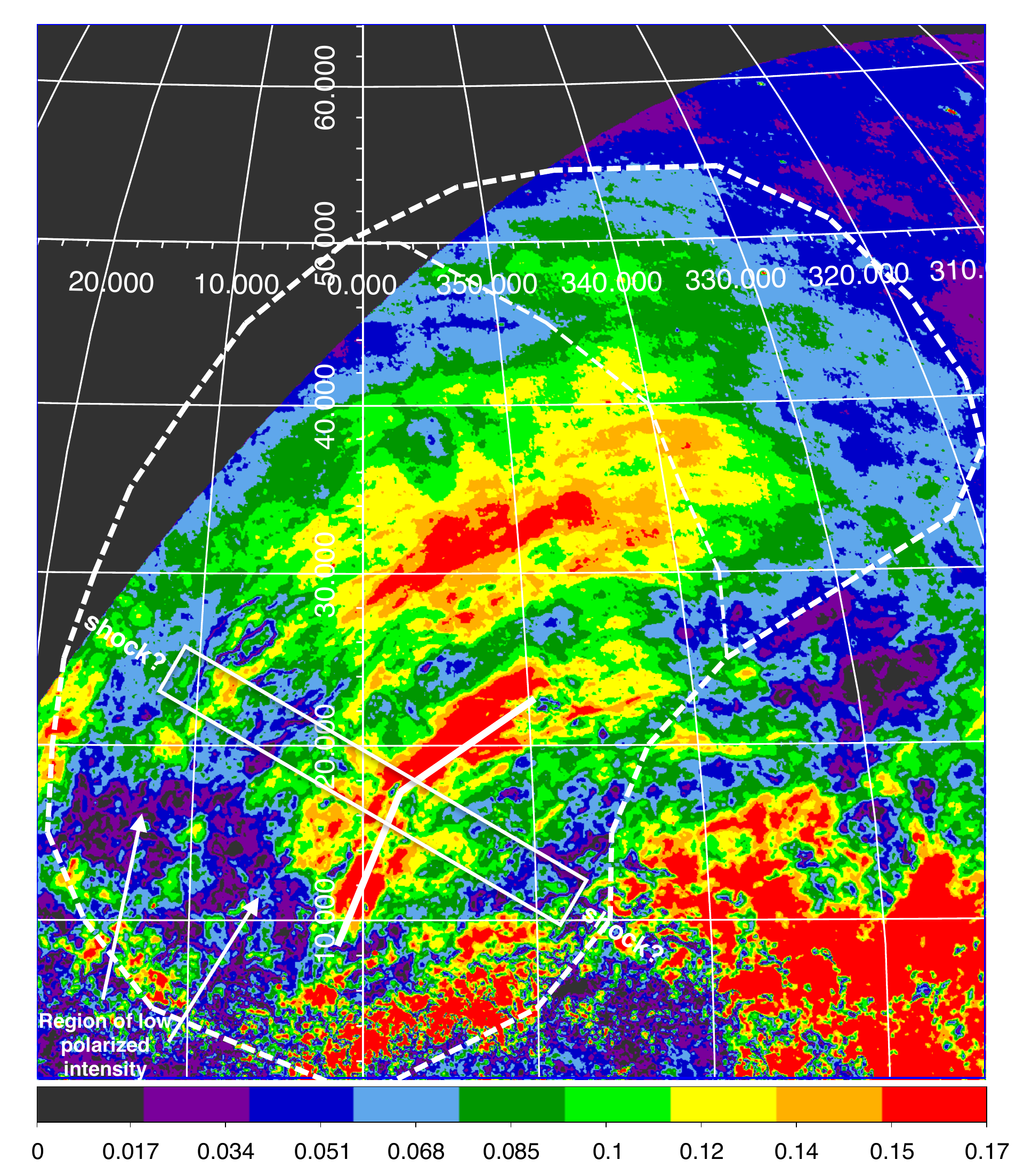}}
\caption{
{\bf Inflexion in the GC Spur in the northern bubble} traced in white.
colors show 2.3 GHz polarisation data from S-PASS  \citep{Carretti2013} in Galactic coordinates with east to the left and the edge of the 
black triangle to upper left corresponding to the instrumental horizon.
The color bar units are Jy/beam (convolved beam is approximately $10'$).
The outer dashed contour delimits the northern S-PASS polarized radio lobe \citep{Carretti2013}; the northern Fermi Bubble \citep{Su2010} edge is contiguous except in the north west where it runs inside the boundary of the radio lobe as shown by the thinner dashed line.
The large, inclined rectangle indicates the position of a faint depolarisation feature, running from north-east (top left) to south-west (bottom right) through the inflexion point, that might be attributed  to a magnetic field aligned with the surface of the giant reverse shock.
A region of low polarized intensity at 2.3 GHz  in the lower south east is indicated; this region is also evident at 23 GHz.
}
\label{fig_GC Spur}
\end{figure}

In the northern bubble there is only a faint feature possibly corresponding to the depolarisation feature described above (see Figure~\ref{fig_GC Spur}).
The north, however, is highly confused and suffers from depolarisation at 2.3 GHz significantly off the plane due to 
$\rho$ Ophiucus and other
local, high latitude gas structures.
There is, however, a feature of the northern bubble that is highly suggestive of a shock.
The Galactic Center Spur  \citep[GCS:][see Figure~\ref{fig_GC Spur}]{Sofue1989,Carretti2013} is a highly-collimated total and polarized intensity radio continuum feature that extends in a north-west direction from just east of the Galactic Center to Galactic latitudes of $b \sim 25^\circ$.
\citet{Carretti2013} have claimed this feature wraps around the forward surface of the northern bubble, an interpretation we adhere to.

On the basis of a geometric analysis of the overall curvature of this structure, \citet{Carretti2013} also claimed, however, that the Spur provides evidence that gas is being driven aloft into the Bubbles at a speed of $\sim 1000$ km/s up to heights of at least a few kpc. 
This analysis was supported by an independent argument that, if the 2.3-GHz-emitting electrons were accelerated (exclusively) in the nucleus, they would have to be transported  at a similar speed in order that they
reach the top of the Bubbles in their loss time.
This high velocity over the entire Bubbles
is not consistent with our present model which suggests that while gas may be injected into the base of the bubbles at $\gsim 500$ km/s, it is slowed to 100-150 km/s upon encountering the giant shocks at heights of $\sim$ 1 kpc.

Moreover, the assumptions leading up to the estimated $\sim 1000$ km/s presented by \citet{Carretti2013} are no longer met in our revised scenario.
In particular, we now expect injection of high energy electrons at large scales in the Bubbles either via direct acceleration at the large scale shocks or as secondaries.
The shock in the north bubble changes our expectation for the geometry of the GCS.
In fact, we expect an inflexion point in the GCS at the shock as the vertical speed of the flow is slowed by $\sim 1/4$ crossing the shock.
Such an inflexion does seem to be present in the GCS and it has the correct geometry: at $b \sim 17^\circ$ 
the angle between the long axis through the GCS changes from $\sim 17^\circ$ to  $\sim 48^\circ$
 (see Figure~\ref{fig_GC Spur}) suggesting a change of speed $\sim \tan[17^\circ]/\tan[48^\circ] = 0.27$, consistent with a strong shock.

\begin{figure}
\centerline{\includegraphics[width=0.7\textwidth]{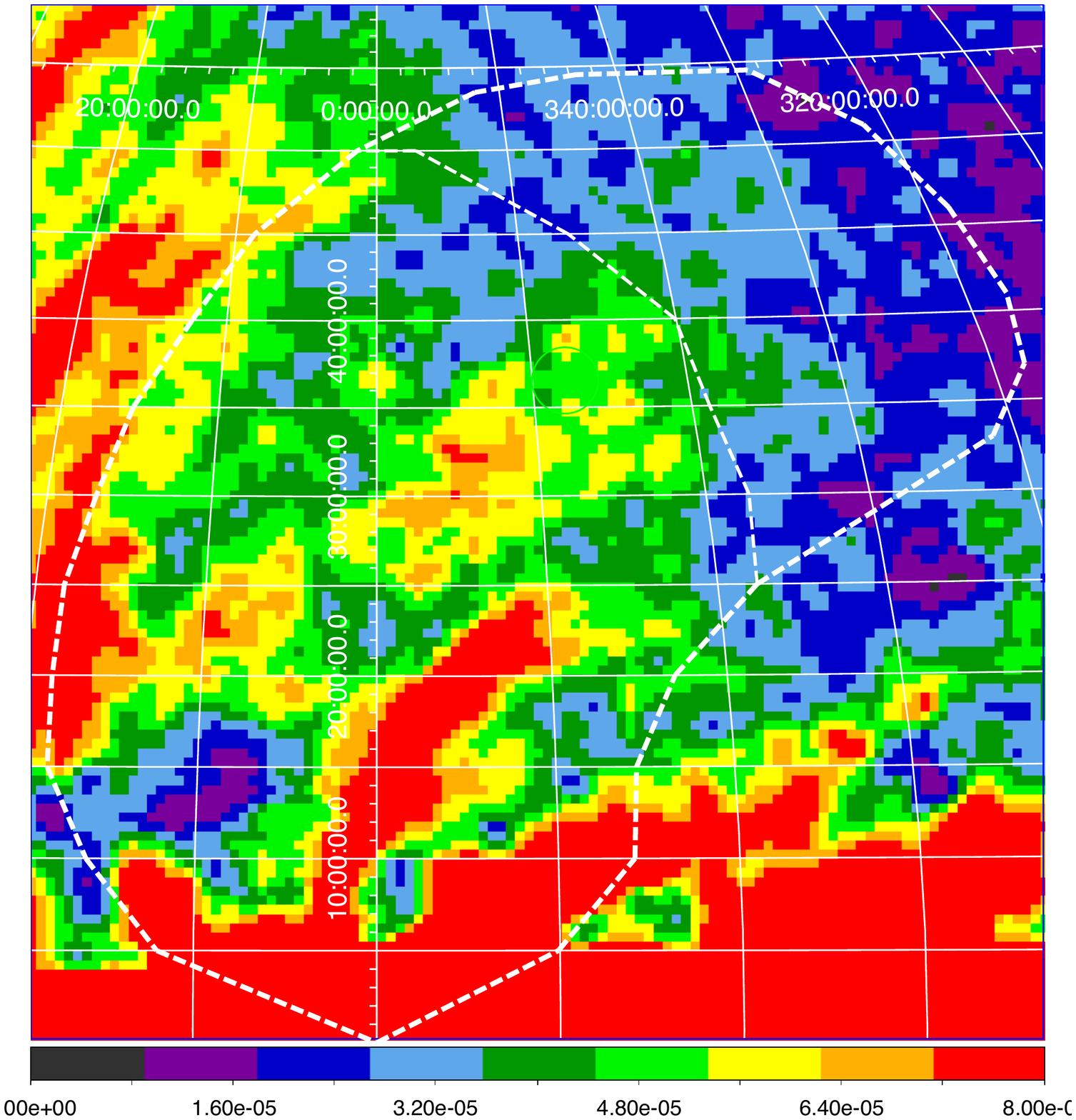}}
\caption{
{\bf Northern radio plume in polarised microwave emission} using 23 GHz data from WMAP\citep{Hinshaw2009} in Galactic coordinates with east to the left.
The color bar is in units of brightness temperature in Kelvin.
The GC Spur is evident, as is the same region of low, polarized intensity in the south east evident at 2.3 GHz that we suspect is below the latitude of the reverse shock.
}
\label{fig_WMAP}
\end{figure}

Furthermore, an independent determination \citep{Vidal2014} of the spectral index of the polarized  emission between 23 and 33 GHz from the GCS reveals a {\it hardening} at higher latitudes for the radio spectral index from
($b \gsim 18^\circ$) from $1.52 \pm 0.51$ to $0.62 \pm 0.44$ (with a similar hardening in a region coincident with part of the eastern edge of the north bubble),  
consistent with a giant shock -- at the correct height above the plane -- reaccelerating electrons as we have described above.

Another piece of corroborative evidence is support of this interpretation of the shock geometry in the north is the following: Consideration of Figure~\ref{fig_GC Spur} 
shows that the south-east region of the north bubble is comparatively dim in 2.3 GHz polarized emission.
While, in general, depolarization due magnetized plasma along the line of sight may affect emission up to latitudes of $\sim 10^\circ$, this region extends to $\sim 20^\circ$ close to the eastern edge.
Moreover, the very same depolarization region is evident also at 23 GHz \citep[cf.~fig,~3 of ][]{Carretti2013}, a frequency high enough to be completely unaffected by Faraday 
depolarization in this region of the sky.
We expect, therefore, that this lack of detected polarized emission betokens a lack of intrinsic emission along this line of sight.
Significantly, this dim region is below the putative shock feature.
This can be naturally explained if, on the one hand, the high-energy primary electrons requisite to produce the emission do not reach these heights from the nucleus and, on the other, are only reaccelerated to microwave-emission energies at the shock that is further from the nucleus.

\subsubsection{Why are the putative shock features not parallel to  the plane?}

Of course, our scenario does not explain why the reverse shock surfaces fail to  parallel  the Galactic plane but two interesting (and not incompatible) explanations for this phenomenology
present themselves:
\begin{enumerate}

\item As already remarked, the long axes of the Bubbles do not run perpendicular to the plane in either hemisphere but rather appear to be pushed towards Galactic west; the radio polarisation data  accentuate this impression with long plumes of emission extending beyond the edge of the $\gamma$-ray bubbles at high latitudes  towards Galactic north-west and south-west. 
Whatever the origin for this apparent distortion, the southern and northern shocks seem to run more-or-less perpendicular to the inferred flow.

\item Currently star-formation activity in the Galactic center is concentrated east of $l = 0$; there is a preponderance of molecular gas on this side of the region and Sagittarius B2, 
one of the Galaxy's most active star-forming giant molecular clouds, resides there \citep{Morris1996}.
Moreover, of the three large stellar clusters in the region, 
one surrounds the super-massive black hole but the other two, Quintuplet and Arches, are located to the east side.
The Quintuplet, at $\sim 4$ Myr old is likely to have already hosted a number of very massive core-collapse supernovae 
and there is tantalising evidence that some such supernovae have  occurred \citep{Sofue2003}.
So it may well be that, even if on the average, injection of mechanical energy is evenly distributed east and west of the GC, it is now predominantly coming from the east, leading to a localised increase in the outflow velocity that tends to push  the shocks further into the halo on this side.
This localization of the  mechanical energy injection associated with star formation may be the origin of the ridge features evident from polarization data on the forward surfaces of both north and south bubbles \citep{Carretti2013}.

\end{enumerate}

We will provide a more detailed and quantitative analysis of this putative shock feature in future work.
Lastly here, we note that the putative southern and northern bubble shock features are remarkably mirror symmetric (as are the overall $\gamma$-ray Bubbles and radio lobes+plumes).

\subsection{Observational Evidence for Condensations Coincident with the Bubbles' Edges} 
\label{ssctn_Cndsnts}

In \S \ref{sctn_intro} we described UV absorption evidence \citep{Keeney2006,Zech2008} for higher altitude warm ionised material in the Galactic bulge that is participating in a nuclear fountain flow.
We believe that this material is  the end state of gas condensing in the Bubbles' shell.
This warm ionised material seems to be concentrated towards the edges of the Bubbles, just as required if the volumetric $\gamma$-ray emissivity is to peak here and as expected for our cooling shell model.
\citet{Keeney2006} determine  that their clouds are likely launched into a conical outflow that is collimated into a cylinder of radius $\sim 1.6$ kpc; with this (somewhat uncertain) radius determination they find that their clouds  reach maximum heights $z_\mathrm{max} \simeq 12 \pm 1$ kpc.
Given the qualitative resemblance between this geometry and that of the Bubbles, we note that, were the radius 
used in the 
  \citet{Keeney2006} cloud trajectory calculations  re-scaled to 3 kpc\footnote{The original radius determination from \citet{Keeney2006} assumes both
  that  the kinematically distinct clumps of absorbing material they detect along the sightline to Mrk 1383 will rise to the same maximum height and 
  that they are on the front and back 
intersections of this sightline with the surface of the assumed cylindrical outflow. 
Relaxing either or both of these assumptions, the radius of the outflow could be different,  
in general, from 1.5 kpc.}, the clouds would be inside the Bubbles or the radio plumes.

\citet{Zech2008} have measured the metallicity of the fountain material found by them to be significantly super-solar metallicity ($\sim 1.6 Z_\odot $).
It is, thus, a very good prospect for having been launched from the GC.
Moreover, the illuminating background UV source is a PAGB star in the Messier 5 globular cluster at known distance (7.5 kpc) and height (5.3 kpc) above the plane which,
 in projection, is coincident with the edge of the northern bubble.
It seems more than plausible that the cloud is  located in the compression shell (R3) and  proffers a lower limit to the radius of the northern Bubble of $\sim$ 2.7 kpc (at z = 5.3 kpc), entirely consistent with the $\sim 3$ kpc radius we have adopted.

\subsection{High Speed Entrained Material at the Base of the Northern Bubble}

\citet{Fox2014} have recently reported on further UV absorption observations along the sightline to the quasar PDS456 $(l,b) =(10.4^\circ, +11.2^\circ)$, which passes through the conically-expanding
 lower part of the northern bubble at a height of $\sim 2.3$ kpc.
These observations find red and blue-shifted warm ionised material  which
kinematic analysis suggests  is moving at $\sim 900$ km/s, 
likely accelerated by the ram pressure of the dynamically dominant `wind fluid', i.e., the volume-filling, hot plasma phase. 
Given this supersonic velocity, this sightline must pass somewhere below the giant shocks.
Indeed, the northern shock position we infer from the radio polarisation observations  discussed in \S\ref{sctn_NB} is safely above the sightline.
The $\sim 900$ km/s speed inferred by \citet{Fox2014} is somewhat higher than the flow speeds indicated by our model 
and the heights somewhat above where we predict the shocks to lie on the basis of modeling the global flow parameters.
It is tantalising, however, that this sightline is on the east side of the north bubble where, observationally, the shock seems to have been pushed furthest from the plane.
This could be a spatially and temporarily-localized upwards excursion in the outflow velocity connected to the recent concentration of star-formation activity on the east side of the Galactic Centre as already discussed.
To obtain a shock $\sim 2.3$ kpc height with an upstream flow requires $\dot{E}_0/\dot{M} \simeq 5 \times 10^{15}$ erg/g or a 
mass loading factor $\beta \simeq 1.7$, only a little smaller than the range of $\sim$2-4 we find for this parameter 
in the allowed parameter space from our modeling of the global outflow.

\subsection{Tensions with X-ray data?}

One possible tension between our model and X-ray observations is the following:
Modeling of data taken by {\it Suzaku} in high latitude scans across the the limbs of both the north and south bubbles \citep{Kataoka2013}
suggest a significantly {\it sub}-solar metallicity $Z \simeq 0.2 Z_\odot$.
This is  contrary to the simple expectation were this material to be supplied by the nucleus (in which case it would be expected to be at least solar in metallicity).
Interestingly, the warm plasma 
found by \citet{Zech2008} along the Messier 5 sightline, relatively close to the {\it Suzaku} northern scan, is super-solar in metallicity as already remarked.
One possible explanation for this apparent tension is that the {\it Suzaku} observations  predominantly reveal emission from  
swept-up halo material on the other side of the contact discontinuity which various pieces of evidence \citep[as summarised by][]{Miller2014} suggest has a metallicity of $\sim 0.3 Z_\odot$.
The northern {\it Suzaku} scan does commence significantly outside the edges of the Bubbles, within the North Polar Spur feature; 
with 6 of the 8 northern pointings outside the $\gamma$-ray edge and 
with the metallicity measured only with confidence in the aggregate (and not for individual pointings), a systematic bias 
towards the metallicity of exterior gas would be expected in our picture that the $\gamma$-ray edge is identified with the contact discontinuity.

We also note here that the $\sim$300 km/s expansion speed of the Bubbles estimated by \citet{Kataoka2013} is also clearly in tension with our claimed expansion speed of only a $\sim$ few $\times$ km/s.
The 300 km/s estimate, however, is not  model-independent: It relies on the interpretation that the increase of the temperature of the plasma in the vicinity of the Bubbles' limbs with respect to the background halo plasma by a factor $\sim3/2$ can be explained by heating at a weak shock sweeping up the halo gas (which is assumed to be initially in hydrostatic equilibrium).
In our model, the plasma {\it inside} the contact discontinuity is heated by the reverse shocks low down in the nuclear outflows into both Galactic hemispheres; we do not specifically deal with the swept-up halo material here (which the Suzaku observations are biased towards, as already mentioned).

Another issue with the X-ray data may be the following:  these have been interpreted by  \citet{Kataoka2013} using a 3-component model incorporating i) an absorbed power-law isotropic background; ii)
a local (unabsorbed) single-temperature, collisional-ionization-equilibrium (CIE) plasma foreground (representing the local plasma bubble); and iii) an absorbed, single-temperature, CIE plasma that represents 
the signal region (the North Polar Spur and the limb regions in either the north or south bubbles).
Such a model may be too simple to adequately describe the putative cooling shell.
Indeed, \citet{Strickland2002} and \citet{Weaver2000} demonstrate that fitting to X-ray plasma emission from starburst outflows often leads to unphysically low metallicity estimates.
Possible circumstances leading to a too-low metallicity estimate are i)
the existence of  neglected multi-temperature plasma distributions; ii) having a large and poorly-constrained 
absorbing column; and/or iii) non-thermal contribution to the continuum (that boosts  apparent free-free emission relative to lines thus making plasma appear to have smaller metallicity).
Any or all of these may be relevant effects for the Bubble plasma, in particular, we expect the gas in the cooling shells to be multi-temperature as already discussed and
catastrophic cooling of the plasma into atomic phase may also generate absorbing material.

\subsection{Forward Shock/Shell?}

Mention of halo material swept-up ahead of the contact discontinuity leads us to consideration of the forward shock in our model.
In \S\ref{s:sctn_BubbleModel} we do not consider cooling in the region between the forward shock and the contact discontinuity.
The implication of this is that, with the attainment of pressure balance with the external atmosphere, the forward shock degenerates into a sound wave.
Even travelling at only the $\sim 280$ km/s sound speed, however, this  would have traveled $\sim 30$ kpc $t/10^8$ yr and, therefore be currently out of the picture given our estimated age for the Bubbles of $\gsim$ few $\times 10^8$ yr.
Yet, observationally, there are, in the northern Galactic hemisphere,  very-large-angular-scale X-ray and non-thermal radio and $\gamma$-ray structures that, at least in projection, 
are external to the Bubbles' edges and appear concentric to them.
One is tempted to identify these as shock or sound waves associated with the Bubbles
(e.g., Figure~4 of \citealt{Su2010}, Figure~1 of \citealt{Jones2012}.)

In  particular, the North Polar Spur \citep[NPS;][]{Bolton1950}
emerges north from the Galactic plane at  $l \sim 20-30^\circ$ and rises to heights above $b \sim 80^\circ$, merging with the Loop I radio structure \citep{Large1962,Haslam1982}.
It parallels the east edge of the north bubble over this extent; indeed, the NPS and Loop I, together, parallel the edge of the northern polarized radio lobe (that subsumes the north bubble) to a remarkable degree.
The NPS exhibits total intensity and polarized, non-thermal radio and microwave emission \citep{Sofue1979,Sun2014,Vidal2014}, thermal X-ray emission \citep{Willingale2003,Snowden1997},  
and $\sim$ GeV $\gamma$-ray emission \citep{Casandjian2009,Ackermann2014}.

It has long been argued on the basis of stellar polarization data that the NPS is a structure of the local ISM \citep[within a few $\times 100$ pc][]{Bingham1967,Berkhuijsen1971}.
However, a number of recent works have claimed  on the basis of various 
independent analyses of X-ray,  radio polarization, and other data
that it 
is actually a 
Galactic-scale feature \citep{Kataoka2013,Fang2014,Sun2014,Sofue2015,Tahara2015} or perhaps a confusion of two structures, one local and one at Galactic scales.
This finding complements the idea that (some part of) the NPS
is a signature of Galactic nuclear activity \citep{Sofue1977,Sofue2000,Bland-Hawthorn2003,Totani2006}
and perhaps even related directly to the Fermi Bubbles \citep{Guo2012,Kataoka2013,Mou2014}.
In addition to the NPS, there are linear $\gamma$-ray and polarized radio arcs that run concentrically between the NPS and the eastern edge of the northern bubble \citep{Su2010,Jones2012}; fainter radio and X-ray spurs  have also been claimed as the north western, south eastern, and south western complements of the NPS\citep{Sofue2000,Sofue2015,Tahara2015}.
We set aside further discussion of this interesting phenomenology  for a future publication.
We do, however,  make  the claim here that accounting, in a fuller model, for cooling in the shell of  material exterior to the contact discontinuity
implies a different channel for dissipation of the flux of internal energy  of  swept-up halo gas.
This retards the advance of the forward surface of this shell
(which, in the absence of cooling, would be the forward shock, degenerating into a sound wave with the attainment of pressure balance) leaving open the possibility
that this feature be identified with the NPS + Loop I feature.
Note that the question of why the north-eastern feature is so much more conspicuous than its north-west, south-east, 
or south-west analogues is not addressed in the current speculation.


\section{Discussion and Conclusion}
\label{sctn_Discussion}

In this paper we have set out what we believe is essentially the correct framework for understanding the  Galactic Center's giant outflows.
These reveal themselves in three disparate non-thermal phenomena: the $\gamma$-ray Fermi Bubbles, the microwave Haze, and the polarized `S-PASS' Radio Lobes.
Requiring only that the Bubbles contain the expected shocks and assuming that 
they are expanding into an atmosphere of finite pressure, $\sim 2$ eV cm$^{-3} \simeq 3.2 \times 10^{-12}$ dyn cm$^{-2}$ consistent with X-ray and other constraints, 
we have scanned over the parameter space for the nuclear energy and mass injection rates, $\{\dot{E}_0, \dot{M} \}$.
There are very few other adjustable parameters in our model:
\begin{enumerate}
\item The effective temperature of the compression shell, $3.5 \times 10^6$ K, is given by X-ray observations.

\item The solid angle of each conical outflow for which we adopt $\Omega = \pi$ as motivated by the observed geometry.

\item The metallicity of the outflowing gas which we have conservatively assumed to be solar but which,  plausibly, may be higher. We have shown that the dynamics around the inflation of twice solar bubbles is not very different to solar metallicity bubbles.

\item The fraction of mechanical energy delivered to the reverse shocks that ends up in cosmic rays, $L_\mathrm{CR}/\dot{E} \equiv  \epsilon_\mathrm{CR}$, 
we find falls in the range $0.05 \lsim \epsilon_\mathrm{CR} \lsim 0.3$ with a preferred value close to $\epsilon_\mathrm{CR} = 0.15$. 
These values are typical for determinations of the cosmic-ray acceleration efficiency of the Galaxy \citep[expressed as a fraction of the input mechanical power from supernovae:][]{Hillas2005,Strong2010},
in giant, star-formation-driven outflows from other systems \citep[e.g., NGC253:][]{Zirakashvili2006}, and for giant shocks in galaxy clusters \citep{Bruggen2012,Pinzke2013}.
 
\item 
The fraction of cosmic-ray luminosity, in turn, that goes to cosmic-ray electrons  which we expect lies in the range 
$0.05 \lsim L_e/L_\mathrm{CR} \lsim 0.2$ from consideration of the synchrotron phenomenology (adopting $\epsilon_\mathrm{CR} = 0.15$).
The $\gamma$-ray phenomenology indicates an even more restrictive range  $L_e/L_\mathrm{CR} \lsim 0.1$ lest the region 
$\lsim 300$ pc downstream of the reverse shocks appear too bright in IC $\gamma$-rays.
This range is, again, entirely consistent with experience from modeling non-thermal emission from cluster scale shocks  \citep[and references therein]{Pinzke2013,Brunetti2014}.

\end{enumerate}
Other inputs to our modeling  are the parameterization of the Galaxy's gravitational potential \citep{Breitschwerdt1991} and the plasma cooling function \citep{Dopita2013}.

Remarkably we have found that most of the $\{\dot{E}_0, \dot{M} \}$ parameter space can be excluded by measurements and/or physically-motivated constraints.
Within the region of parameter space that is {\it not} excluded, our model reproduces 
\begin{enumerate}

\item the $\gamma$-ray luminosity,  spectrum, and surface brightness distribution of the Bubbles, mostly via hadronic emission, but with a non-negligible IC contribution;

\item the luminosity, spectrum, and extent of the hard spectrum, total intensity microwave Haze;

\item the luminosity, spectrum, and extent of the steep spectrum, polarized $\sim$GHz S-PASS Lobes;

\item the extent, temperature, and density of the Bubbles' plasma;

\item the size of the Bubbles without fine-tuning;

\item and the mass flux and initial radius of the nuclear outflow and the plasma number density in the injection zone.

\end{enumerate}
This same, residual region of parameter space 
presents a compelling fit to a star-formation-driven system with  parameters that match the conditions in the nucleus.
From our modeling, presented in \S \ref{s:sctn_BubbleModel}, of a radiative bubble expanding into an atmosphere of pressure similar to that in the Galactic halo, 
for values of $\dot{E}_0$ and $\dot{M}$ selected by our reverse shock analysis, 
the asymptotic radius, temperature, and bolometric cooling luminosity of such a bubble are close to the observed values for the Fermi Bubbles.

Our model explains both the hard spectral indices of the CR electron population responsible for the Haze synchrotron emission and the CR proton and ion population responsible for the (hadronic) Bubble $\gamma$-rays and why, within uncertainties, {\it these are the same} at $\gamma \simeq 2.1$: these populations are both (re)accelerated at the giant shocks which, our modeling shows, are required to have  Mach numbers in the range 6-9 within the allowable $\{\dot{E}_0,\dot{M}\}$ parameter space.

Simultaneously, we explain the steep-spectrum, polarized radio emission that subsumes the Bubbles.
This  has not been adequately addressed in other models: this emission is coming from an aged ($\gsim 3 \times 10^7$ year old) and mixed population of electrons that is difficult to explain in models that postulate activity $\sim 1$ Myr ago to explain the Bubbles.
These electrons survive long enough to travel from the internal shocks out to the edge of the Bubbles where they radiate on the strong and rather regular magnetic field compressed into the cooling shell from the inside, supplemented by secondary electrons created in situ.
The phenomenon of mixing of low-energy electrons radiating at 2.3 GHz of  different ages implies a rather uniform $F_\nu \propto \nu^{-1}$ over the solid angle of the Bubbles.
Our model also allows, however, that for low-energy electrons that break out of the Bubbles a further, distance-dependent steepening is incurred because a systematic flow away from the Bubbles re-establishes a one-to-one correspondence between (reset) electron age and distance.
This detailed phenomenology is indeed reproduced by the polarized radio lobes which exhibit a rather constant $F_\nu \propto \nu^{-1.1}$ spectrum within the boundaries of the Bubbles but show a systematic steepening going from the Bubble interior into the radio plumes that extend beyond them \citep[cf.~Figure~S4 for the Supplementary Material of][]{Carretti2013}.

For fiducial parameters, our modeling from  \S \ref{s:sctn_BubbleModel} suggests that the Bubbles are at least $\sim 5 \times 10^8$ years old, very similar to the cooling time of plasma at inferred Bubble temperature and number density.
The Bubbles may be related to whatever Galaxy-scale process led to the formation of the central molecular zone gas torus which, at $\sim 3 \times 10^7 \msun$, 
has a very similar $H_2$ mass to the plasma content of the Bubbles. 
\citet{Sjouwerman1998} has claimed, on the basis of counting OH/IR stars in the inner 50 pc around the GC, evidence for a starburst in the GC $\gsim 1 $ Gyr ago 
which may be a credible event to initiate the inflation of the Bubbles (though we emphasise that their current non-thermal emission and apparent slow growth can be supported by the {\it current or recent} nuclear star-formation rate).

In theory, the Bubbles would reach their asymptotic final radius after $\sim$Gyr, but they may not survive this long.
Indeed, they may already have begun to blow-out as evidenced by the polarized radio plumes that extend significantly beyond their upper $\gamma$-ray edges \citep{Carretti2013}.
This process may be meditated by the Rayleigh Taylor instability with the buoyant fluid of cosmic rays and hot plasma breaking through the overlying shell of relatively denser shell material at the top of the Bubbles.
If so, the plumes might be expected to grow at the expansion speed, $v_\mathrm{exp} \sim 5$ km/s, implying a length scale $t_\mathrm{cool} \times v_\mathrm{exp} \sim 2.6$ kpc which is, indeed, similar to their observed extension.

In any case, the plume phenomenology is consistent with a hadronic origin for the Bubbles' $\gamma$-ray emission: in this scenario, although (lower energy) electrons, lower density plasma, 
and magnetic fields can leak out of the Bubbles and synchrotron radiate beyond them, leaking hadrons will not radiate (or only weakly) 
beyond the cooling shell/contact discontinuity because of a lack of dense target gas there.
(Given the ubiquitous CMB, TeV electrons would also radiate IC from the plumes were they present there, but, as emphasised, in our model these will be concentrated into a zone much further down in the 
Bubbles, close to their acceleration site in the giant shocks).
Before moving on from this topic,
 we  again 
flag the possibility that the overall extension of the Bubbles/radio lobes in a  direction west and away from the plane is essentially due to the motion of the Galaxy through the local group medium towards the group barycenter in the direction of Andromeda.

On the basis of our modeling we predict there should be significant mass in $\sim 10^4 - 10^6$ K gas -- likely exceeding $10^7 \msun$ -- concentrated towards the Bubbles' edges.
The cooler and denser phases of this gas will be raining down under gravity.

\subsection{Future Work}

In subsequent publications we will present or examine in more detail
the detailed spectral morphology of the synchrotron emission in the Bubbles,
the observational evidence for the giant reverse shocks we have inferred,  a detailed treatment of the dynamical production of the dense condensations, 
and a full treatment of the dynamic importance of cooling in the zone {\it forward} of the contact discontinuity and the possibility to identify the North Polar Spur
with the forward surface of a cooling shell of swept-up halo gas.
We will also examine the origin of the $\gamma$-ray spectral downturn at $\sim$ 100 GeV \citep{Ackermann2014} and the hardening of the
$\gamma$-ray spectrum at high latitudes \citep{Yang2014,Selig2014}. 
The former is not explicitly dealt with in our current model but may signify the energy where the Bubbles themselves cease to trap  cosmic-ray protons effectively; the latter may signify an important role 
for diffusive transport of cosmic rays in the upper reaches of the Bubbles \citep{Yang2014} which is, again, not accounted for in the current model.
Also awaiting a detailed treatment is the dynamical importance of the strong and coherent 
magnetic field that appears to be wrapped over the surface of the Bubble, particularly in stabilising this surface.

\subsection{Further Speculations}

For fiducial parameters, the giant shocks are capable of accelerating cosmic-ray protons to energies around the knee in the observed cosmic ray spectrum.
Given the very hard spectrum, a total power in these particles that approaches $10^{39}$ erg/s, and that the particles are delivered into the halo but not too far above the disk,
these shocks are interesting candidates \citep[cf.][]{Parizot2014} for the main accelerators of Galactic cosmic rays in this energy range  \citep[cf.][]{Cheng2012,Lacki2013,Taylor2014}.

Our model selects a region of the $\{\dot{E}_0,\dot{M}\}$  parameter space implying that, if star formation driven, the
outflow is mass loaded at a level $\beta \sim 2-4$, broadly consistent with observations of external, star-formation-driven outflows \citep[][and references threrein]{Strickland2009}.
Further, it is remarkable that 
our results indicate that the mass efflux in the outflow is $\dot{M} \sim 0.03 - 0.1 \msun$/yr, rather close to the 
{\it current} rate of nuclear star formation \citep[][and references therein]{Longmore2013,Crocker2012}, in turn close to the
long-time-averaged nuclear star formation 
rate\footnote{This can be rather directly and robustly determined from the fact that the total stellar mass of the nuclear bulge (the stellar population coincident with the central molecular zone) is $\sim 10^9 \msun$ which has been formed over a timescale $10^{10}$ years \citep[][]{Serabyn1996,Figer2004}.}.
Moreover, on the basis of the semi-analytic model presented in \S \ref{s:sctn_BubbleModel} we can say that the Fermi Bubble system is rather delicately poised in terms of allowable values of the ratio 
$\dot{E}_0/\dot{M}$: 
systems with $\dot{E}_0/\dot{M}$ ratios half those favored by our analysis asymptote to final radii within $\sim 10^8$ year that are smaller than those observed (and contain plasma that is too cold);
systems with twice the favoured value for this ratio (of $\sim 1.6 \times 10^{15}$ erg/g) do not asymptote to a final radius until $\sim 10^{10}$ year (and this radius is 10 kpc, significantly larger than the observed Bubbles) and spend only a brief time in the `vicinity' of the observed size (implying a return to a "Why now?'' fine-tuning); cf. Figure~\ref{f:radius}.

These observations suggest the operation of a negative feedback process to maintain the required delicate balance.
An obvious mechanism to achieve this -- completing the cycle --
is that material fountaining back to the plane in the vicinity of the nucleus will trigger further star-formation in this region.
Star-formation that is too vigorous or that produces a too large value of $\dot{E}_0/\dot{M}$ in the outflow (which may currently be happening on the east side of the nucleus) 
will tend to switch itself off because the outflowing plasma takes too long to condense and fall back from the downstream of the Bubbles.
The matter cycle will not be 100\% efficient, of course, but the system is not closed:
some amount of accretion to the nucleus will continue through the plane and, moreover, 
interaction and mixing between Bubble matter and the surrounding halo gas around the contact discontinuity may catalyse the condensation of halo gas
so that, overall, more gas fountains back to the plane than is lofted upwards \citep{Marinacci2010}, some of it low metallicity.

\section{Acknowledgements}

RMC is the recipient of an Australian Research Council Future Fellowship (FT110100108). 
RMC warmly acknowledges enlightening conversations and/or correspondence with Felix Aharonian, Joss Bland-Hawthorn,
Luke Drury, Christoph Federrath, Ed Jenkins, Federico Marinacci, Matt Miller, Christoph Pfrommer,  Robin Shelton,  Heinz V{\"o}lk, and Alex Wagner and thanks Doug Finkbeiner 
for first mentioning to him the possible implications of the Milky Way's motion through the local medium for distorting the Bubbles.
Ralph Sutherland is particularly thanked for supplying plasma cooling data generated by MAPPINGS IV code prior to public release, 
Alex Hill for providing information about H-$\alpha$ emission in the southern Bubble as observed by the Wisconsin H-Alpha Mapper (WHAM) southern sky survey \citep{Haffner2010}, 
and Ruizhi Yang for providing processed, preliminary maps of the {\it Fermi}-detected emission in the southern bubble.
We are grateful to the referee for a considered and helpful report.




\appendix

\section{Model of a spherical, radiative bubble}
\label{s:appendix}

In this appendix we develop our model for a spherical bubble (illustrated in Fig. \ref{f:bubble}) driven by the energy input provided by a star formation driven super-wind in the Galactic Center. In this model we do not assume that the bubble is highly over-pressured with respect to the ISM and we also incorporate cooling. We consider equations for energy, mass and momentum conservation in that order.

\paragraph{Energy equation.} Let $\epsilon_{\rm b}$ and $p_{\rm b} = (\gamma-1) \ ,\epsilon_{\rm b}$ be the internal energy density and pressure inside the bubble, where $\gamma = 5/3$ is the polytropic index. Also let $\rho_b$ be the bubble density between the reverse shock and the contact discontinuity (assumed spatially constant), $\bf v$ the internal velocity and 
$\dot{E}$ (in ergs s$^{-1}$) the rate of energy injection into the bubble. The cooling rate of gas with temperature $T$, is $\rho^2 \Lambda_\rho(T) = n^2 \Lambda[T] \> \rm ergs \> cm^{-3} \> s^{-1}$ where $\Lambda_\rho(T) = (\mu m)^{-2} \Lambda[T]$. $\Lambda[T]$ is the cooling function based on total number density $n$ and has been evaluated from the MAPPINGS emission code \citep{sutherland13a}. In the interior of the bubble there is a reverse shock where the ram pressure of the Galactic Center wind is of order the internal pressure of the bubble. The energy equation is derived by integrating the total energy equation over the volume, whose bounding surface comprises the reverse shock and the outwardly moving contact discontinuity. The result is:
\begin{equation}
\dt \int_V \left(\epsilon_b+ \frac{1}{2} \rho v^2 \right) \> d^3x + \int_S p_b u_i n_i \> dS 
= \dot E - \int_V \rho^2 \Lambda_\rho (T) \> d^3x
\label{e:bubble_energy}
\end{equation}
In this equation we make the usual assumption of neglecting the kinetic energy of shocked gas inside the bubble in comparison to the internal energy ($\rho v^2 \ll \epsilon$). Let $R_{\rm b}$ be the radius of the bubble, with volume $V_{\rm b} = 4\pi/3 R_{\rm b}^3$. We ignore the volume of the bubble inside the reverse shock. The energy equation (\ref{e:bubble_energy}) can then be cast as an equation for the bubble pressure:
\begin{equation}
\td{p_b}{t} = - 3 \gamma p_b \frac {1}{R_b} \td {R_b}{t} + \frac {3(\gamma-1)}{4 \pi} \frac {\dot E}{R_b^3} - (\gamma-1) \rho^2 \Lambda_\rho(T)
\label{e:pressure}
\end{equation}

\paragraph{Mass}
Let $\dot M$ be the rate of injection of mass into the bubble and let $\Delta \dot M$ be the rate of mass dropout associated with cooling. We estimate $\Delta \dot M$ by equating the enthalpy flux of hot gas associated with $\Delta \dot M$  with the energy loss via cooling. That is,
\begin{equation}
\Delta \dot M \times \frac {5}{2} \frac {kT}{\mu m} = \rho^2 \Lambda_\rho \times V_b
\end{equation}
Hence, the mass equation becomes:
\begin{equation}
\td {M_b}{t} = \dot M - \frac{2}{5} \frac {\mu m}{k T} \, \rho^2 \, \Lambda_\rho (T) \times \frac {4 \pi}{3} R_b^3 
\label{e:mass}
\end{equation}
The bubble density is given by:
\begin{equation}
\rho_b = \frac{3}{4 \pi} \frac {M_b}{R^3}
\label{e:density}
\end{equation}
 
\paragraph{Momentum.} Cooling in the hot, tenuous background medium is slow so that, in the early stages of expansion, the bubble does not form a thin shell, as in the classic case of an interstellar bubble \citep{castor75a,weaver77a}. We assume that the pressure and velocity in the shocked ambient medium are approximately constant. In this case the expansion of the contact discontinuity (i.e. the surface of the bubble) and the forward shock are governed by the conditions at the forward shock and matching the pressure of the shocked ISM to the bubble pressure. The resulting equations for the velocity of expansion of the bubble and the velocity of the external shock (radius $R_{\rm s}$) are, respectively:
\begin{eqnarray}
\td{R_{\rm b}}{t} &=& \left( \frac {p_{\rm a}}{\rho_{\rm a}} \right)^{1/2} \, 
\frac {(p_{\rm b}-p_{\rm a})/p_{\rm a}}{\sqrt{\gamma + (\gamma+1)p_{\rm b}-p_{\rm a})/2 p_{\rm a}}} 
\label{e:dRbdt} \\
\td{R_{\rm s}}{t} &=& \left( \frac {\gamma p_{\rm a}}{\rho_{\rm a}} \right)^{1/2} \> 
\left[ 1 + \frac {\gamma+1}{2 \gamma} \left( \frac {p_{\rm b}-p_{\rm a}}{p_{\rm a}} \right) \right]^{1/2}
\label{e:dRsdt}
\end{eqnarray}
For a large pressure differential $\Delta p = p_{\rm b} - p_{\rm a} \gg p_{\rm a}$,
\begin{equation}
\td {R_{\rm b}}{t} \sim \left( \frac {2}{\gamma+1} \frac {\Delta p}{\rho_{\rm a}} \right)^{1/2} 
\end{equation}
and
\begin{equation}
\td {R_{\rm s}}{t} \sim \left( \frac {\gamma+1}{2} \frac {\Delta p}{\rho_{\rm a}} \right)^{1/2}
=  \frac {\gamma+1}{2} \td {R_{\rm b}}{dt}
\end{equation}
When the pressure difference approaches zero,
\begin{equation}
\td {R_{\rm b}}{t} \rightarrow 0
\end{equation}
and
\begin{equation}
\td {R_{\rm s}}{t} \rightarrow \left( \frac {\gamma p_{\rm a}}{\rho_{\rm a}} \right)^{1/2}
\end{equation}
so that as the pressure differential tends to zero, the bubble stalls and the ambient shock propagates away from the bubble as a sound wave.

In developing a numerical solution for an expanding bubble, we integrate equations (\ref{e:pressure}), (\ref{e:mass}), (\ref{e:density}), (\ref{e:dRbdt}) and (\ref{e:dRsdt})  with input parameters the mass injection rate, $\dot M$, the energy flux $\dot E$, and the metallicity $Z$ (which affects the cooling). The solution starts at a time $t=10^5 \rm yr$ with initial values defined by the solution of these equations for a highly over-pressured bubble, for which negligible mass-loss and cooling has occurred. This solution is defined by:
\begin{eqnarray}
R_{\rm b} &=&  \left[ \frac{125}{264 \pi} \right]^{1/5}\, t^{3/5} \\
R_{\rm s} &=& \frac {4}{3} R_{\rm b} \\
p_{\rm b} &=& \frac {12}{25} \, \left[ \frac{125}{264 \pi} \right]^{2/5} \, t^{-4/5} \\
M &=& \dot M \, t \\
E_{\rm tot} &=& \frac {5}{11} F_{\rm E} \, t
\end{eqnarray}
where $E_{\rm tot}$ is the total energy within the bubble.

\end{document}